\documentclass[aps,reprint]{revtex4-2}
\usepackage{etoolbox} %
\usepackage{blindtext}
\usepackage{graphicx}
\usepackage{subcaption}
\usepackage{amssymb}
\usepackage{xcolor}
\usepackage{graphicx}
\usepackage{floatrow}
\usepackage[fleqn]{amsmath}
\usepackage[font=small,skip=3pt]{caption}
\usepackage{makecell}
\usepackage[version=4]{mhchem} %
\usepackage{hyperref}
\usepackage{cleveref}
\usepackage{url}
\usepackage{siunitx}
\makeatletter
\appto{\appendix}{%
  \@ifstar{\def\theequation@prefix{A.}}%
          {}%
}
\makeatother

\begin{document}
\title{Tritium-Lean Fusion Power Plants with Asymmetric Deuterium-Tritium Transport and Pumping}
\author{J. F. Parisi$^{1}$}
\email{jparisi@pppl.gov}
\author{S. Meschini$^{2,3}$}
\author{A. Rutkowski$^{4}$}
\author{A. Diallo$^1$}
\affiliation{$^1$Princeton Plasma Physics Laboratory, Princeton University, Princeton, NJ, USA}
\affiliation{$^2$Plasma Science and Fusion Center, Massachusetts Institute of Technology, Cambridge, MA, USA}
\affiliation{$^3$Polytechnic University of Turin, Turin, Piedmont, Italy}
\affiliation{$^4$Department of Astrophysical Sciences, Princeton University, Princeton, NJ, USA}

\begin{abstract}
Asymmetries in deuterium-tritium (D-T) particle transport and divertor pumping speeds are shown to enhance tritium self-sufficiency in fusion power plants. Using a diffusive particle transport model that links the plasma core, separatrix, and divertor regions, it is shown that reducing tritium transport while increasing deuterium transport improves both tritium burn efficiency and overall fusion power. By selectively increasing deuterium transport, tritium burn efficiency can be further optimized, assuming the availability of asymmetric D-T fueling and advanced divertor technologies. These asymmetries become especially beneficial at high tritium burn efficiency. In one example, by increasing the D-T particle diffusivity ratio and decreasing the D-T divertor pump speed ratio, each by a factor of five, the tritium burn efficiency increases eleven-fold from 0.026 to 0.29 at fixed fusion power. We propose a novel approach to achieve asymmetric D-T pumping using either an isotope separation and divertor re-injection approach or a partial ionization plasma centrifuge. In an ARC-class power plant, this approach could yield an order-of-magnitude improvement in tritium burn efficiency and/or increases in fusion power output. These findings motivate the development of techniques and technologies to reduce core tritium transport and increase tritium divertor pumping speeds.
\end{abstract}
\maketitle

\section{Introduction}

Forthcoming burning plasma experiments such as ITER and SPARC \cite{Aymar2002,Creely2020} mark significant milestones in demonstrating the feasibility of magnetic confinement fusion. However, these devices have not prioritized tritium fuel economy—an essential factor for commercial fusion power plants. In fact, ITER is projected to achieve a tritium burn fraction of only 0.36\% \cite{Abdou2021}. By contrast, a tritium self-sufficient deuterium–tritium (D–T) fusion power plant will require—and stand to benefit from—a substantially higher tritium burn efficiency (TBE) \cite{Abdou2021,Whyte2023,Meschini2023}.

TBE is defined as the ratio of the tritium burn rate to the tritium fueling rate \cite{Whyte2023}. Achieving a high TBE requires solving problems in multiple domains, including core, edge, and divertor plasma physics, the design and operation of fueling systems, plasma exhaust and pumping, and the fuel cycle. Optimizing tritium economy presents both physics and engineering challenges, where advances in one area can compensate for lack of progress in another.

Higher TBE offers several advantages. A key benefit is the reduction of on-site tritium inventory, which eases fuel cycle engineering requirements, thereby lowering capital costs and shortening development timelines. Simplified site licensing and regulatory processes are also expected with a smaller tritium inventory, and the reduced tritium inventory lowers the potential for accidental releases into the environment, improving public safety and minimizing ecological contamination risks. Additionally, lower startup inventories would mitigate the risk of global tritium shortages and could accelerate the deployment of fusion power \cite{Girard_2007,Humrickhouse_2018,Strikwerda_2024}. Higher TBE can reduce the required tritium breeding ratio (TBR) for tritium self-sufficiency \cite{Abdou2021,Meschini2023}, while lower tritium particle densities in the plasma core may decrease tritium retention in plasma-facing components and allow a wider range of materials to be considered for plasma-facing components.

In this work, we argue that is desirable for large asymmetries between deuterium and tritium to exist for particle fueling, transport, and pumping. We show how deliberately degrading deuterium particle transport and improving tritium particle transport improves the tritium economy of a fusion power plant, provided that sufficient deuterium fueling and pumping is possible in relatively high deuterium particle transport configurations. Therefore, usefully exploiting advances in particle confinement physics relies on advances in divertor pumping technology.

We also argue that departing from a 50:50 D-T fuel ratio -- typical for burning D-T fusion system designs \cite{Tanabe2013,Meschini2023} -- can be advantageous, if not necessary to obtain reasonably high TBE values. It has been suggested that operating a tritium-lean plasma can increase the TBE \cite{Boozer2021}, albeit with a loss in fusion power. Recently, it was suggested that tritium-lean plasmas with spin-polarized fuel could access high TBE without loss of fusion power \cite{Parisi_2024}. In this work, we take another approach, showing simultaneously high TBE and fusion power could be achieved by creating and exploiting asymmetries in D-T particle transport and divertor pumping speeds.

Studies of D-T plasmas in experiments \cite{Balet1993,scott1995isotopic,Keilhacker1999,Maggi2018,Horvath2021,Tala2023} and modeling \cite{Estrada-Mila:2005aa,Belli2020,Belli2021,Garcia2022} have shown that the deuterium and tritium heat and particle transport can be asymmetric, although the results for particle transport are far from definitive. For turbulent ion \textit{heat} transport, the gyroBohm heat diffusivity scaling -- $\chi_{\mathrm{s}} \sim \sqrt{m_s}$ for a species $s$ with mass $m_s$ -- has additional dependencies that scale as $1/\sqrt{m_s}$ once kinetic electron effects are included \cite{Belli2020}. For turbulent ion \textit{particle} transport, it was demonstrated that not only could the deuterium and tritium particle diffusivities differ, $D_{\mathrm{D}} \neq D_{\mathrm{T}}$, but under certain conditions they even have opposite signs, $D_{\mathrm{D}} D_{\mathrm{T}} < 0$ \cite{Belli2021}.

Fast ions, electromagnetic effects, $E \times B$ flow shear, and alpha channelling could also play an important role in D-T particle transport asymmetries \cite{Fisch1995,Belli2020,White2021,Garcia2022}; only in a very limited physical regime -- collisionless, electrostatic, single ion, adiabatic electron -- does the gyroBohm scaling hold \cite{Belli2020}. Given that burning plasmas will have significant fast and impurity ion populations and may have larger electromagnetic fluctuation amplitudes, it is not unreasonable to assume that significant differences in D-T particle transport could arise. Previous work optimized the D-T fuel composition for minimal outflow and exhaust of tritium \cite{Tokar2011}, finding a 2:1 D-T fuel mix was optimal. Recent work has suggested using deuterium flows to sweep out impurities from the plasma core and to provide high tritium confinement \cite{Boozer2024stellarators}, thereby enhancing TBE.

In this work, however, we are not concerned with the mechanisms giving rise to D-T asymmetries, but the consequences and opportunities should such asymmetries exist, leveraging results from \cite{Estrada-Mila:2005aa,Belli2021}. In order to benefit from improved tritium burn efficiency, it is necessary to have a divertor pumping system that preferentially removes tritium. This allows the deuterium divertor particle density to become much larger than the tritium density, which is correlated with TBE. These results build on those described in \cite{Whyte2023}, where it was shown that TBE improves significantly as the volumetric pumping speed for helium increases relative to hydrogen.

Asymmetries in pumping naturally arise from the working principles of the pumps. The difference in molecular masses between D$_2$ and T$_2$ leads to distinct mean thermal velocities in a D-T gas at temperature $T$ due to the $1/\sqrt{m_s}$ dependence. Since torus pumping operates across a range of flow regimes, from fluid to molecular, asymmetries in pumping speed between different species may emerge. Further asymmetries are introduced by the technological choices in pumping systems. For example, the differing sublimation pressures of D$_2$ and T$_2$ impact cryocondensation-based technologies \cite{souers1986hydrogen}. Additionally, when cryosorption is used, the use of sorbent materials introduces another asymmetry due to differences in sticking probability \cite{nalwa2001handbook}. Isotopic effects have also been observed in other pumping techniques, such as proton conductor pumps \cite{iwahara1999hydrogen} and vapor diffusion pumps \cite{teichmann2021particle}, although in these cases deuterium is generally faster to pump than tritium due to its lower mass. Given the benefits of asymmetric D-T pumping, there is strong incentive to explore additional techniques for its enhancement.

\begin{figure*}[bt]
    \centering
    \includegraphics[width=0.7\textwidth]{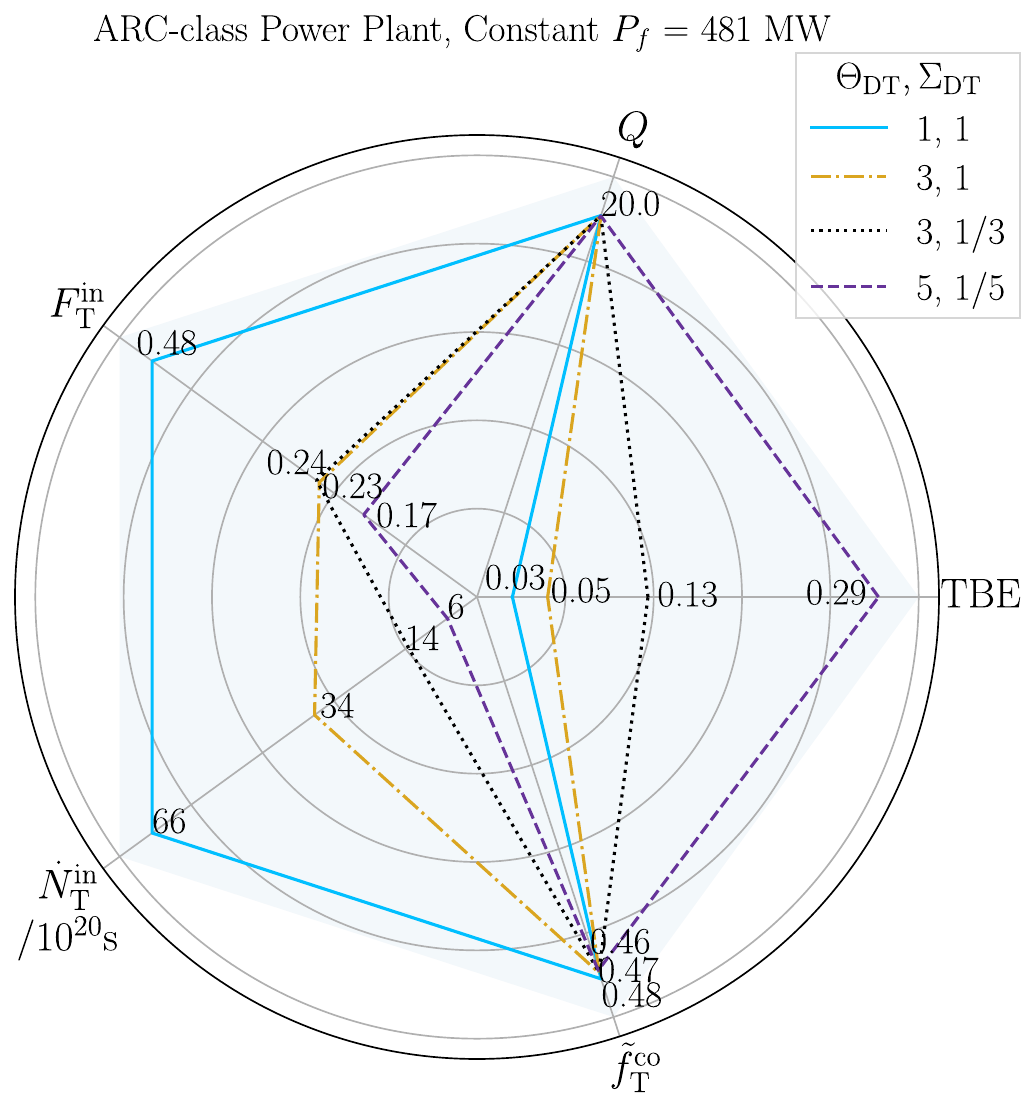}
    \caption{Tritium self-sufficiency and fusion power parameters of an ARC-class device for four cases with high fusion power $P_f = 481$ MW: The D-T transport ratio is $\Theta_{\mathrm{DT}}$ (\Cref{eq:ThetaDTratio}), the D-T pumping speed ratio is $\Sigma_{\mathrm{DT}}$ (\Cref{eq:SigmaDT}) the tritium flow rate injection fraction is $F_{\mathrm{T}}^{\mathrm{in}}$ (\Cref{eq:fueling_ratio}), the tritium fueling rate is $\dot{N}_{\mathrm{T} }^{\mathrm{in} }$, the plasma gain is $Q$, and the typical tritium core density fraction is $\tilde{f}_{\mathrm{T}}^{\mathrm{co}}$ (\Cref{eq:fTcotilde}). All axes have a value of zero at the origin and are linearly scaled.}
    \label{fig:arcclass_spider}
\end{figure*}

In \Cref{fig:arcclass_spider}, we summarize the main effects of variable D-T particle transport and D-T divertor pumping for an ARC-class device with a fusion power of $P_f = 481$ MW. Increasing the D-T particle transport ratio $\Theta_{\mathrm{DT}}$ and decreasing the D-T divertor volumetric pumping speed ratio $\Sigma_{\mathrm{DT}}$ broadly have similar benefits for increasing the fusion power -- this is shown by the progressively increasing $\Theta_{\mathrm{DT}}$ values and decreasing $\Sigma_{\mathrm{DT}}$ values in \Cref{fig:arcclass_spider}.  $\Theta_{\mathrm{DT}}$ and $\Sigma_{\mathrm{DT}}$ are defined in \Cref{sec:diff_part_transp,sec:diff_part_pumping} respectively. Going from the nominal case ($\Theta_{\mathrm{DT}} = 1, \Sigma_{\mathrm{DT}} = 1$) to the efficient case ($\Theta_{\mathrm{DT}} = 5, \Sigma_{\mathrm{DT}} = 1/5$) in \Cref{fig:arcclass_spider}, the tritium fueling rate decreases by a factor of 11 from $\dot{N}_{\mathrm{T} }^{\mathrm{in} } = 66 \times 10^{20} \SI{}{s^{-1}}$ to $6 \times 10^{20} \SI{}{s^{-1}}$ while maintaining fusion power. Overall, the TBE increases from 0.03 to 0.29.

The main takeaway from \Cref{fig:arcclass_spider} is that very high TBE plasmas could be enabled by physics and technology advances in improving and controlling $\Theta_{\mathrm{DT}}$ and $\Sigma_{\mathrm{DT}}$. This is discussed in detail in \Cref{sec:ultrahighTBE_regimes}.

The layout of this paper is as follows: In \Cref{sec:diff_part_transp}, we introduce the model and notation that includes the effects of asymmetric D-T transport. In \Cref{sec:diff_part_pumping}, we extend the model to also include the effects of asymmetric divertor pumping. In \Cref{sec:vacuum_pumping} we consider constraints from  vacuum technology and in \Cref{sec:current_technology} we consider the results in the context of current technology. We apply the model to an ARC-class power plant in \Cref{sec:ARC}, focusing on two distinct operating regimes with either high TBE values or high fusion power. We describe potential avenues for pumping technologies that could facilitate the D-T asymmetric scenarios in \Cref{sec:future_tech_develop}. We summarize in \Cref{sec:DISCUSS}.

\section{Differential Particle Transport} \label{sec:diff_part_transp}

In this section, we introduce the notation for different deuterium and tritium particle fluxes. We assume a three species plasma with deuterium, tritium, and electrons. In this work, we draw extensively from the model originally introduced in \cite{Whyte2023} and extended to variable tritium fraction in \cite{Parisi_2024}. Our analysis is applicable to the steady-state operation of magnetic confinement devices such as tokamaks and stellarators. The main quantities used in this work are listed in \Cref{tab:tab0}, in the appendix.

\subsection{Core Fuel Mix}

We first consider the effect of different deuterium and tritium particle transport on the core D-T fuel mix. The tritium particle flow rate through a flux surface is
\begin{equation}
\dot{N}_{\mathrm{T}}^{\mathrm{co} } = A \Gamma_{\mathrm{T}},
\label{eq:NdotTco}
\end{equation}
where $A$ is the surface area of a flux surface and $\Gamma_{\mathrm{T}}$ is the particle flux. We use a diffusive model for the particle flux where fluxes are driven by the density gradient,
\begin{equation}
\Gamma_{\mathrm{T}} = - D_\mathrm{T} \nabla n_{\mathrm{T}}^{\mathrm{co}}.
\label{eq:diffusivetransp}
\end{equation}
Here, $D_\mathrm{T}$ is a diffusion coefficient and $n_{\mathrm{T}}^{\mathrm{co}}$ is the tritium density. The tritium flow rate $\dot{N}_{\mathrm{T}  }^{\mathrm{co} }$ is therefore
\begin{equation}
\dot{N}_{\mathrm{T}  }^{\mathrm{co} } = - A D_\mathrm{T} \nabla n_{\mathrm{T}}^{\mathrm{co}}.
\label{eq:NTdot_co}
\end{equation}
It should be noted that non-diffusive processes are often important for particle transport \cite{Weiland1989,Mordijck2020}, which we have neglected in this analysis but consider in a forthcoming work. The total hydrogen particle flux is
\begin{equation}
\begin{aligned}
\dot{N}_{\mathrm{Q}  }^{\mathrm{co} } & = \dot{N}_{\mathrm{T}  }^{\mathrm{co} } + \dot{N}_{\mathrm{D}  }^{\mathrm{co} } \\ & = - A \left( D_\mathrm{T} \nabla n_{\mathrm{T}}^{\mathrm{co}} + D_\mathrm{D} \nabla n_{\mathrm{D}}^{\mathrm{co}} \right).
\label{eq:NQdot_co}
\end{aligned}
\end{equation}
Here, $\dot{N}_{\mathrm{D}  }^{\mathrm{co} }$, $D_\mathrm{D}$, $n_{\mathrm{D}}^{\mathrm{co}}$ are the core flow rate, diffusivity, and density for deuterium. Substituting $n_{\mathrm{D}}^{\mathrm{co}} = n_{\mathrm{T}}^{\mathrm{co}} \left(1-f_{\mathrm{T}  }^{\mathrm{co} }\right)/f_{\mathrm{T}  }^{\mathrm{co} }$, in \Cref{eq:NQdot_co}, where the tritium core density fraction is
\begin{equation}
f_{\mathrm{T}  }^{\mathrm{co} } \equiv n_{\mathrm{T}}^{\mathrm{co}}/n_{Q}^{\mathrm{co}},
\label{eq:ft_core_dens}
\end{equation}
where $n_{\mathrm{Q}}^{\mathrm{co}} = n_{\mathrm{D}}^{\mathrm{co}} + n_{\mathrm{T}}^{\mathrm{co}}$ is the hydrogen fuel density, gives
\begin{equation}
\dot{N}_{\mathrm{Q}  }^{\mathrm{co} } = - A D_\mathrm{T} \nabla n_{\mathrm{T}}^{\mathrm{co}} \left( 1 + \Theta_{\mathrm{DT} } \frac{\Lambda - f_{\mathrm{T}  }^{\mathrm{co} }}{f_{\mathrm{T}  }^{\mathrm{co} }}  \right).
\label{eq:NQdot_co_approx}
\end{equation}
Here, the core D-T particle diffusivity ratio is
\begin{equation}
\Theta_{\mathrm{DT} } \equiv D_\mathrm{D}/D_\mathrm{T},
\label{eq:ThetaDTratio}
\end{equation}
and the core tritium-to-hydrogen logarithmic density gradient ratio is
\begin{equation}
\Lambda \equiv L_{n,T}/L_{n,Q},
\label{eq:Lambda_main}
\end{equation}
where the logarithmic density gradient is $L_{n,T} \equiv - \left( \partial_r \ln n_{\mathrm{T} } \right)^{-1}$, which is a typical quantity used in gyrokinetic stability and transport analysis \cite{Catto1978,Frieman1982,Parra2008,Abel2013}. Another useful quantity is the tritium core flow rate fraction
\begin{equation}
F_{\mathrm{T}  }^{\mathrm{co} } \equiv \dot{N}_{\mathrm{T}}^{\mathrm{co}}/\dot{N}_{Q}^{\mathrm{co}}.
\label{eq:FTco_definit}
\end{equation} 
Using \Cref{eq:NTdot_co,eq:NQdot_co_approx}, \Cref{eq:FTco_definit} becomes
\begin{equation}
F_{\mathrm{T}  }^{\mathrm{co} } = \frac{ f_{\mathrm{T}  }^{\mathrm{co} }}{ \Theta_{\mathrm{DT}} \left(\Lambda - f_{\mathrm{T}  }^{\mathrm{co}} \right) + f_{\mathrm{T}  }^{\mathrm{co}} }.
\label{eq:FTco_ftco}
\end{equation}
In the limit of equal deuterium and tritium diffusivity and $\Lambda = 1$, \Cref{eq:FTco_ftco} reduces to $F_{\mathrm{T}  }^{\mathrm{co} } = f_{\mathrm{T}  }^{\mathrm{co} } $, which was the limit used in \cite{Parisi_2024}. In this work, we also use $\Lambda = 1$, but explicitly retain $\Lambda$ in subsequent equations. We show the effect of different $\Lambda$ values in \Cref{sec:approximation1}.

We do not explicitly impose the constraint from ambipolarity of the turbulent particle fluxes: $\sum_s Z_s \Gamma_s = 0$ for a sum over species $s \in [\mathrm{e}, \mathrm{D}, \mathrm{T}]$. Because ambipolarity must always be satisfied, we assume that $\Gamma_e$ may vary as we consider variations in $\Gamma_D$ and $\Gamma_T$. We discuss ambipolarity further in \Cref{sec:TBE_confinement}.

\begin{figure}[!tb]
    \centering
    \begin{subfigure}[t]{0.9\textwidth}
    \centering
    \includegraphics[width=1.0\textwidth]{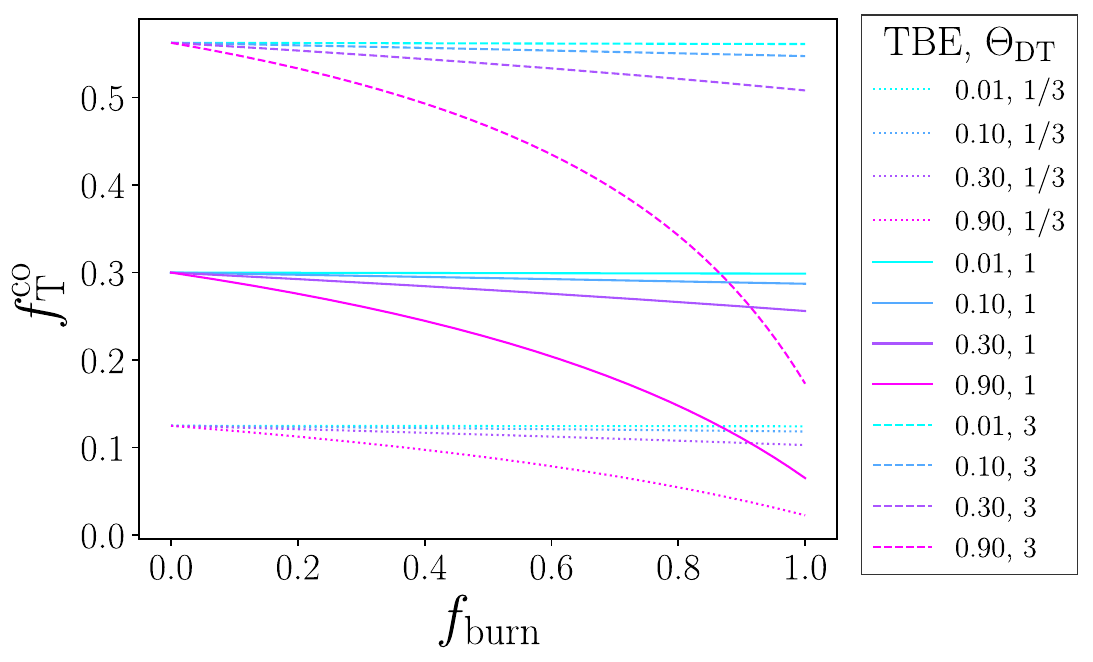}
    \caption{}
    \end{subfigure}
     ~
    \begin{subfigure}[t]{0.9\textwidth}
    \centering
    \includegraphics[width=1.0\textwidth]{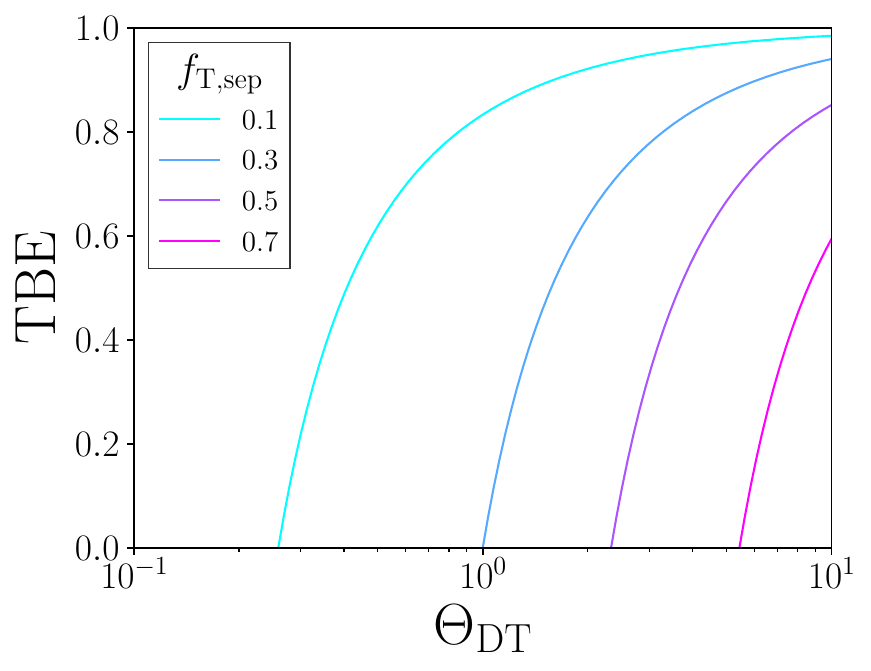}
    \caption{}
    \end{subfigure}
    \caption{(a) Tritium core fraction $f_{\mathrm{T}}^{\mathrm{co}}$ (\Cref{eq:fTco_def}) versus enclosed burn fraction $f_{\mathrm{burn}}$ for different tritium burn efficiency (TBE, \Cref{eq:TBE_early}) TBE versus $\Theta_{\mathrm{DT} }$ values. (b) TBE versus $\Theta_{\mathrm{DT} }$. Here, $\Lambda = 1$ and $F_{\mathrm{T}}^{\mathrm{in}} = 0.3$.}
    \label{fig:fTco_v_burn}
\end{figure}

\subsection{Tritium Burn Efficiency}

In this subsection, we relate core transport to tritium burn efficiency. 

The tritium fraction $F_{\mathrm{T}}^{\mathrm{in}}$ of the fueling rate is
\begin{equation}
F_{\mathrm{T}}^{\mathrm{in}} \equiv \dot{N}_{\mathrm{T}}^{\mathrm{in}}/\dot{N}_{Q}^{\mathrm{in}},
\label{eq:fueling_ratio}
\end{equation}
where $\dot{N}_{\mathrm{T}}^{\mathrm{in}}$ and $\dot{N}_{Q}^{\mathrm{in}}$ are the particle fueling rates for tritium and total hydrogen fuel. The tritium fraction $F_{\mathrm{T}}^{\mathrm{div}}$ of the unburned hydrogen fuel removal rate in the divertor is
\begin{equation}
F_{\mathrm{T}}^{\mathrm{div}} \equiv \dot{N}_{\mathrm{T}}^{\mathrm{div}}/\dot{N}_{Q}^{\mathrm{div}},
\label{eq:divertor_ratio}
\end{equation}
where $\dot{N}_{\mathrm{T}}^{\mathrm{div}}$ and $\dot{N}_{Q}^{\mathrm{div}}$ are the number of tritium and total unburned fuel particle removal flow rates. The tritium enrichment $H_{\mathrm{T}}$ measures the relative tritium flow rate in the divertor compared with the core,
\begin{equation}
H_{\mathrm{T} } \equiv F_{\mathrm{T}}^{\mathrm{div}}/F_{\mathrm{T}}^{\mathrm{in}}.
\label{eq:HTfirst}
\end{equation}
We now define $\dot{N}_{\mathrm{T}}^{\mathrm{co}}$ (\Cref{eq:NdotTco}) as the tritium flow rate through a flux surface that encloses a fraction $f_{\mathrm{burn}}$ of the total fusion power,
\begin{equation}
\dot{N}_{\mathrm{T}}^{\mathrm{co}} \equiv \dot{N}_{\mathrm{T}}^{\mathrm{in}} - f_{\mathrm{burn}} \dot{N}_{\mathrm{T}}^{\mathrm{burn}},
\label{eq:NTco_2}
\end{equation}
where $\dot{N}_{\mathrm{Q}}^{\mathrm{co}}$ also satisfies
\begin{equation}
\dot{N}_{\mathrm{Q}}^{\mathrm{co}} \equiv \dot{N}_{\mathrm{Q}}^{\mathrm{in}} - 2 f_{\mathrm{burn}} \dot{N}_{\mathrm{T}}^{\mathrm{burn}}.
\label{eq:NQco_2}
\end{equation}
\Cref{eq:NTco_2,eq:NQco_2} are derived in Appendix B of \cite{Parisi_2024}. Here, $\dot{N}_{\mathrm{T}}^{\mathrm{burn}}$ is the tritium burn rate across the entire plasma. By introducing the radial coordinate $f_{\mathrm{burn}}$, we have introduced a radial dependence to the problem with $f_{\mathrm{burn}}$: on the magnetic axis $f_{\mathrm{burn}} = 0$ and on the separatrix $f_{\mathrm{burn}} = 1.0$. \Cref{eq:NTco_2,eq:NQco_2} describe a model where all of the deuterium and tritium fuel is injected on the magnetic axis. We ignore other particle sources such as particles re-entering the plasma core from the scrape-off layer.
Dividing \Cref{eq:NTco_2} by \Cref{eq:NQco_2} gives
\begin{equation}
F_{\mathrm{T}  }^{\mathrm{co} } = F_{\mathrm{T}  }^{\mathrm{in} } \frac{ 1 - f_{\mathrm{burn}} \mathrm{TBE} }{1 - 2 f_{\mathrm{burn}} F_{\mathrm{T}  }^{\mathrm{in} } \mathrm{TBE} }.
\label{eq:FTco_new}
\end{equation}
Because $F_{\mathrm{T}  }^{\mathrm{in} }$ and TBE are scalars and $f_{\mathrm{burn}}$ is a flux function, $F_{\mathrm{T}  }^{\mathrm{co} }$ is also a flux function. The tritium burn efficiency (TBE) \cite{Whyte2023} is defined as 
\begin{equation}
\mathrm{TBE} \equiv \frac{\dot{N}_{\mathrm{T}}^{\mathrm{burn}}}{\dot{N}_{\mathrm{T}}^{\mathrm{in}}} = \left( \frac{\dot{N}_{\mathrm{T}}^{\mathrm{div}}}{\dot{N}_{\mathrm{He}}^{\mathrm{div}}} +1 \right)^{-1},
\label{eq:TBEzeroth}
\end{equation}
which physically represents the probability of a tritium atom undergoing fusion between its injection into the plasma and its removal through the divertor exhaust. Here, $\dot{N}_{\mathrm{He}}^{\mathrm{div}}$ is the helium ash removal rate.
Equating \Cref{eq:FTco_ftco,eq:FTco_new}, $f_{\mathrm{T}}^{\mathrm{co}}$ is
\begin{equation}
\begin{aligned}
& f_{\mathrm{T}  }^{\mathrm{co} } = \frac{\Lambda \Theta_{\mathrm{DT} } (f_{\mathrm{burn}} \mathrm{TBE} -1)}{ f_{\mathrm{burn}} \mathrm{TBE} \left(1 + \Theta_{\mathrm{DT} } \right) + 1 - \Theta_{\mathrm{DT} } - 1/ F_{\mathrm{T}}^{\mathrm{in}}}.
\end{aligned}
\label{eq:fTco_def}
\end{equation}
We plot solutions to \Cref{eq:fTco_def} in \Cref{fig:fTco_v_burn}(a), showing how $f_{\mathrm{T}}^{\mathrm{co}}$ decreases with increasing $f_{\mathrm{burn}}$, increasing TBE, and decreasing $\Theta_{\mathrm{DT}}$ (at fixed $F_{\mathrm{T}}^{\mathrm{in}}$).

\begin{figure*}[tb]
    \centering
    \includegraphics[width=0.97\textwidth]{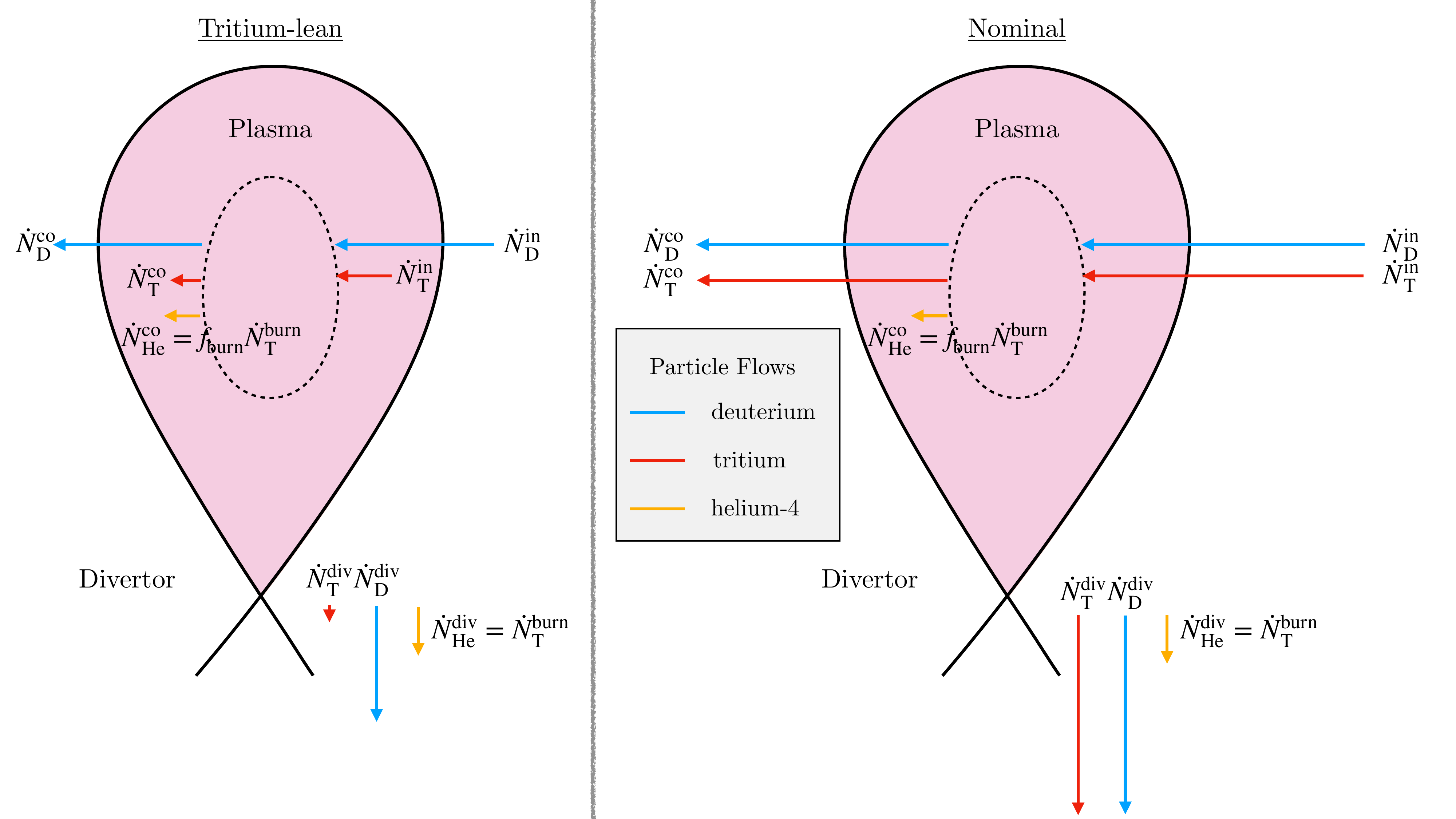}
    \caption{Schematic of the particle flows in a tritium-lean and a nominal configuration, both at equal total fusion power. The dashed flux surface in the plasma is described by the flux coordinate $f_{\mathrm{burn} }$ (\Cref{eq:NTco_2}). The arrows' magnitudes represent the size of the particle flows. In the tritium-lean case, the tritium flows are significantly smaller than the deuterium flows.}
    \label{fig:schematic_flows}
\end{figure*}

Defining the tritium density fraction at the separatrix,
\begin{equation}
f_{\mathrm{T}}^{\mathrm{sep}} \equiv f_{\mathrm{T}}^{\mathrm{co}}(f_{\mathrm{burn}}=1),
\label{eq:fTsep_bc}
\end{equation}
and rearranging \Cref{eq:fTco_def} with $f_{\mathrm{burn}} = 1$ for the TBE, we find
\begin{equation}
\mathrm{TBE} = 1 + \frac{f_{\mathrm{T}}^{\mathrm{sep}} \left( 1 - 2 F_{\mathrm{T}}^{\mathrm{in}} \right)  }{f_{\mathrm{T}}^{\mathrm{sep}} F_{\mathrm{T}}^{\mathrm{in}} \left( 1 + \Theta_{\mathrm{DT} } \right) - F_{\mathrm{T}}^{\mathrm{in}} \Lambda \Theta_{\mathrm{DT} } }.
\label{eq:TBE_early}
\end{equation}
Solutions to the TBE in \Cref{eq:TBE_early} are shown in \Cref{fig:fTco_v_burn}(b): decreasing $f_{\mathrm{T}}^{\mathrm{sep}}$ and increasing $\Theta_{\mathrm{DT} }$ are beneficial to the TBE.

In our model, the tritium particle diffusivity scales as
\begin{equation}
    |D_{\mathrm{T}}| \sim \frac{1}{A}\frac{\dot{N}_{\mathrm{T}}^{\mathrm{co}}}{\nabla n_{\mathrm{{T}}}^{\mathrm{co}}} = \frac{1}{A}\frac{\dot{N}_{\mathrm{T}}^{\mathrm{in}} - f_{\mathrm{burn}} \dot{N}_{\mathrm{T}}^{\mathrm{burn}}}{\nabla n_{\mathrm{{T}}}^{\mathrm{co}}}.
\end{equation}
Using the definition of the TBE,
\begin{equation}
    |D_{\mathrm{T}}| \sim \frac{1}{A}\frac{ \dot{N}_{\mathrm{T}}^{\mathrm{burn}} \left( \frac{1}{\mathrm{TBE}} - f_{\mathrm{burn}} \right)}{\nabla n_{\mathrm{{T}}}^{\mathrm{co}}},
    \label{eq:DTscaling}
\end{equation}
the relative diffusivities scale as
\begin{equation}
    \Theta_{\mathrm{DT}} \sim \left(1 + \frac{1}{F_{\mathrm{T}}^{\mathrm{in}}} \frac{1 -2 F_{\mathrm{T}}^{\mathrm{in}}}{1 - f_{\mathrm{burn}} \mathrm{TBE}} \right) \frac{L_{n_{\mathrm{D}}}}{L_{n_{\mathrm{T}}}} \frac{f_{\mathrm{T}}^{\mathrm{co}}}{1-f_{\mathrm{T}}^{\mathrm{co}}}.
    \label{eq:DTratio_TBE}
\end{equation}
Given that we are generally interested in $F_{\mathrm{T}}^{\mathrm{in}} < 1/2$ (in order to obtain high TBE), higher TBE with all other variables constant in \Cref{eq:DTratio_TBE} requires higher $\Theta_{\mathrm{DT}}$.
This is shown in \Cref{fig:fTco_v_burn}(b) where for fixed $f_{\mathrm{T}}^{\mathrm{sep}}$, $\Lambda = 1$, and $F_{\mathrm{T}}^{\mathrm{in}} = 0.3$, the value of $\Theta_{\mathrm{DT}}$ increases with TBE.
We will show in \Cref{sec:ThetaDTSigmaDTscan} that for fixed fusion power, increases in $\Theta_{\mathrm{DT}}$ that maximize the TBE come from $D_{\mathrm{T}}$ decreasing and $D_{\mathrm{D}}$ increasing by equal factors. %

Using the TBE definition in \Cref{eq:TBEzeroth}, we relate the TBE to $H_{\mathrm{T} }$, $F_{\mathrm{T}}^{\mathrm{div}}$, and $F_{\mathrm{T}}^{\mathrm{in}}$,
\begin{equation}
H_{\mathrm{T} } = \frac{1 - \mathrm{TBE} }{ 1- 2 F_{\mathrm{T}  }^{\mathrm{in} } \mathrm{TBE}  },
\label{eq:TBE_zerop5}
\end{equation}
where conservation of particles requires
\begin{equation}
\dot{N}_{Q}^{\mathrm{in}} = 2 \dot{N}_{\mathrm{T}}^{\mathrm{burn}} + \dot{N}_{Q}^{\mathrm{div}}
\label{eq:particle_conservation_Q}
\end{equation}
and 
\begin{equation}
\dot{N}_{T}^{\mathrm{in}} = \dot{N}_{\mathrm{T}}^{\mathrm{burn}} + \dot{N}_{\mathrm{ T}}^{\mathrm{div}}.
\label{eq:particle_conservation_T}
\end{equation}
A schematic for this system of equations is shown in \Cref{fig:schematic_flows}. The right figure corresponds to a nominal case where the deuterium and tritium flows are equal. The left figure corresponds to a tritium-lean case where the tritium flows are significantly reduced. Notably, we will see that the total deuterium flows also decrease in tritium-lean configurations.

\section{Differential Pumping} \label{sec:diff_part_pumping}

In this section, we couple core transport and divertor pumping to find expressions for the TBE and fusion power.

\subsection{Tritium Burn Efficiency}

The divertor flow rate for a species $x$ is given by the neutral gas density $n_{x}^{\mathrm{div}}$ and effective pumping speed $S_x$ \cite{Whyte2023},
\begin{equation}
\dot{N}_{x}^{\mathrm{div}} = n_{x}^{\mathrm{div}} S_x.
\label{eq:neutralpump}
\end{equation}
It will be helpful to define the helium-to-tritium pumping speed,
\begin{equation}
    \Sigma_{\mathrm{HeT}} = S_{\mathrm{He}}/S_{\mathrm{T}},
    \label{eq:Sigma_HeT}
\end{equation}
and the divertor density ratio
\begin{equation}
    f_{\mathrm{HeT,div}} = n_{\mathrm{He}}^{\mathrm{div}}/n_{\mathrm{T}}^{\mathrm{div}}.
    \label{eq:fHeTdiv}
\end{equation}
Using \Cref{eq:TBEzeroth,eq:Sigma_HeT,eq:fHeTdiv}, we obtain a simple expression for the TBE,
\begin{equation}
    \mathrm{TBE} = \frac{f_{\mathrm{HeT,div}} \Sigma_{\mathrm{HeT}}}{1 + f_{\mathrm{HeT,div}} \Sigma_{\mathrm{HeT}}}.
    \label{eq:TBEnew}
\end{equation}
\Cref{eq:TBEnew} shows that if $\Sigma_{\mathrm{HeT}}$ is held constant, the only way to change the TBE is to change the helium to tritium density fraction. As explained in \cite{Whyte2023}, the TBE increases monotonically with $f_{\mathrm{HeT,div}} \Sigma_{\mathrm{HeT}}$. In this paper, we hold $\Sigma_{\mathrm{HeT}}$ constant, and therefore any increase in TBE comes from increasing $f_{\mathrm{HeT,div}}$. This paper describes new ways to increase $f_{\mathrm{HeT,div}}$. Techniques for increasing $\Sigma_{\mathrm{HeT}}$ are left for future works, although the techniques for changing $\Sigma_{\mathrm{DT}}$ will likely also apply to $\Sigma_{\mathrm{HeT}}$.

Why does increasing $f_{\mathrm{HeT,div}} \Sigma_{\mathrm{HeT}}$ also increase the tritium burn efficiency? If the fusion power is constant,
\begin{equation}
    \begin{aligned}
        & f_{\mathrm{HeT,div}} \Sigma_{\mathrm{HeT}} \sim \frac{1}{\dot{N}_{\mathrm{T}}^{\mathrm{div}}} = \frac{1}{\dot{N}_{\mathrm{T}}^{\mathrm{in}} - \dot{N}_{\mathrm{T}}^{\mathrm{burn}}} \sim \frac{1}{\dot{N}_{\mathrm{T}}^{\mathrm{co}}} \\ 
        & \sim \frac{1}{D_{\mathrm{T}}} \sim \mathrm{TBE}.
    \end{aligned}
    \label{eq:usefulfHeTdivSigma}
\end{equation}
Therefore, increasing $f_{\mathrm{HeT,div}} \Sigma_{\mathrm{HeT}}$ corresponds to a decrease in tritium particle flux $\dot{N}_{\mathrm{T}}^{\mathrm{co}}$ due to a lower $D_{\mathrm{T}}$ value. Thus, increasing $f_{\mathrm{HeT,div}} \Sigma_{\mathrm{HeT}}$ does not causally increase the TBE, but rather, is a necessary divertor condition for reduced tritium particle transport in the plasma core.

\subsection{Fusion Power}

We now calculate the effect of differential D-T pumping speeds on the fusion power. The helium-to-electron core density ratio \cite{Whyte2023} is,
\begin{equation}
f_{\mathrm{dil}} \equiv n_{\alpha}^{\mathrm{co}}/n_{e}^{\mathrm{co}},
\label{eq:dil1}
\end{equation}
where $n_{\alpha}^{\mathrm{co}}$ and $n_{e}^{\mathrm{co}}$ are the core alpha and electron densities. Using quasineutrality without impurities,
\begin{equation}
n_{e}^{\mathrm{co}} = n_{Q}^{\mathrm{co}} + 2 n_{\alpha}^{\mathrm{co}},
\end{equation}
the power density at the flux surface $f_{\mathrm{burn} }=1/2$ is
\begin{equation}
\begin{aligned}
& \tilde{p}_f = 4 \tilde {f}_{\mathrm{T}}^{\mathrm{co}} (1- \tilde {f}_{\mathrm{T}}^{\mathrm{co}}) (1-2 f_{\mathrm{dil}})^2 \left( n_e^{\mathrm{co}} \right)^2 \langle v \overline{ \sigma} \rangle \frac{E}{4},
\end{aligned}
\label{eq:pfform2}
\end{equation}
where $n_{e}^{\mathrm{co}}$ is the electron density, $\langle v \overline{ \sigma} \rangle$ is the fusion reactivity, and we have defined
\begin{equation}
\tilde{f}_{\mathrm{T}}^{\mathrm{co}} \equiv {f}_{\mathrm{T}}^{\mathrm{co}} \left( f_{\mathrm{burn}} \equiv 1/2  \right),
\label{eq:fTcotilde}
\end{equation}
so $\tilde{f}_{\mathrm{T}}^{\mathrm{co}}$ describes the flux surface that encloses half of the power fusion power, $f_{\mathrm{burn} } = 1/2$. Thus, at fixed $n_{e}^{\mathrm{co}}$ and temperature, the fusion power density $\tilde{p}_f$ relative to its maximum value $p_{f,max}$ where $p_{f,max}$ has $\tilde{f}_{\mathrm{T}}^{\mathrm{co}} = 1/2, f_{\mathrm{dil}} = 0$, is given by the power multiplier $p_{\Delta}$
\begin{equation}
p_{\Delta} \equiv \frac{\tilde{p}_f}{p_{f,max}} = 4 \tilde{f}_{\mathrm{T}}^{\mathrm{co}} (1-\tilde{f}_{\mathrm{T}}^{\mathrm{co}}) (1-2 f_{\mathrm{dil}})^2.
\label{eq:pDeltaform0}
\end{equation}

We now find an expression for $f_{\mathrm{dil}}$. An important quantity is the helium enrichment, the ratio of the helium-to-fuel density ratios in the divertor and core,
\begin{equation}
\eta_{\mathrm{He}} \equiv f_{\mathrm{He},\mathrm{div}} /f_{\alpha,\mathrm{co}},
\label{eq:etaHe}
\end{equation}
where the helium-to-fuel divertor density ratio is
\begin{equation}
f_{\mathrm{He},\mathrm{div}} \equiv \frac{n_{\mathrm{He}}^{\mathrm{div}}}{n_{\mathrm{Q}}^{\mathrm{div}}} = \frac{f_{\mathrm{HeT,div}}}{ 1 + f_{\mathrm{DT,div}} },
\label{eq:ashtofuel}
\end{equation}
the helium-to-fuel density ratio in the core is
\begin{equation}
f_{\alpha}^{\mathrm{co}} \equiv n_{\alpha}^{\mathrm{co}}/n_{Q}^{\mathrm{co}},
\label{eq:falphaco}
\end{equation}
and the deuterium-to-tritium divertor density ratio is
\begin{equation}
    f_{\mathrm{DT,div}} =n_{\mathrm{D}}^{\mathrm{div}}/n_{\mathrm{T}}^{\mathrm{div}}.
    \label{eq:fDTdiv}
\end{equation}
Combining \Cref{eq:pDeltaform0,eq:etaHe,eq:ashtofuel,eq:falphaco} we obtain
\begin{equation}
p_{\Delta} = 4 \tilde{f}_{\mathrm{T}}^{\mathrm{co}} (1-\tilde{f}_{\mathrm{T}}^{\mathrm{co}}) \left[1 - \left(1 + \frac{\eta_{\mathrm{He}}}{ 2 f_{\mathrm{He},\mathrm{div}} }   \right)^{-1}  \right]^{2},
\label{eq:pDeltaform1}
\end{equation}
where we used
\begin{equation}
f_{\mathrm{dil}} = \frac{f_{\mathrm{He},\mathrm{div}}/\eta_{\mathrm{He}}}{1 + 2 f_{\mathrm{He},\mathrm{div}}/\eta_{\mathrm{He}}}.
\label{eq:fdil}
\end{equation}
We use
\begin{equation}
    f_{\mathrm{HeT,div}} = \frac{1}{\Sigma_{\mathrm{HeT}}} \frac{1}{\frac{1}{\mathrm{TBE}} -1},
\end{equation}
$f_{\mathrm{He,div}}$ (\Cref{eq:ashtofuel}), and $p_{\Delta}$ (\Cref{eq:pDeltaform1}) to find
\begin{equation}
p_{\Delta} = \frac{ 4 \tilde{f}_{\mathrm{T}}^{\mathrm{co}} (1-\tilde{f}_{\mathrm{T}}^{\mathrm{co}})}{ \left[1 -  \frac{2}{ \eta_{\mathrm{He}} \Sigma_{\mathrm{HeT}} \left( 1 + f_{\mathrm{DT},\mathrm{div}} \right) \left( 1 - \frac{1}{\mathrm{TBE}} \right) }  \right]^{2}}.
\label{eq:pDeltaformG}
\end{equation}

\begin{figure}[!tb]
    \centering
    \begin{subfigure}[t]{0.9\textwidth}
    \centering
    \includegraphics[width=1.0\textwidth]{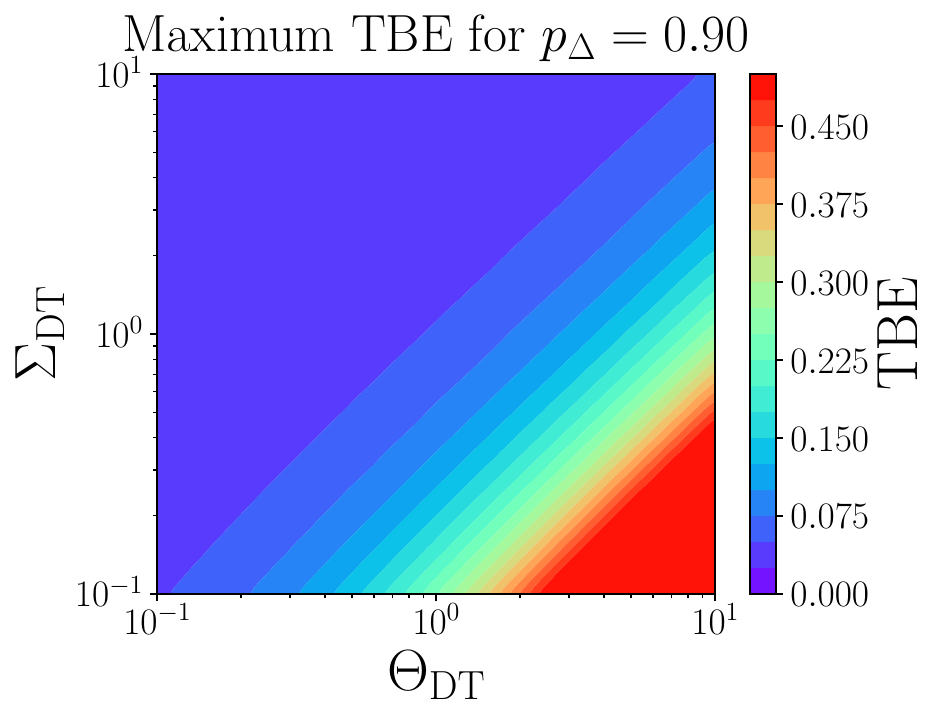}
    \caption{TBE.}
    \end{subfigure}
     ~
    \begin{subfigure}[t]{0.9\textwidth}
    \centering
    \includegraphics[width=1.0\textwidth]{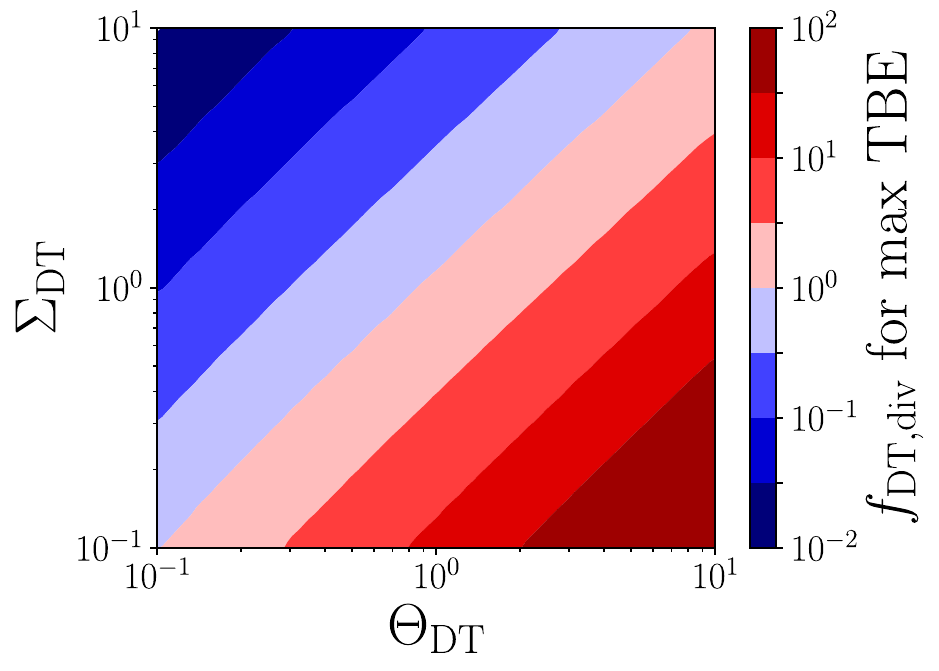}
    \caption{$f_{\mathrm{DT,div}}$.}
    \end{subfigure}
     ~
    \begin{subfigure}[t]{0.9\textwidth}
    \centering
    \includegraphics[width=1.0\textwidth]{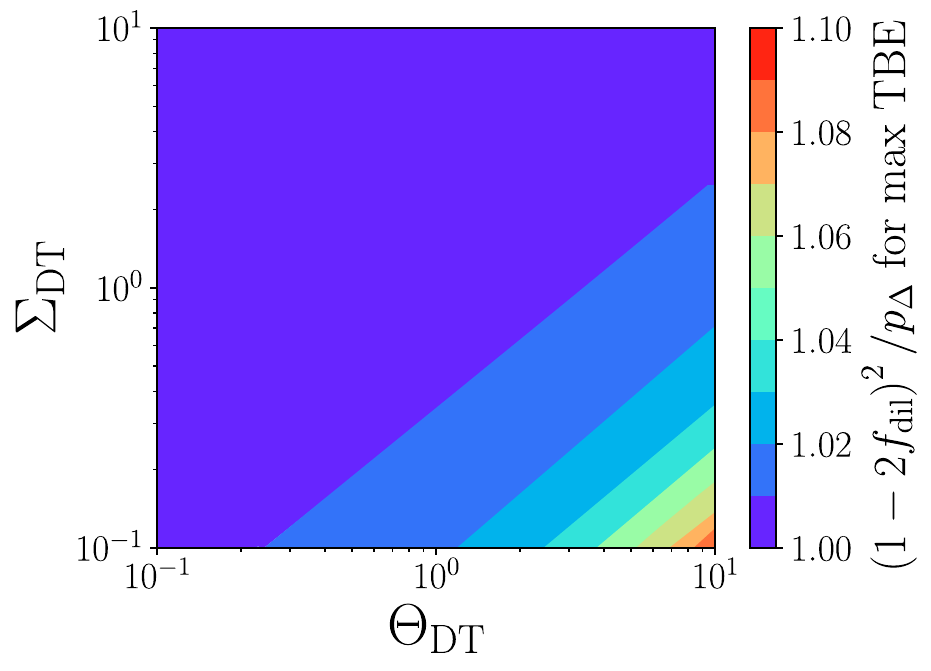}
    \caption{$\left( 1 - 2 f_{\mathrm{dil}} \right)^2 / p_{\Delta}$.}
    \end{subfigure}
    \caption{Results of numerical scheme to maximize TBE for fixed fusion power multiplier $p_{\Delta} = 0.90$. (a): Maximum achievable TBE versus $\Theta_{\mathrm{DT} }$ and $\Sigma_{\mathrm{DT} }$. (b): corresponding $f_{\mathrm{DT, div} }$. (c): corresponding $\left( 1 - 2 f_{\mathrm{dil}} \right)^2 / p_{\Delta}$.}
    \label{fig:maxTBEfixedpDelta}
\end{figure}

\begin{figure}[!tb]
    \centering
    \begin{subfigure}[t]{0.9\textwidth}
    \centering
    \includegraphics[width=1.0\textwidth]{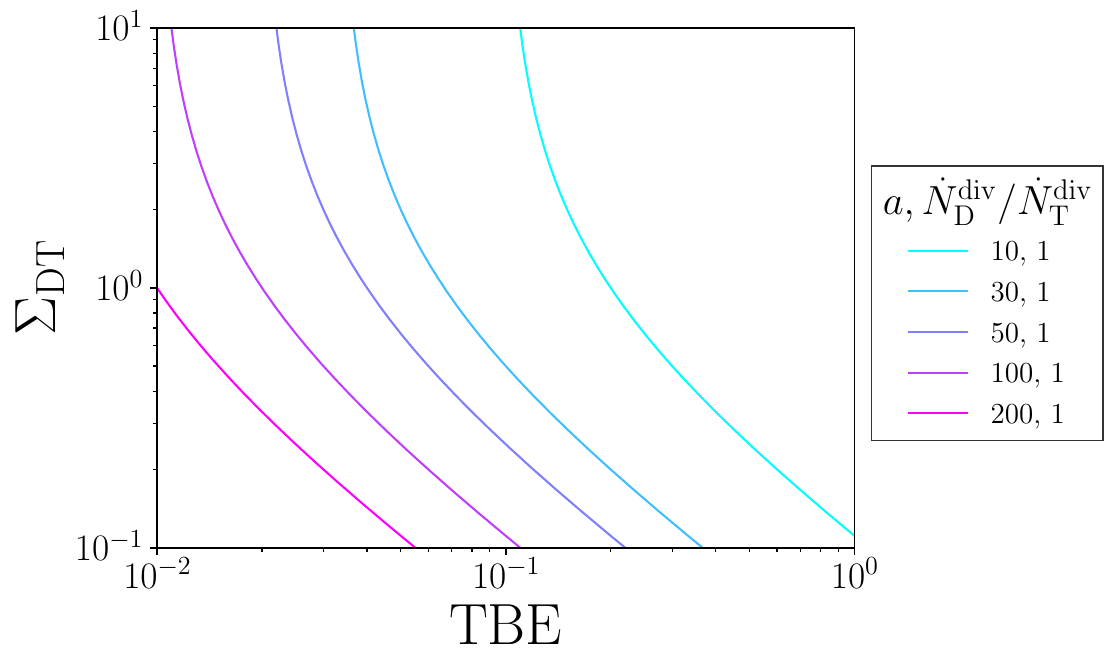}
    \caption{$\Sigma_{\mathrm{DT}}$ varying TBE and $a$.}
    \end{subfigure}
     ~
    \begin{subfigure}[t]{0.9\textwidth}
    \centering
    \includegraphics[width=1.0\textwidth]{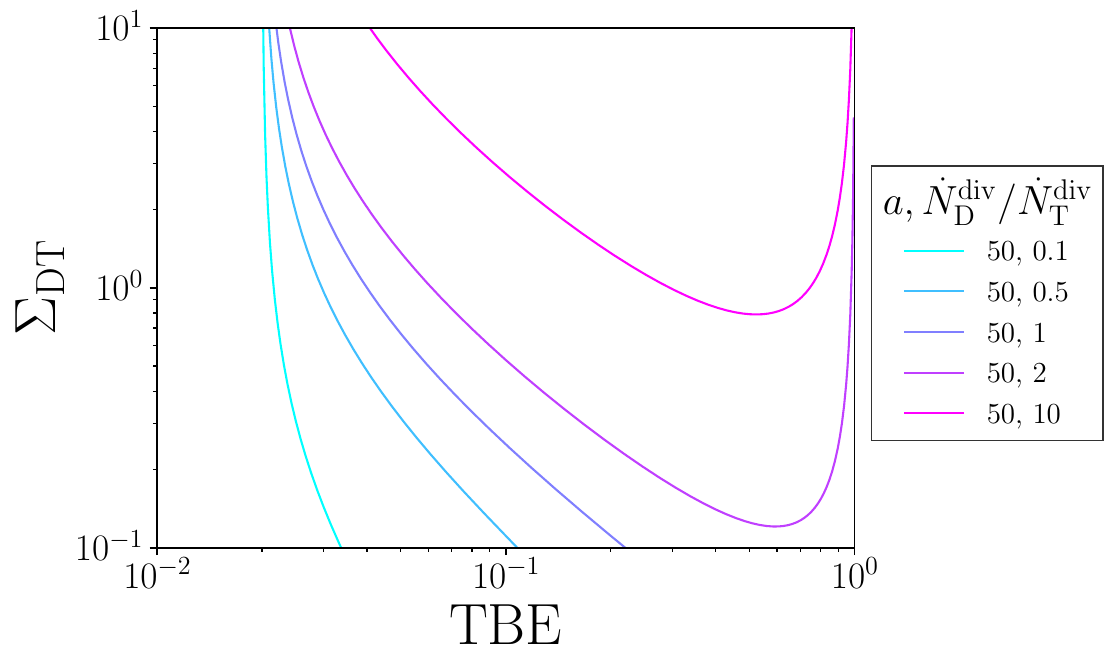}
    \caption{$\Sigma_{\mathrm{DT}}$ varying TBE and $\dot{N}_{\mathrm{D}}^{\mathrm{in}}/\dot{N}_{\mathrm{T}}^{\mathrm{in}}$.}
    \end{subfigure}
    \caption{Required $\Sigma_{\mathrm{DT}}$ versus TBE using approximations in  \Cref{eq:sigmaDT_TBE,eq:fDTdiv_approx} for fixed fusion power and assuming TBE$\ll 1$.}
    \label{fig:SigmaDT_intuition}
\end{figure}

\begin{figure*}[!tb]
    \centering
    \begin{subfigure}[t]{0.32\textwidth}
    \centering
    \includegraphics[width=1.0\textwidth]{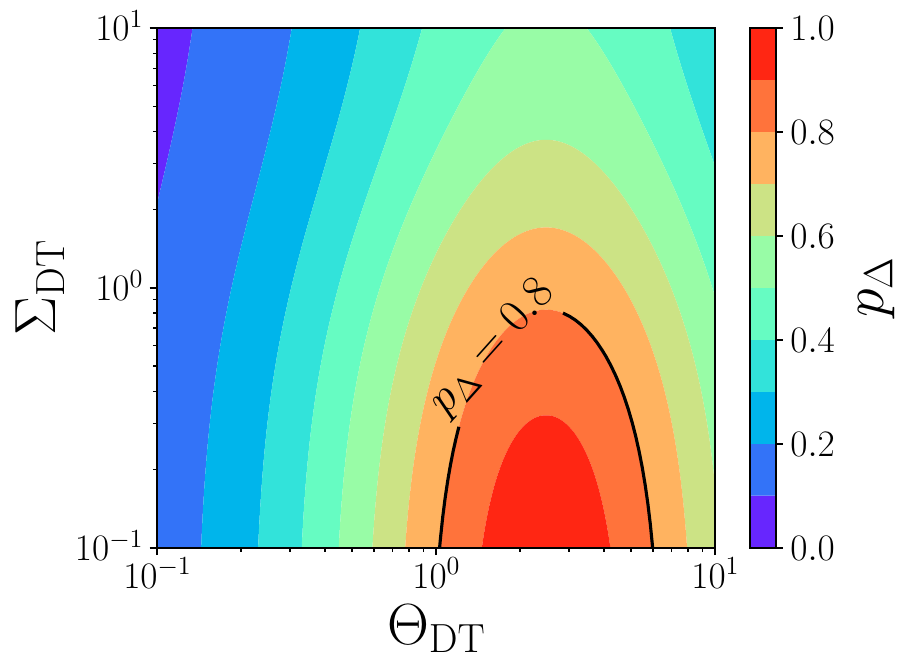}
    \caption{$F_{\mathrm{T} }^{\mathrm{in} } = 0.3$.}
    \end{subfigure}
     ~
    \begin{subfigure}[t]{0.32\textwidth}
    \centering
    \includegraphics[width=1.0\textwidth]{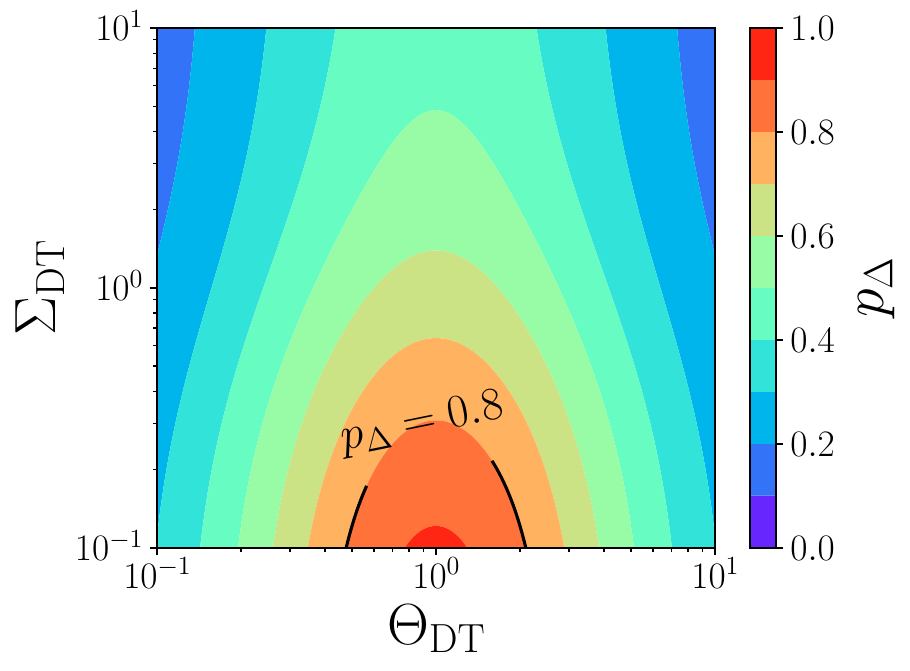}
    \caption{$F_{\mathrm{T} }^{\mathrm{in} } = 0.5$.}
    \end{subfigure}
     ~
    \begin{subfigure}[t]{0.32\textwidth}
    \centering
    \includegraphics[width=1.0\textwidth]{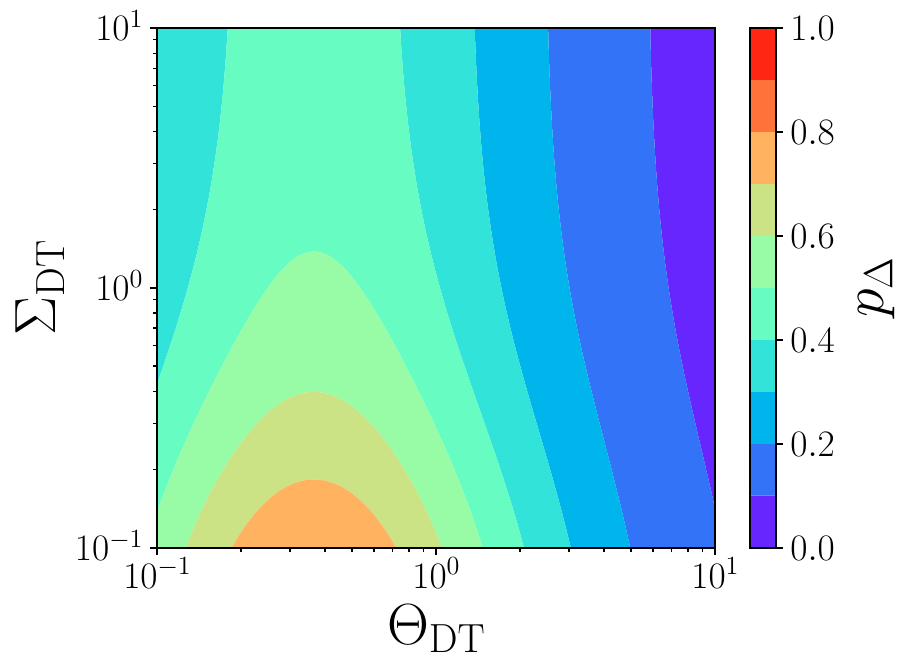}
    \caption{$F_{\mathrm{T} }^{\mathrm{in} } = 0.7$.}
    \end{subfigure}
    \caption{Fusion power multiplier $p_{\Delta}$ versus $\Theta_{\mathrm{DT} }$ and $\Sigma_{\mathrm{DT} }$ for three $F_{\mathrm{T} }^{\mathrm{in} }$ values. In all figures, $\Lambda = \Sigma_{\mathrm{HeT}} =\eta_{\mathrm{He} } = 1.0$, TBE = $0.2$.}
    \label{fig:pDelta_plots_FTin_scan}
\end{figure*}

Applying the boundary condition for equal flow rates at the separatrix and divertor,
\begin{equation}
F_{\mathrm{T}}^{\mathrm{sep}} = F_{\mathrm{T}}^{\mathrm{div}} = \frac{1}{1 + f_{\mathrm{DT,div} }\Sigma_{\mathrm{DT} }},
\label{eq:BC_sepdiv}
\end{equation}
gives (using \Cref{eq:FTco_ftco})
\begin{equation}
\frac{ f_{\mathrm{T}  }^{\mathrm{sep} }}{ \Theta_{\mathrm{DT}} \left(\Lambda - f_{\mathrm{T}  }^{\mathrm{sep}} \right) + f_{\mathrm{T}  }^{\mathrm{sep}} } = \frac{1}{1 + f_{\mathrm{DT,div} }\Sigma_{\mathrm{DT} }},
\label{eq:intermediate}
\end{equation}
where the deuterium-to-tritium pumping speed ratio is
\begin{equation}
    \Sigma_{\mathrm{DT}} \equiv S_{\mathrm{D}}/S_{\mathrm{T}}.
    \label{eq:SigmaDT}
\end{equation}
Rearranging \Cref{eq:intermediate} gives
\begin{equation}
 f_{\mathrm{DT,div} } = \frac{\Theta_{\mathrm{DT}}}{\Sigma_{\mathrm{DT} }} \frac{  \Lambda - f_{\mathrm{T}  }^{\mathrm{sep}}  }{f_{\mathrm{T}  }^{\mathrm{sep}}}.
\label{eq:littlefTdiv}
\end{equation}
$f_{\mathrm{DT,div} }$ is an important quantity: by increasing $f_{\mathrm{DT,div} }$ through smaller $\Sigma_{\mathrm{DT} }$ and larger $\Theta_{\mathrm{DT}}$ values, the fusion power can be maintained while the TBE is increased. When $p_{\Delta}$ is fixed, per \Cref{eq:pfform2}, this means that $\tilde {f}_{\mathrm{T}}^{\mathrm{co}} (1- \tilde {f}_{\mathrm{T}}^{\mathrm{co}}) (1-2 f_{\mathrm{dil}})^2$ is also fixed. At relatively low TBE values and fixed power, the terms $\tilde {f}_{\mathrm{T}}^{\mathrm{co}} (1- \tilde {f}_{\mathrm{T}}^{\mathrm{co}})$ and $(1-2 f_{\mathrm{dil}})^2$ individually vary little with $\Sigma_{\mathrm{DT} }$ and larger $\Theta_{\mathrm{DT}}$. This is shown in \Cref{fig:maxTBEfixedpDelta} (c). Therefore, assuming $f_{\mathrm{dil}}$ is fixed, from \Cref{eq:ashtofuel}
\begin{equation}
    \mathrm{TBE}/\Sigma_{\mathrm{HeT}} \sim f_{\mathrm{HeT,div}} = f_{\mathrm{He,div}} \left( 1 + f_{\mathrm{DT,div}} \right),
    \label{eq:HeTdiv_pDeltafixed}
\end{equation}
where $f_{\mathrm{He,div}}$ is constant and independent of $f_{\mathrm{DT,div}}$. Given that $f_{\mathrm{DT,div}}$ increases strongly with smaller $\Sigma_{\mathrm{DT} }$ and larger $\Theta_{\mathrm{DT}}$ (see \Cref{fig:maxTBEfixedpDelta} (b)), \Cref{eq:HeTdiv_pDeltafixed} implies that $f_{\mathrm{HeT,div}}$ also increases. According to \Cref{eq:TBEnew}, the TBE only depends on $f_{\mathrm{HeT,div}}$ and $\Sigma_{\mathrm{HeT}}$, the latter we hold fixed. Therefore, by increasing $f_{\mathrm{DT,div}}$, $f_{\mathrm{HeT,div}}$ also increases, increasing the TBE. This argument holds until relatively large TBE values, when holding $p_{\Delta}$ constant requires both $\tilde {f}_{\mathrm{T}}^{\mathrm{co}} (1- \tilde {f}_{\mathrm{T}}^{\mathrm{co}})$ and $(1-2 f_{\mathrm{dil}})^2$ to individually vary.

It is important to distinguish the effect of $\Theta_{\mathrm{DT}}$ from $\Sigma_{\mathrm{DT} }$. Using \Cref{eq:usefulfHeTdivSigma} that $f_{\mathrm{HeT,div}} \Sigma_{\mathrm{HeT}} \sim 1 / \dot{N}_{\mathrm{T}}^{\mathrm{co}}$ and \Cref{eq:HeTdiv_pDeltafixed},
\begin{equation}
    1 + \frac{1}{\Sigma_{\mathrm{DT} }} \Theta_{\mathrm{DT}} \frac{  \Lambda - f_{\mathrm{T}  }^{\mathrm{sep}}  }{f_{\mathrm{T}  }^{\mathrm{sep}}} \sim \frac{1}{\dot{N}_{\mathrm{T}}^{\mathrm{co}}} \sim \frac{1}{D_{\mathrm{T}} \nabla n_{\mathrm{T}}^{\mathrm{co}}}.
\end{equation}
Therefore, at fixed power, $\Theta_{\mathrm{DT}} \left( \Lambda - f_{\mathrm{T}  }^{\mathrm{sep}} \right)/ f_{\mathrm{T}  }^{\mathrm{sep}}$, and $n_{\mathrm{T}}^{\mathrm{co}}$, any decreases in $D_{\mathrm{T}}$ must have a corresponding decrease in  $\Sigma_{\mathrm{DT} }$ as the TBE increases. Physically, as the D-T particle diffusivities decrease, the power plant operator may choose to decrease the D-T fueling, which decreases the divertor particle flow rate. Because this increases the TBE at fixed power due to lower required fueling (for both D and T), the D-T divertor density $f_{\mathrm{DT,div}}$ will increase. However, the ratio of tritium-to-deuterium flow rates at the divertor has not changed significantly i.e. $f_{\mathrm{DT,div}} \Sigma_{\mathrm{DT}}$ is approximately constant. Thus, $\Sigma_{\mathrm{DT}}$ must decrease. The required $\Sigma_{\mathrm{DT}}$ value can also be written as
\begin{equation}
    \Sigma_{\mathrm{DT}} = \frac{1}{f_{\mathrm{DT,div}}} \frac{\dot{N}_{\mathrm{D}}^{\mathrm{in}}/\dot{N}_{\mathrm{T}}^{\mathrm{in}} - \mathrm{TBE}}{1 - \mathrm{TBE}}.
    \label{eq:sigmaDT_TBE}
\end{equation}
Using the arguments surrounding \Cref{eq:HeTdiv_pDeltafixed} to write $f_{\mathrm{DT,div}}$ at fixed fusion power as
\begin{equation}
    f_{\mathrm{DT,div}} = a \; \mathrm{TBE} - 1,
    \label{eq:fDTdiv_approx}
\end{equation}
we can approximately capture the behavior of $\Sigma_{\mathrm{DT}}$ in \Cref{eq:sigmaDT_TBE}. In \Cref{eq:fDTdiv_approx}, $a = 1/  f_{\mathrm{He,div}} \Sigma_{\mathrm{HeT}} \gg 1$. \Cref{fig:SigmaDT_intuition} shows solutions to \Cref{eq:sigmaDT_TBE,eq:fDTdiv_approx} versus TBE. For large regions of parameter space, access to high TBE regimes without decreasing fusion power requires low values of $\Sigma_{\mathrm{DT}}$.

To summarize: decreasing $\Sigma_{\mathrm{DT} }$ and increasing $\Theta_{\mathrm{DT}}$ increases the TBE while maintaining fusion power. This increases the relative amount of deuterium to tritium in the divertor, $f_{\mathrm{DT,div}}$, which also increases the amount of helium relative to tritium in the divertor $f_{\mathrm{HeT,div}}$. As assumed throughout this paper, the helium-to-tritium pumping speed $\Sigma_{\mathrm{HeT}}$ is fixed. Therefore, the increase in $f_{\mathrm{HeT,div}}$ increases the TBE (\Cref{eq:TBEnew}). Using $\Sigma_{\mathrm{DT} }$ and $\Theta_{\mathrm{DT}}$ allows us to simultaneously increase $f_{\mathrm{HeT,div}}$ and $f_{\mathrm{DT,div}}$, keeping $f_{\mathrm{He,div}}$ and $f_{\mathrm{dil}}$ roughly constant. If we could not vary $\Sigma_{\mathrm{DT} }$ and $\Theta_{\mathrm{DT}}$, increases in $f_{\mathrm{HeT,div}}$ (and therefore TBE) are much larger than increases in $f_{\mathrm{DT,div}}$, meaning that the plasma becomes more diluted through $f_{\mathrm{dil}}$ (\Cref{eq:fdil}) and therefore the power decreases. With technology that allows smaller $\Sigma_{\mathrm{DT}}$ values, the fusion power can be sustained as the TBE increases through smaller $f_{\mathrm{DT,div}}$.

We summarize numerical results to support the above arguments. Shown in \Cref{fig:maxTBEfixedpDelta}(a), we plot the maximum TBE versus $\Theta_{\mathrm{DT}}$ and $\Sigma_{\mathrm{DT} }$ for fixed $p_{\Delta} = 0.90$ (\Cref{eq:pDeltaformG}). The maximization in the TBE is performed by searching over all possible $F_{\mathrm{T} }^{\mathrm{in}}$ and selecting the $F_{\mathrm{T} }^{\mathrm{in}}$ that gives the highest TBE value, subject to all other quantities in the system being physical. As expected, smaller $\Sigma_{\mathrm{DT} }$ and larger $\Theta_{\mathrm{DT}}$  values increase the TBE. This is achieved by order of magnitude increases in $f_{\mathrm{DT,div} }$, shown in \Cref{fig:maxTBEfixedpDelta}(b). In \Cref{fig:maxTBEfixedpDelta}(c), we plot the power dilution term $(1-2 f_{\mathrm{dil}})^2 / p_{\Delta}$ -- apart from exceptionally large $\Theta_{\mathrm{DT}}$ and very small $\Sigma_{\mathrm{DT} }$, the quantity $(1-2 f_{\mathrm{dil}})^2 / p_{\Delta}$ is very close to 1, indicating that $f_{\mathrm{dil}}$ changes slowly across $\Theta_{\mathrm{DT}}$, $\Sigma_{\mathrm{DT} }$ space -- and therefore the assumption of fixed $f_{\mathrm{He,div}}$ at fixed power (used in \Cref{eq:HeTdiv_pDeltafixed}) is often valid. Only when the TBE no longer satisfies TBE$\ll 1$ is $(1-2 f_{\mathrm{dil}})^2 / p_{\Delta}$ slightly larger than 1.

\begin{figure*}[!tb]
    \centering
    \begin{subfigure}[t]{0.280\textwidth}
    \centering
    \includegraphics[width=1.0\textwidth]{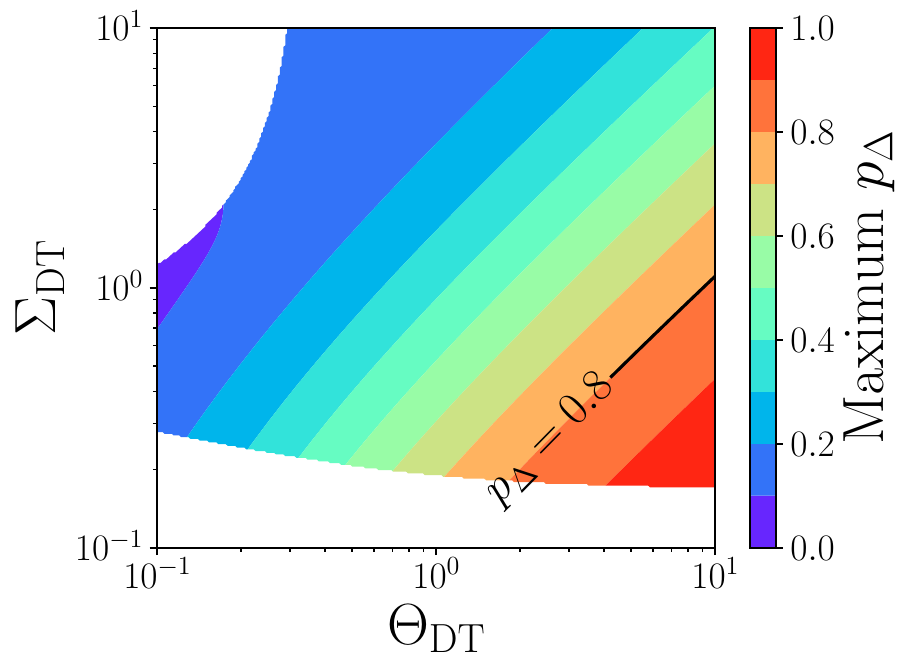}\hfill%
    \caption{TBE = 0.50.}
    \end{subfigure}
     ~
    \begin{subfigure}[t]{0.280\textwidth}
    \centering
    \includegraphics[width=1.0\textwidth]{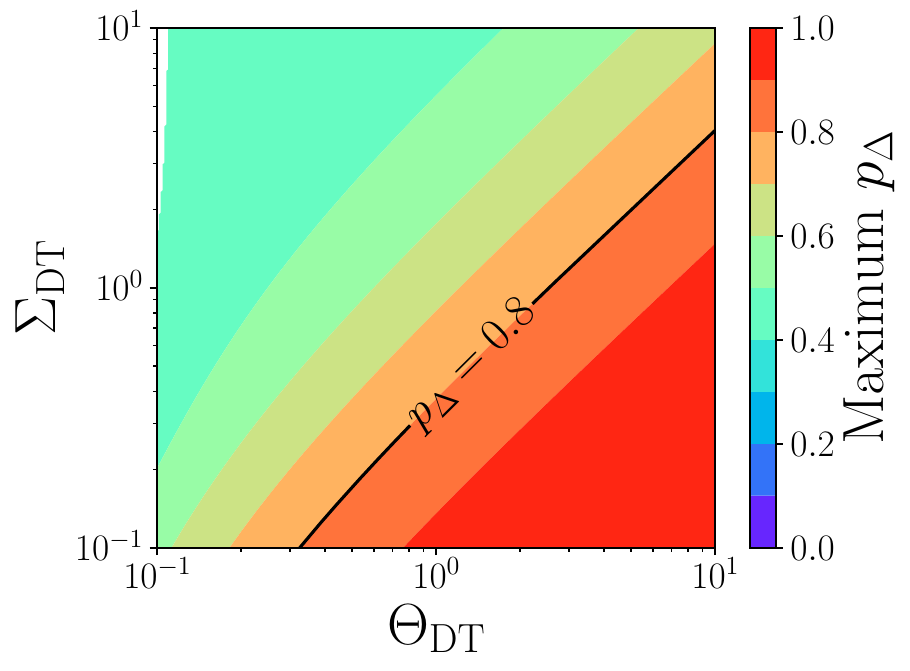}
    \caption{TBE = 0.20.}
    \end{subfigure}
     ~
    \begin{subfigure}[t]{0.280\textwidth}
    \centering
    \includegraphics[width=1.0\textwidth]{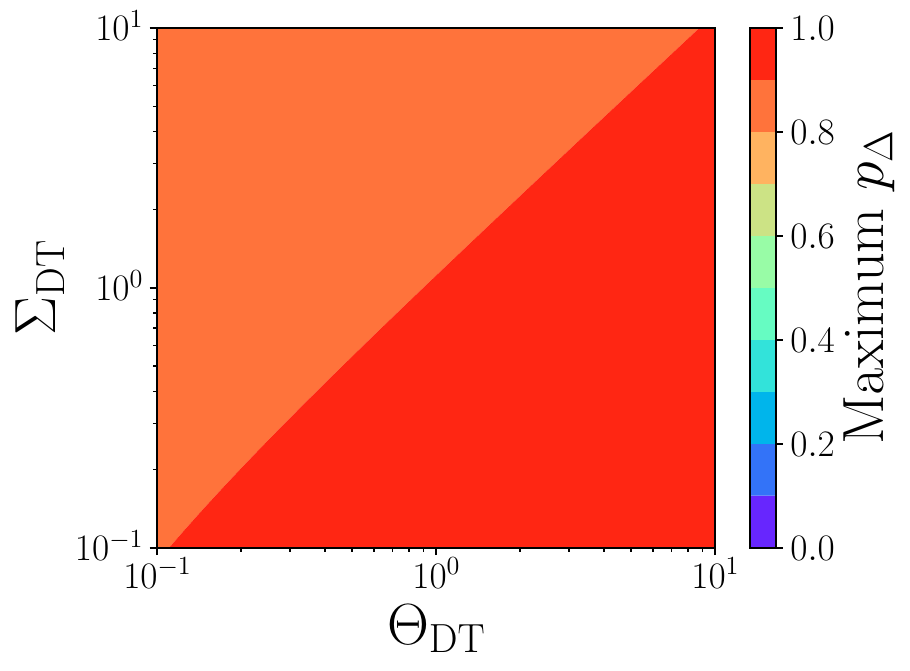}
    \caption{TBE = 0.05.}
    \end{subfigure}
    \centering
    \begin{subfigure}[t]{0.280\textwidth}
    \centering
    \includegraphics[width=1.0\textwidth]{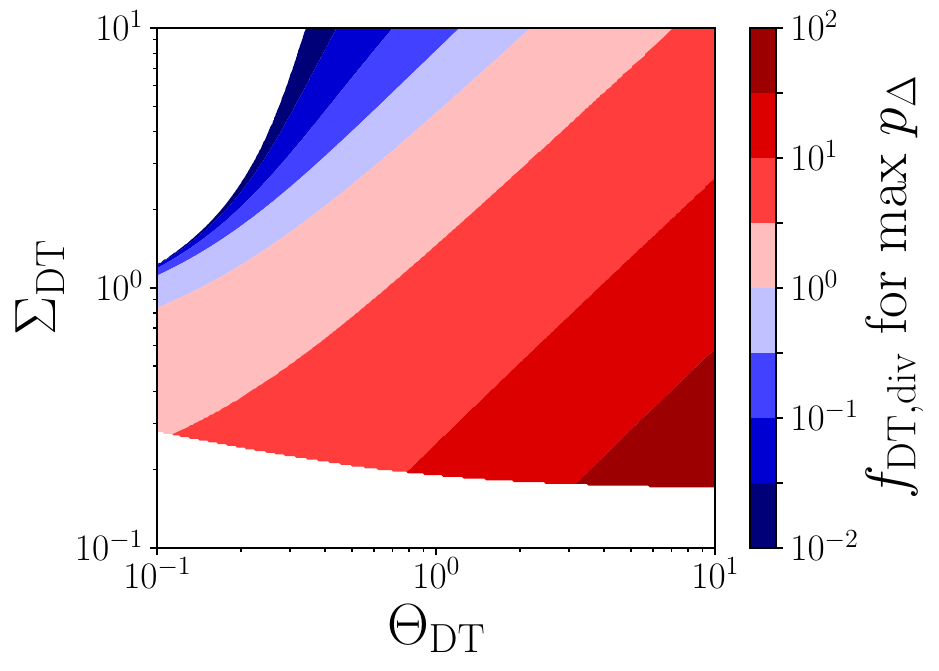}
    \caption{TBE = 0.50.}
    \end{subfigure}
     ~
    \begin{subfigure}[t]{0.280\textwidth}
    \centering
    \includegraphics[width=1.0\textwidth]{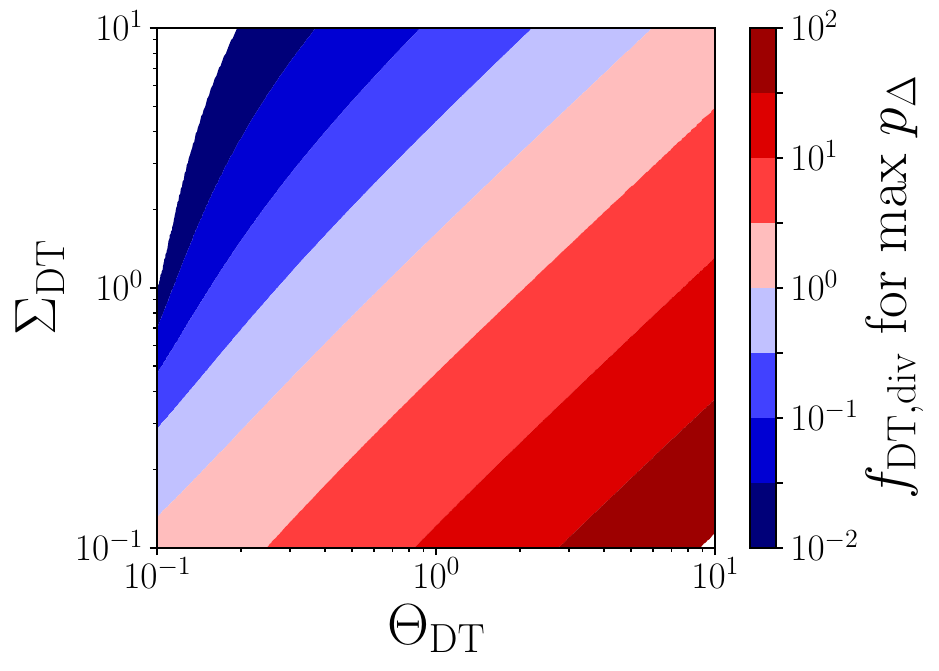}
    \caption{TBE = 0.20.}
    \end{subfigure}
     ~
    \begin{subfigure}[t]{0.280\textwidth}
    \centering
    \includegraphics[width=1.0\textwidth]{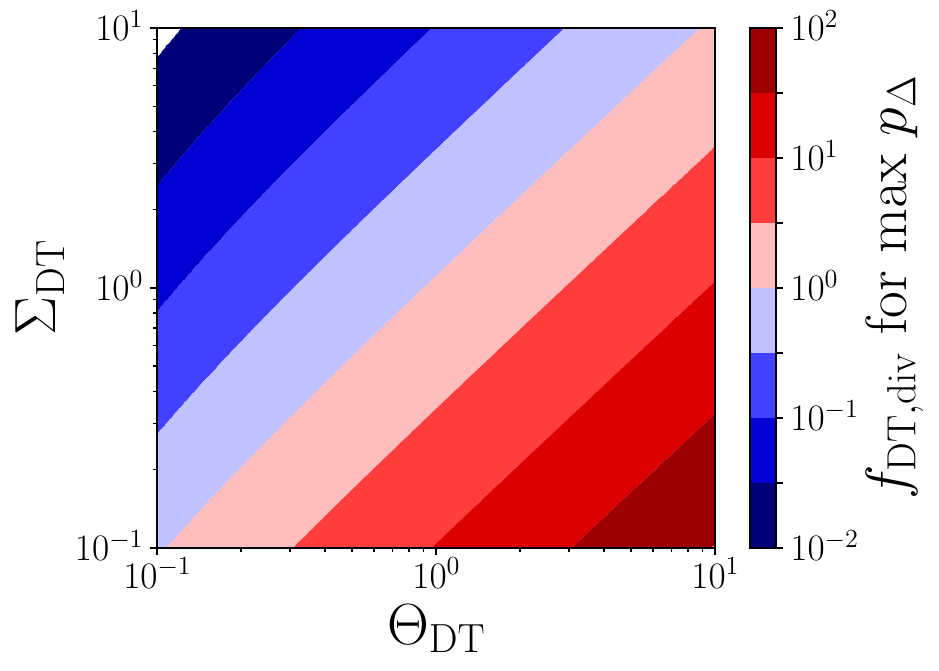}
    \caption{TBE = 0.05.}
    \end{subfigure}
    \centering
    \begin{subfigure}[t]{0.280\textwidth}
    \centering
    \includegraphics[width=1.0\textwidth]{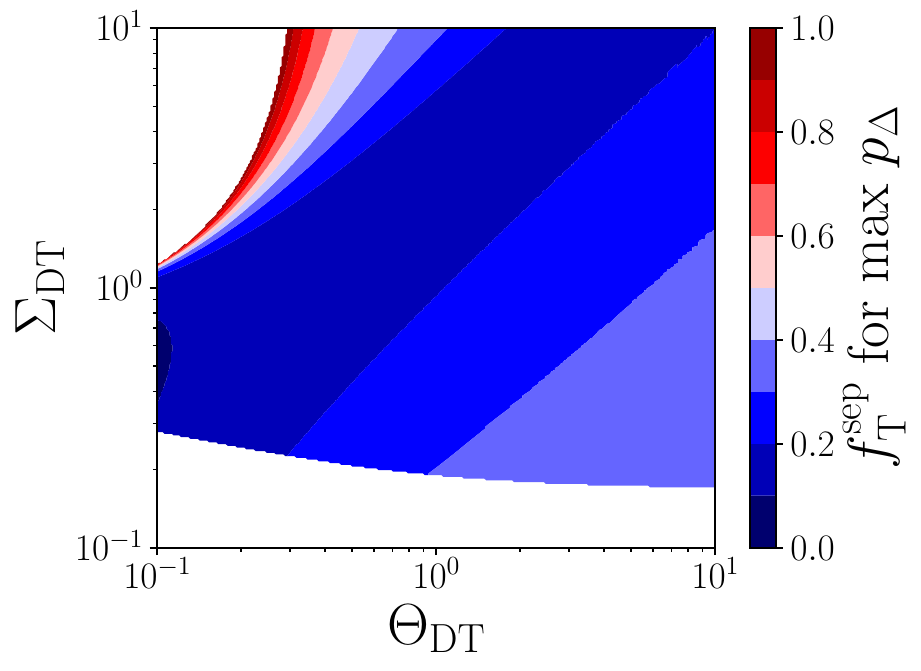}
    \caption{TBE = 0.50.}
    \end{subfigure}
     ~
    \begin{subfigure}[t]{0.280\textwidth}
    \centering
    \includegraphics[width=1.0\textwidth]{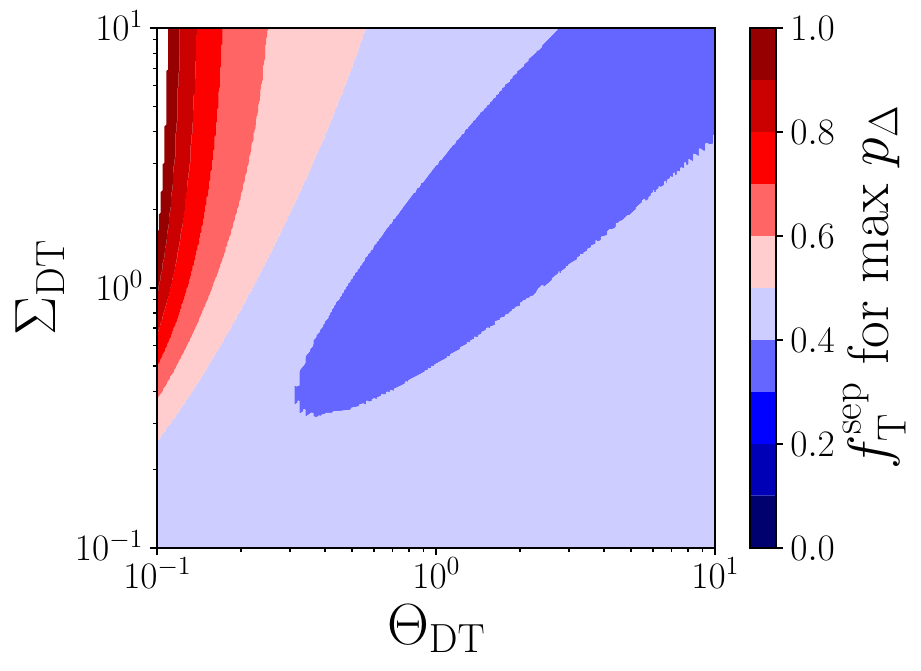}
    \caption{TBE = 0.20.}
    \end{subfigure}
     ~
    \begin{subfigure}[t]{0.280\textwidth}
    \centering
    \includegraphics[width=1.0\textwidth]{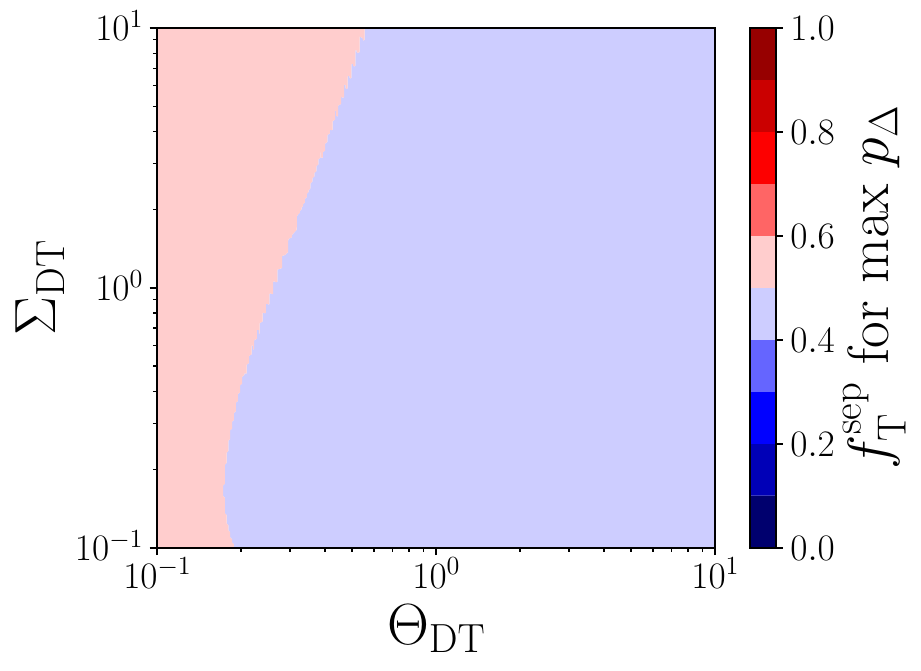}
    \caption{TBE = 0.05.}
    \end{subfigure}
    \centering
    \begin{subfigure}[t]{0.280\textwidth}
    \centering
    \includegraphics[width=1.0\textwidth]{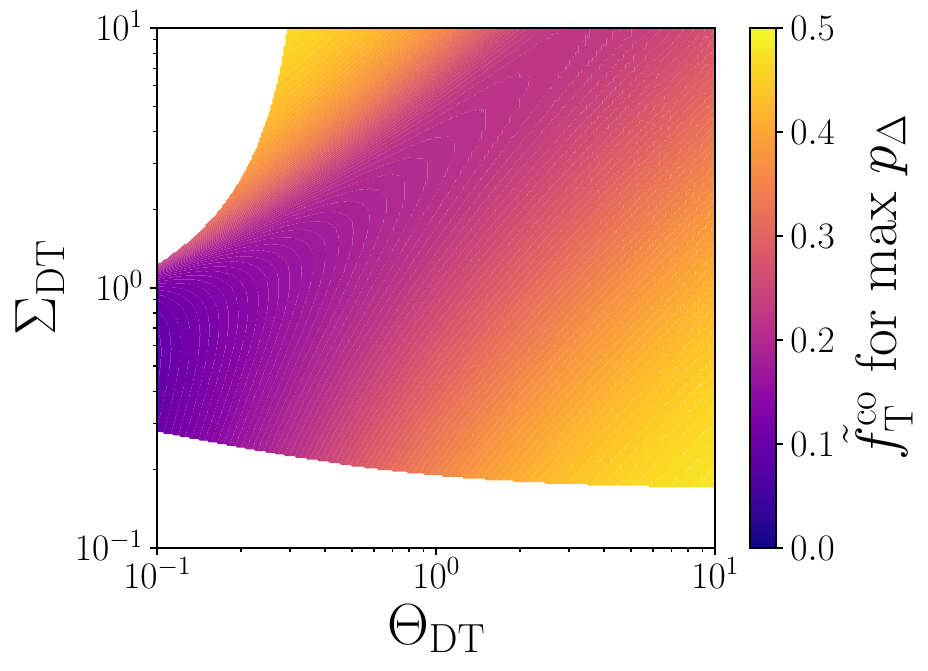}
    \caption{TBE = 0.50.}
    \end{subfigure}
     ~
    \begin{subfigure}[t]{0.280\textwidth}
    \centering
    \includegraphics[width=1.0\textwidth]{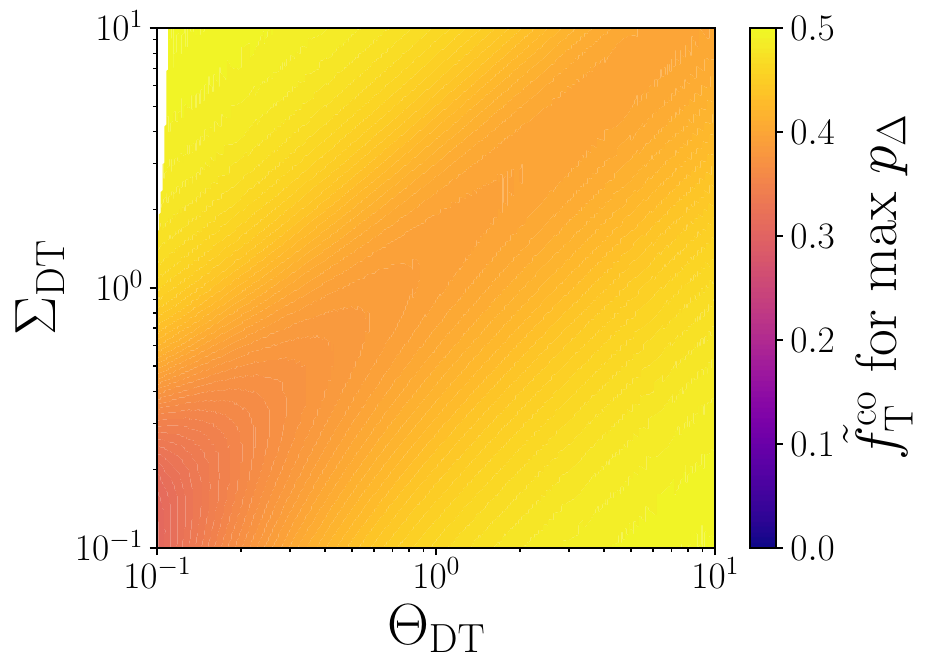}
    \caption{TBE = 0.20.}
    \end{subfigure}
     ~
    \begin{subfigure}[t]{0.280\textwidth}
    \centering
    \includegraphics[width=1.0\textwidth]{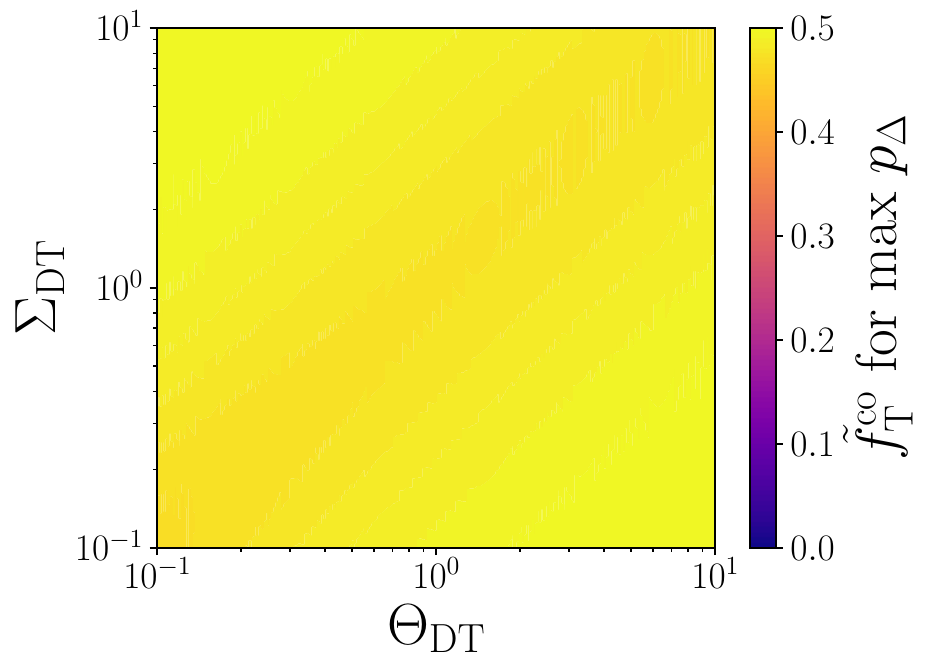}
    \caption{TBE = 0.05.}
    \end{subfigure}
    \centering
    \begin{subfigure}[t]{0.280\textwidth}
    \centering
    \includegraphics[width=1.0\textwidth]{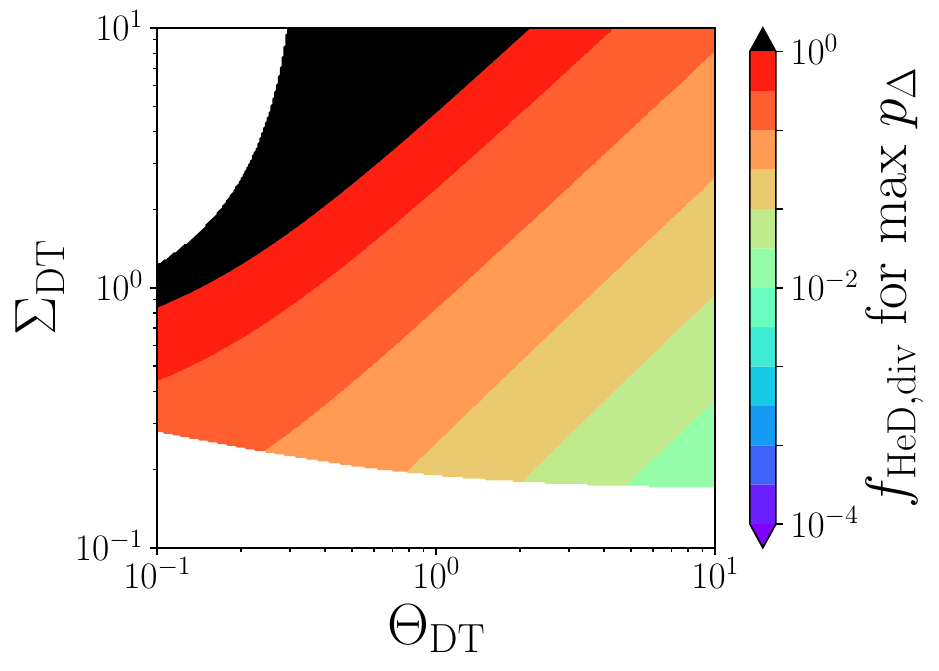}
    \caption{TBE = 0.50.}
    \end{subfigure}
     ~
    \begin{subfigure}[t]{0.280\textwidth}
    \centering
    \includegraphics[width=1.0\textwidth]{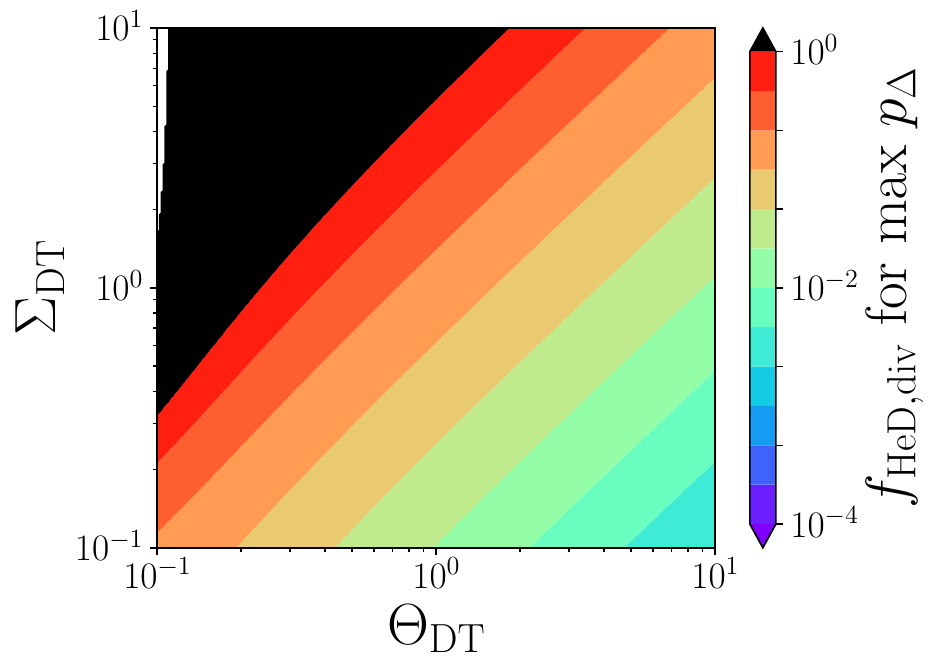}
    \caption{TBE = 0.20.}
    \end{subfigure}
     ~
    \begin{subfigure}[t]{0.28\textwidth}
    \centering
    \includegraphics[width=1.0\textwidth]{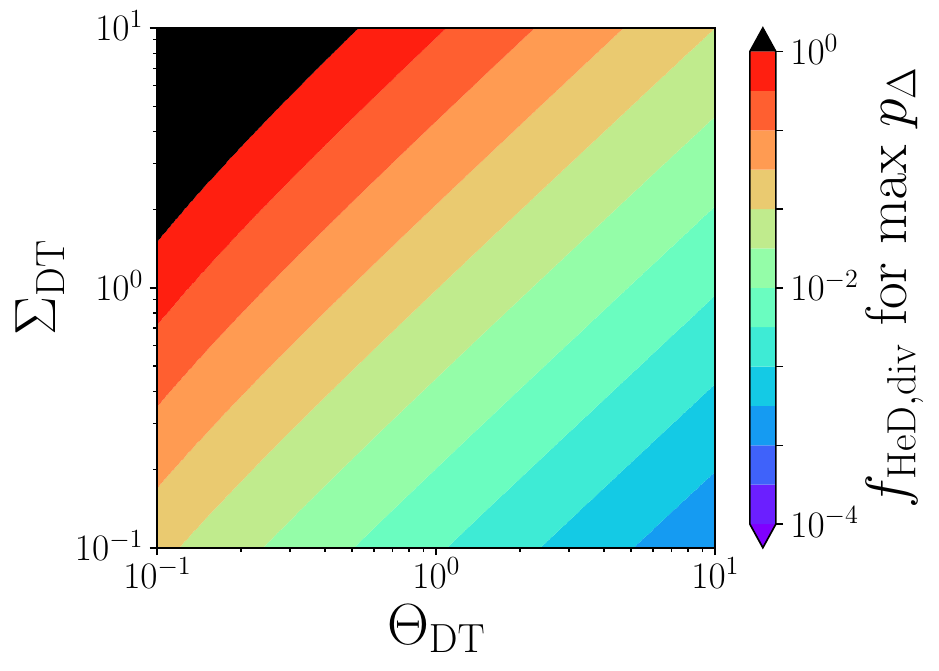}
    \caption{TBE = 0.05.}
    \end{subfigure}
    \caption{Figures of merit versus $\Theta_{\mathrm{DT}}$ and $\Sigma_{\mathrm{DT}}$. First row: fusion power multiplier $p_{\Delta}$, second row: $f_{\mathrm{DT},\mathrm{div}}$, third row $f_{\mathrm{T}}^{\mathrm{sep}}$, fourth row $\tilde{ f}_{\mathrm{T}}^{\mathrm{co}}$, fifth row $f_{\mathrm{HeD,div}}$: In all plots, $\Sigma_{\mathrm{HeT}} \eta_{\mathrm{He}} = 0.5$, and $F_\mathrm{T}^{\mathrm{in}}$ is the value that maximizes $p_{\Delta}$.}
    \label{fig:pDelta_diff_DTpumpingtransport}
\end{figure*}

Solutions to $p_{\Delta}$ in \Cref{eq:pDeltaformG} are shown in \Cref{fig:pDelta_plots_FTin_scan}: each subplot corresponds to a different $F_{\mathrm{T} }^{\mathrm{in}}$ value. For all subplots, TBE, $\Lambda$, and $\Sigma_{\mathrm{HeT} }$ are fixed. For these parameters, decreasing $\Sigma_{\mathrm{DT} }$ always allows a high power, and the optimal $\Theta_{\mathrm{DT} }$ value depends on $F_{\mathrm{T} }^{\mathrm{in}}$. However, smaller $F_{\mathrm{T} }^{\mathrm{in}}$ values are more desirable if larger $\Theta_{\mathrm{DT} }$ values are achievable. Given that the TBE in these plots is fixed, for a given $p_{\Delta}$, the tritium injection flow rate $\dot{N}_{\mathrm{T} }^{\mathrm{in} }$ is also the same for all plots. Therefore, when varying $F_{\mathrm{T} }^{in}$ at fixed TBE and fixed $p_{\Delta}$, $\dot{N}_{\mathrm{D} }^{\mathrm{in} }$ is changing but $\dot{N}_{\mathrm{T} }^{\mathrm{in} }$ is fixed. This explains why at lower $F_{\mathrm{T} }^{\mathrm{in}}$ (higher $\dot{N}_{\mathrm{D} }^{\mathrm{in} }$), a larger $\Theta_{\mathrm{DT} }$ is necessary to more quickly remove deuterium from the core with the benefit of higher fusion power.

We now consider key quantities for an optimization scheme where the maximum power is found for a fixed TBE value. The optimization is performed over $F_{\mathrm{T} }^{\mathrm{in} }$. In the top row of \Cref{fig:pDelta_diff_DTpumpingtransport}, we plot $p_{\Delta}$ as a function of $\Theta_{\mathrm{DT}}$ and $\Sigma_{\mathrm{DT}}$ for fixed $\eta_{\mathrm{He}}$, $\Sigma_{\mathrm{HeT}}$, for three TBE values. Increasing $\Theta_{\mathrm{DT}}$ and decreasing $\Sigma_{\mathrm{DT}}$ gives higher $p_{\Delta}$; the effect on $p_{\Delta}$ is particularly strong at higher TBE values, indicating the importance of $\Theta_{\mathrm{DT}}$ and $\Sigma_{\mathrm{DT}}$ in tritium-efficient regimes. In the remaining four rows, we show $f_\mathrm{DT,\mathrm{div}}$, $f_\mathrm{T}^{\mathrm{sep}}$, $f_\mathrm{T}^{\mathrm{co}}$, and $f_\mathrm{He,\mathrm{div}}$. Writing the TBE in \Cref{eq:TBEnew} to leading order as
\begin{equation}
    \mathrm{TBE} \approx f_{\mathrm{HeT,div}} \Sigma_{\mathrm{HeT}} = f_{\mathrm{HeD,div}} f_{\mathrm{DT,div}} \Sigma_{\mathrm{HeD}} \Sigma_{\mathrm{DT}},
    \label{eq:TBEnew_expanded}
\end{equation}
and using $f_{\mathrm{DT,div}}$ in \Cref{eq:littlefTdiv} we find
\begin{equation}
    \mathrm{TBE} \approx f_{\mathrm{HeD,div}} \frac{\Sigma_{\mathrm{HeT}}}{\Sigma_{\mathrm{DT}}} \frac{\Theta_{\mathrm{DT}} \left(\Lambda - f_{\mathrm{T}  }^{\mathrm{sep}} \right) }{f_{\mathrm{T}  }^{\mathrm{sep}}}.
    \label{eq:TBEnew_expanded2}
\end{equation}
From \Cref{eq:TBEnew_expanded2}, given that $\mathrm{TBE} \geq 0$, we must enforce $\Lambda \geq f_{\mathrm{T}  }^{\mathrm{sep}}$. Assuming that $\Sigma_{\mathrm{HeD}}$, $\Theta_{\mathrm{DT}}$, and $\Lambda$ are fixed, TBE increases by increasing $f_{\mathrm{HeD,div}}$ and decreasing $f_{\mathrm{T}  }^{\mathrm{sep}}$. Comparing the three columns in the third row of \Cref{fig:pDelta_diff_DTpumpingtransport}, at higher TBE the $f_{\mathrm{T}  }^{\mathrm{sep}}$ values are smaller. Additionally, at fixed TBE we generally see that increasing $\Theta_{\mathrm{DT}}$ corresponds to decreasing $f_{\mathrm{T}  }^{\mathrm{sep}}$ (unless TBE is very large), in agreement with \Cref{eq:TBEnew_expanded2}; this only fails to hold when $\Theta_{\mathrm{DT}}$ is small and $\Sigma_{\mathrm{DT}}$ is large (and therefore $\Sigma_{\mathrm{HeD}} = \Sigma_{\mathrm{HeT}}/\Sigma_{\mathrm{DT}}$ is small at fixed $\Sigma_{\mathrm{HeT}}$). Comparing $f_{\mathrm{HeD,div}}$ in the final row of \Cref{fig:pDelta_diff_DTpumpingtransport} shows how higher values of $f_{\mathrm{HeD,div}}$ at higher TBE, as expected from \Cref{eq:TBEnew_expanded2}.

It is important to keep in mind the consequences of fixed $\Sigma_{\mathrm{HeT}}$ when varying $\Sigma_{\mathrm{DT}}$. Because
\begin{equation}
    \Sigma_{\mathrm{HeT}} = \Sigma_{\mathrm{HeD}} \Sigma_{\mathrm{DT}},
\end{equation}
the product $\Sigma_{\mathrm{HeD}} \Sigma_{\mathrm{DT}}$ must be fixed. We find advantageous regimes for $\Sigma_{\mathrm{DT}} < 1$, requiring faster volumetric pumping for the heavier isotope, $\mathrm{T}$. However, as $\Sigma_{\mathrm{DT}}$ is decreased there must be a corresponding increase in $\Sigma_{\mathrm{HeD}}$, which again, requires the heavier element, He, to be pumped faster than deuterium.

\subsection{Fusion Gain}

In this section, we study the effects of differential deuterium-tritium transport and pumping on the fusion plasma gain $Q$. 

Power balance in the core approximately requires
\begin{equation}
p_{\alpha} \left( 1 + 5 / Q \right)  = \frac{w_{\mathrm{th} }}{\tau_E},
\label{eq:corepower}
\end{equation}
where $p_{\alpha}$ is the alpha heating power density, $Q = p_f / p_{\mathrm{heat} }$ is the plasma fusion gain on a flux surface, $p_{\mathrm{heat} }$ is the heating power density absorbed by the plasma, $\tau_E$ is the energy confinement time and $w_{\mathrm{th} }$ is the stored thermal energy density. Expressing the core thermal energy density as
\begin{equation}
w_{\mathrm{th} } = \frac{3}{2} \left(n_e^{\mathrm{co}} + n_{Q}^{\mathrm{co}} \right) k_B T + \frac{3}{2} n_{\alpha}^{\mathrm{co}} k_B \langle T_{\alpha} \rangle,
\end{equation}
where the alpha particle pressure is
\begin{equation}
w_{\mathrm{th} , \alpha} = \frac{3}{2} n_{\alpha} k_B \langle T_{\alpha} \rangle,
\label{eq:w1}
\end{equation}
and the temperature $\langle T_{\alpha} \rangle$ is an average over alpha particle energies. Here, electron and ion temperatures have an equal value $T$. Using quasineutrality, $n_{Q}^{\mathrm{co}} = n_e^{\mathrm{co}} - 2n_{\alpha}^{\mathrm{co}}$, \Cref{eq:w1} is
\begin{equation}
w_{\mathrm{th} } = 3 k_B T M,
\end{equation}
where the dilution factor is
\begin{equation}
M \equiv \left( 1 + f_{\mathrm{dil}} \left( \frac{1}{2} \frac{\langle T_{\alpha} \rangle}{T} -1 \right)  \right),
\end{equation}
which is equal to 1 when there are no helium dilution effects. We obtain
\begin{equation}
n_e \tau_E (1 + 5/Q) = \frac{15 k_B T}{\langle v \overline{\sigma} \rangle E /4} C,
\label{eq:fusiongainform1}
\end{equation}
where
\begin{equation}
C \equiv \frac{ 1 + f_{\mathrm{dil}} \left( \frac{1}{2} \frac{\langle T_{\alpha} \rangle}{T} -1 \right)  }{4 \tilde{f}_{\mathrm{T}}^{\mathrm{co}} (1-\tilde{f}_{\mathrm{T}}^{\mathrm{co}}) (1-2f_{\mathrm{dil}})^2},
\label{eq:C}
\end{equation}
is the required multiplier to keep $Q$ constant at fixed $T$. For simplicity, in this work we use $\langle T_{\alpha} \rangle = T$ as in \cite{Whyte2023}, although the effect of different $\langle T_{\alpha} \rangle$ values is presented in Appendix D3 in \cite{Parisi_2024}.

\begin{figure}[!tb]
    \centering
    \begin{subfigure}[t]{0.8\textwidth}
    \centering
    \includegraphics[width=1.0\textwidth]{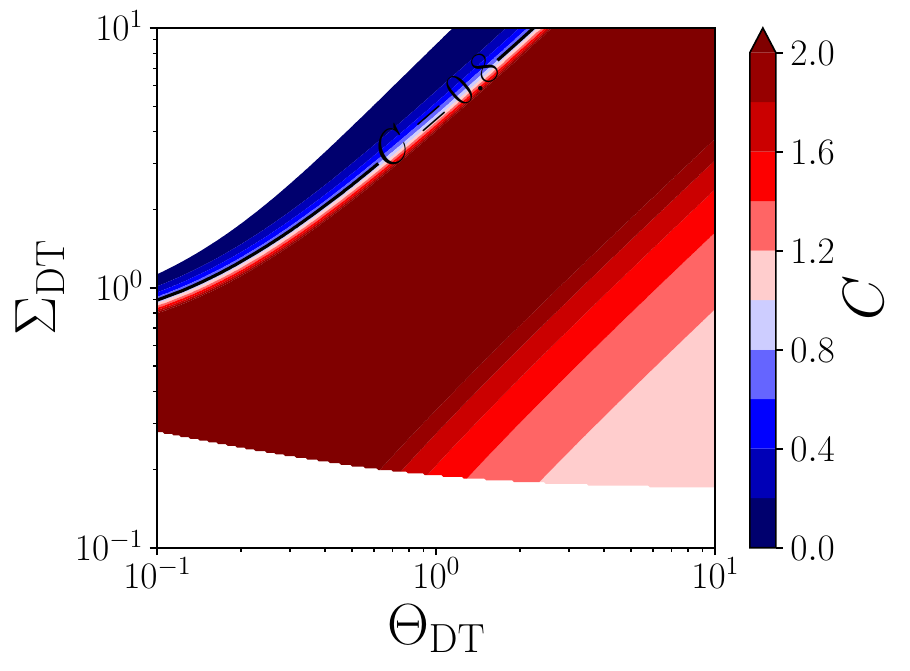}
    \caption{TBE=0.50.}
    \end{subfigure}
     ~
    \begin{subfigure}[t]{0.8\textwidth}
    \centering
    \includegraphics[width=1.0\textwidth]{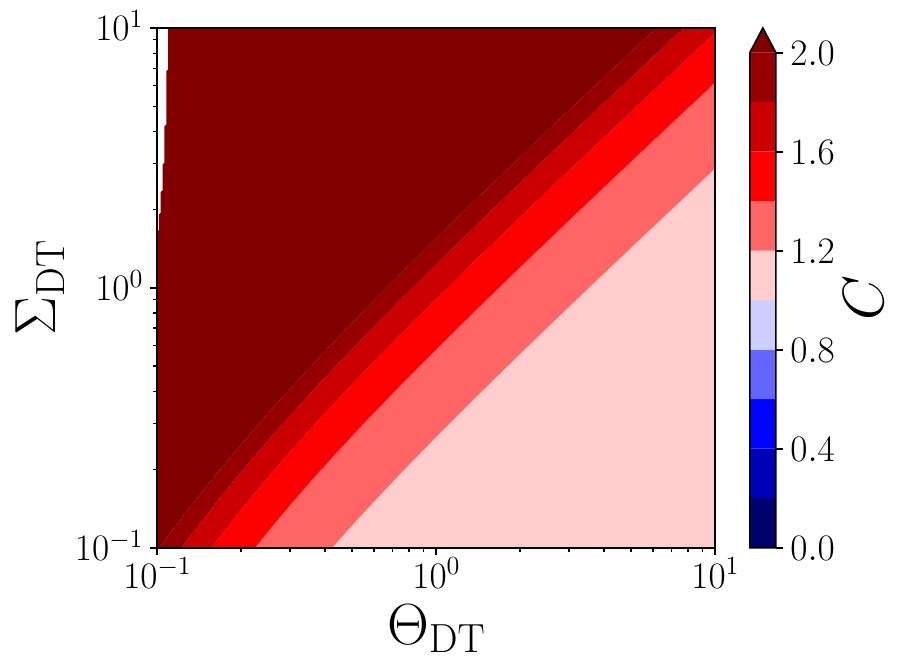}
    \caption{TBE=0.20.}
    \end{subfigure}
     ~
    \begin{subfigure}[t]{0.8\textwidth}
    \centering
    \includegraphics[width=1.0\textwidth]{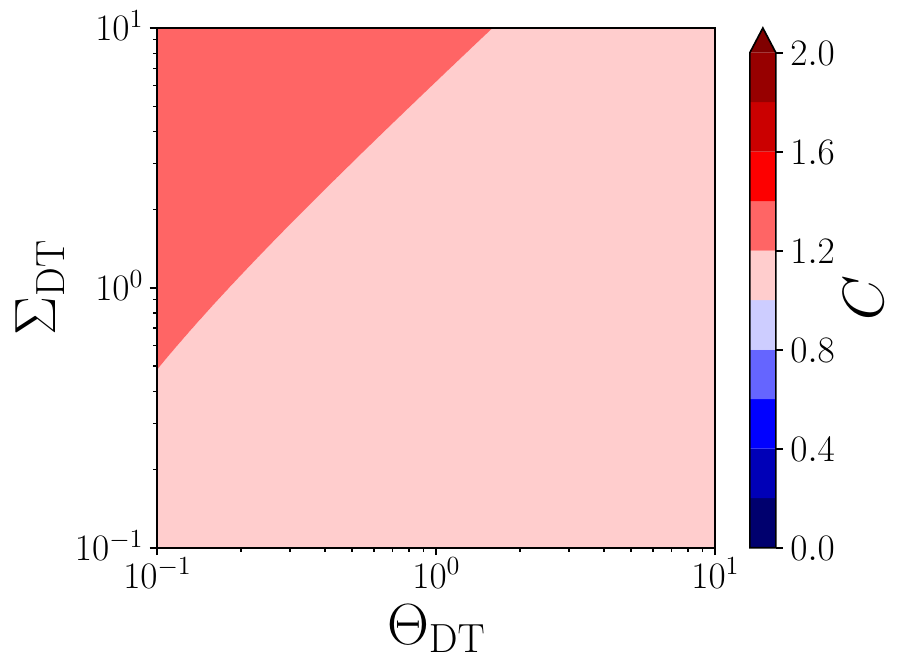}
    \caption{TBE=0.05.}
    \end{subfigure}
    \caption{Required $n_e \tau_E$ multiplication factor $C$ (\Cref{eq:C}) versus $\Theta_{\mathrm{DT}}$, $\Sigma_{\mathrm{HeT}}$ to maintain a fixed plasma gain. Lower $C$ values indicate a higher fusion gain.}
    \label{fig:required_C_vals}
\end{figure}

We now show the effects of $\Theta_{\mathrm{DT}}$ and $\Sigma_{\mathrm{DT}}$ on $C$. In \Cref{fig:required_C_vals}, we plot the $C$ values corresponding to the three TBE values in \Cref{fig:pDelta_diff_DTpumpingtransport} (TBE=0.50,0.20,0.05) versus $\Theta_{\mathrm{DT}}$ and $\Sigma_{\mathrm{DT}}$ for three TBE values. At higher TBE values, TBE = 0.50 in \Cref{fig:required_C_vals}(a), there are two regions for a $C$ minimum with the $\Theta_{\mathrm{DT}}$ and $\Sigma_{\mathrm{DT}}$ values we consider (however, the upper spatial region in \Cref{fig:required_C_vals}(a) is undesirable because the fusion power is very low), but at lower TBE values, there is a single $C$ minimum region in $\Theta_{\mathrm{DT}}$ and $\Sigma_{\mathrm{DT}}$. The results in \Cref{fig:required_C_vals} showing higher $\Theta_{\mathrm{DT}}$ and lower $\Sigma_{\mathrm{DT}}$ increasing the fusion gain further reinforce the trend we find of higher $\Theta_{\mathrm{DT}}$ and lower $\Sigma_{\mathrm{DT}}$ to be desirable for fusion power.

\section{Vacuum pumping} \label{sec:vacuum_pumping}

In this section, we briefly review vacuum pumping and relate pumping speed to the main divertor quantities introduced above.

The throughput $Q_{\mathrm{pump}}$ of a vacuum pump is
\begin{equation}
    Q_{\mathrm{pump}} \equiv p S = \dot{N}k_B T
    \label{eq:pumpthroughput}
\end{equation}
where $p$ is the pressure at the pump inlet, $S$ is the volumetric pumping speed, $\dot{N}$ is the atomic or molecular flow rate, $k_B$ is the Boltzmann constant and $T$ is the temperature. Note that the pump throughput is typically reported in units of \SI{}{Pa.m^3.s^{-1}} instead of \SI{}{W}, and is usually referenced to the standard temperature $T_\mathrm{st} = \SI{273.15}{K}$.

The ideal pumping speed of a pump is
\begin{equation}\label{eq:pumping_speed_ideal}
    S_\mathrm{id} = A_p \frac{\Bar{c}}{4},
\end{equation}
where $\Bar{c}$ is the mean gas velocity and $A_p$ is the cross sectional area of the pump. Assuming a Maxwell-Boltzmann distribution, $\Bar{c} = \sqrt{8 RT/\pi M}$ where $M$ is the molar mass of the gas and $R$ is the ideal gas constant. This (kinetic) approach is valid in the free molecular flow regime. 

In a real vacuum system, the gas experiences different flow regimes (free molecular, transitional, and viscous), and can be described by the Knudsen number $\mathrm{Kn} \equiv \lambda / d$ where $\lambda$ is the gas particle mean free path and $d$ is the effective molecular diameter. A high Knudsen number ($\mathrm{Kn} \gtrsim 10$) indicates a free molecular flow, while a low Knudsen number ($\mathrm{Kn} \lesssim 0.01$) indicates a viscous flow. Between these two regimes, a transitional flow occurs.

Torus pumping systems have been extensively characterized and analyzed using kinetic models that cover all the regimes, with typical dome pressures ranging from \qty{1}{Pa} to \qty{10}{Pa} during the burn phase. The complexity of divertor neutral flows due to intricate geometries requires advanced computational models to accurately account for the conductance of all channels and piping \cite{day2014towards, hauer2015iter, vasileiadis2016modeling}. For simplicity, we assume a purely molecular flow regime, allowing us to derive idealized relationships between the pumping speeds of different species (e.g., species $x$ and $y$).
Assuming an equal temperature for $x$ and $y$, \Cref{eq:pumping_speed_ideal} gives
\begin{equation}\label{eq:pump_speed_ratio}
    \frac{S_x}{S_y} = \sqrt{\frac{M_y}{M_x}}.
\end{equation}
Considering that hydrogenic species are found as molecules ($Q_2$ in the divertor neutrals region), while helium is in atomic form, from \Cref{eq:pump_speed_ratio}:
\begin{equation}\label{eq:pumping_speed_D_He}
    \frac{S_\mathrm{D_2} }{S_\mathrm{He}} = 1,
\end{equation}
and
\begin{equation}\label{eq:pumping_speed_T_D}
    \frac{S_\mathrm{T_2}}{S_\mathrm{He}} = \frac{S_\mathrm{T_2}}{S_\mathrm{D_2}} = \sqrt{\frac{2}{3}}.
\end{equation}
Neglecting for simplicity the differences in capture coefficient, the relation between the pumping speeds of different species is given by Eqs. \eqref{eq:pumping_speed_D_He} and \eqref{eq:pumping_speed_T_D}. The total pumping speed provided by $N_\mathrm{p}$ pumps is:
\begin{equation}
    \begin{aligned}
        S_\mathrm{tot} = \frac{N_\mathrm{p}Q_{\mathrm{pump}}}{p_\mathrm{div}} = N_\mathrm{p}(n_\mathrm{T}^\mathrm{div}S_\mathrm{T,div} + \\ + n_\mathrm{D}^\mathrm{div} S_\mathrm{D,div} + n_\mathrm{He}^\mathrm{div}S_\mathrm{He,div}) \frac{k_B T}{p_\mathrm{div}},
    \end{aligned}
\end{equation}
that can be rewritten as:
\begin{equation}
\begin{aligned}
S_\mathrm{tot} = & N_\mathrm{p} \dot{N}_\mathrm{T}^\mathrm{div} \frac{k_B T}{p_\mathrm{div}} \times \\
& \bigg{(}1 + f_\mathrm{DT, div} \Sigma_\mathrm{DT} + f_\mathrm{He,div}\Sigma_\mathrm{HeT} \bigg{)},
\end{aligned}
\label{eq:total_S}
\end{equation}
again assuming that all species have an equal temperature.

\begin{figure}
    \centering
    \includegraphics[width=\linewidth]{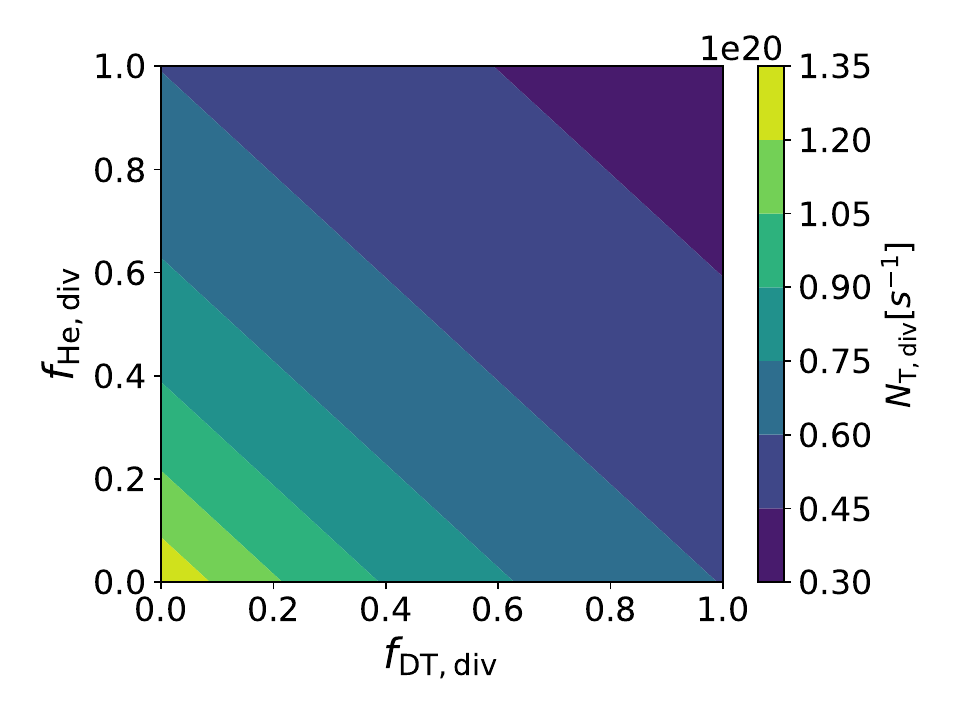}
    \caption{Maximum exhausted tritium flow rate versus $f_\mathrm{DT, div}$ and $f_\mathrm{He,div}$. Here, $S_\mathrm{tot} = \SI{50}{m^3.s^{-1}}$ and $p_\mathrm{div} = \SI{5}{Pa}$.}
    \label{fig:N_Tdiv_fHe_fDT}
\end{figure}

Using these results, we can find $\dot{N}_\mathrm{T}^\mathrm{div}$ as a function of just two variables: $f_\mathrm{DT, div}$ and $f_\mathrm{He,div}$. Considering that: $S_\mathrm{tot}$ is limited by the pump performance; \Cref{eq:pumping_speed_D_He,eq:pumping_speed_T_D} prescribe the relative pumping speeds; $k_B T/p_\mathrm{div}$ can be considered a constant ($T = T_\mathrm{st}$ and $p_\mathrm{div} \sim 1 - \SI{10}{Pa}$); and the number of pumps $ N_\mathrm{p}$ is constrained by geometrical considerations;
\Cref{eq:total_S} provides a relation between ($\dot{N}_\mathrm{T}^\mathrm{div}$, $f_\mathrm{DT, div}$, $f_\mathrm{He,div}$) and engineering parameters. \Cref{fig:N_Tdiv_fHe_fDT} shows the maximum allowable tritium flux through the divertor by using 6 pumps with $S_\mathrm{tot} = \SI{50}{m^3.s^{-1}}$ (typical pumping speed of vapor diffusion pumps) as a function of $f_\mathrm{DT, div}$, $f_\mathrm{He,div}$.

\Cref{eq:total_S} can be related to core transport by using
\begin{equation}
 \dot{N}_\mathrm{T}^\mathrm{div} \simeq \dot{N}_\mathrm{T}^\mathrm{sep} = - A D_\mathrm{T} \nabla n_\mathrm{T}^{\mathrm{co}}\bigg|_{sep},
\end{equation}
allowing one to link the radial profile of $n_\mathrm{T}^{\mathrm{co}}$ with the pumping speed and the divertor densities of D, T, and He. %

The approach presented in this section is useful for deriving idealized relationships between the pumping speeds of different species. However, as explained previously, it is not sufficient for a detailed modeling of the torus pumping system, as a more comprehensive framework accounting for the full range of flow regimes is required \cite{day2014towards,hauer2015iter,vasileiadis2016modeling}.

\section{COMPARISON WITH EXISTING TECHNOLOGY} \label{sec:current_technology}

In this section, we briefly review the new results in this paper in the context of current fuel cycle technologies. Most fuel cycle designs have focused on operations with a 50:50 D-T ratio, as this ratio maximizes fusion power density and, at low TBE, results in an exhaust stream with a similar ratio, facilitating direction internal recycling (DIR) operations \cite{Day_2013}. However, in this work, we demonstrated that operating at a low $F_\mathrm{T}^\mathrm{in}$ and leveraging asymmetric deuterium-tritium transport and pumping lead to (a) a significant enhancement in TBE, and (b) a much broader operating space for D-T fractions, both in the core and in the divertor. It is important to evaluate how these novel operational schemes affect fuel cycle design and how existing technologies constrain the parameter space presented in \Cref{sec:diff_part_transp}.

\subsection{Fueling system}\label{subsec:fueling_system}
Fueling schemes anticipated for ITER and modern fusion devices primarily rely on fuel pellet injection. Fuel pellets can be produced by pipe guns, which also provide pellet acceleration, or by extruders, with the latter being preferred for their ability to produce a larger number of pellets over extended periods. The pellets produced by the extruder are then accelerated using gas guns (single or two-stage), mechanical centrifuges, electro-thermal accelerators, or electro-magnetic rail guns \cite{combs2018pellet}.
No significant issues are expected when operating at different D-T injection ratios. Extruders are capable of producing pellets with any D-T mixture. Similarly, the performance of pellet injectors is not affected by the D-T mixture. While operating with a rapidly changing D-T fraction may be challenging due to the timescales associated with the extrusion process (on the order of minutes), maintaining operation around a steady-state value, even at very low tritium fractions, appears feasible with current technologies \cite{baylor2005pellet}.

An important constraint in the fueling system comes from the maximum amount of tritium contained in a given volume \cite{baylor2016pellet}, e.g., the extruder. Therefore, operations at low $F_\mathrm{T}^\mathrm{in}$ are beneficial for tritium management and safety.  
\Cref{tab:tab1} shows the relevant fuel cycle parameters for an ARC-class FPP. A design implementing differential transport and pumping ($\Theta_\mathrm{DT} = 5$, $\Sigma_\mathrm{DT} = 1/5$, $F_\mathrm{T}^\mathrm{in} = 0.17$) achieves a TBE of 29\%, and a total particle flow rate $\dot{N}_\mathrm{tot,in} = \dot{N}_\mathrm{T}^\mathrm{in} + \dot{N}_\mathrm{D}^\mathrm{in} = \SI{3.5e21}{s^{-1}}$, or $Q_{\mathrm{tot,in}} = \SI{13.1}{Pa.m^3.s^{-1}}$. For a similar FPP operating without differential transport and pumping ($\Theta_\mathrm{DT} = 1$, $\Sigma_\mathrm{DT} = 1$, $F_\mathrm{T}^\mathrm{in} = 0.48$),  TBE = 3\%, the total particle flow rate is much higher $\dot{N}_\mathrm{tot,in} = \SI{1.4e22}{s^{-1}}$, or $Q_{\mathrm{tot,in}} = \SI{51.6}{Pa.m^3.s^{-1}}$. Thanks to the TBE improvements achieved by D-T asymmetric transport and pumping, both the total flow rate and the tritium flow rate are lower, despite the fractional increase in D flow rate. Thus, tritium-lean plasmas also tend to be deuterium-lean.

\subsection{Exhaust pumping}\label{subsec:exhaust_pumping}
In \Cref{sec:vacuum_pumping} we computed $\dot{N}_\mathrm{T}^\mathrm{div}$ for an ideal vacuum pump. It should be noted that in a real pump additional factors can affect the pumping speed. The non-ideal behaviour of a real pump is accounted for by the capture coefficient, $c_\mathrm{capt}$:
\begin{equation}
    c_\mathrm{capt} = \frac{S_\mathrm{real}}{S_\mathrm{id}},
\end{equation}
\noindent where $S_\mathrm{real}$ is the real pump speed. The capture coefficient is therefore computed from experimental tests or detailed simulations of the pump operations. For a cryopump, in addition to the geometry and the flow regime, $c_\mathrm{capt}$ depends on the sticking coefficient ($c_\mathrm{stick}$) of the different species in the exhaust stream. For a typical charcoal cryosorbent (e.g., the one used in ITER), $c_\mathrm{stick, He} = 0.1$, $c_\mathrm{stick, D_2} = 0.88$, and $c_\mathrm{stick, T_2} = 1$ \cite{day2005performance}, resulting in higher pumping speed for D and T than He. For a vapor diffusion pump, the capture coefficient depends on the interaction between the gas with the vapor jet, with lighter elements showing a lower capture coefficient due to increased backflow through the jet. The combined effect of the $m^{-1/2}$ dependence of the pumping speed and of the caputure coefficient still results in He (and D) achieving higher pumping speeds than T2 \cite{teichmann2021particle}. These simple considerations highlight how differential pumping is strongly driven by the technology (and the underlying physical principles) employed in the torus pumping, and the limitations of an ideal pumping approach.

Common pumping speeds for candidate primary pumping technology are 50 - \SI{80}{m^3.s^{-1}} for cryopumps, 50 - \SI{70}{m^3.s^{-1}} for vapor diffusion pumps, and 1 - \SI{3}{m^3.s^{-1}} for turbomolecular pumps. However, practical considerations (e.g., tritium safety) limits the throughput of cryopumps, resulting in a much lower effective pumping speed. For instance, in ITER cryopumps the surface saturation of charcoal is reached at a gas load of $\SI{0.5}{Pa.m^3}$ per $\SI{}{cm^2}$ of sorbent surface. Each cryopump has 28 baffles coated with charcoal, for a total adsorbing surface of $\SI{11.2}{m^2}$ per pump \cite{mack2002design}. Considering a throughput of $\sim \SI{200}{Pa.m^3.s^{-1}}$ \cite{abdou2020physics}, and 6 cryopumps, the time before regeneration due to saturation of the sorbent is:

\begin{equation}\label{eq:pump_saturation}
    \Delta t_\mathrm{sat} = 0.5 \frac{N_\mathrm{p}A_\mathrm{sorbent}}{Q_{\mathrm{pump}}} \sim \SI{1100}{s},
\end{equation}

\noindent while the time before regeneration due to hydrogen flammability limit (in a volume $V \sim \SI{6}{m^3}$ as per ITER cryopumps \cite{day2007basics}, and a maximum allowable partial pressure of $p_\mathrm{H_2,max} = \SI{1.7}{kPa}$ \cite{graham1979cryopump}) is:

\begin{equation}\label{eq:pump_flammability}
    \Delta t_\mathrm{flamm} = \frac{p_\mathrm{H_2,max}V}{Q_{\mathrm{pump}}} \sim \SI{50}{s},
\end{equation}

\noindent or, stated in another way, for $\Delta t$ of interest for FPPs, the throughput of a cryopump is $\sim 20\times$ less than the one predicted by considering the cryosorbent saturation as limiting factor. Note also that $\Delta t_\mathrm{flamm}$ depends on the partial pressure of all the hydrogenic species, not just tritium. Increasing $f_\mathrm{He,div}$ or $\Sigma_\mathrm{HeT}$ is an effective way to reduce the throughput and, consequently, to increase the operating time of the cryopump before regeneration is needed. Instead, decreasing the tritium fraction relative to the other hydrogen species does not provide benefits unless the hydrogen throughput decreases, as the bottleneck for operations is the flammability limit and not the tritium inventory in the cryopump.

For the values in \Cref{tab:tab1}, referring to the same cases of \Cref{subsec:fueling_system}, for a device using differential transport and pumping the total flow rate through the exhaust system is $\dot{N}_\mathrm{tot}^\mathrm{div} = \dot{N}_\mathrm{T}^\mathrm{div} + \dot{N}_\mathrm{D}^\mathrm{div} + \dot{N}_\mathrm{He}^\mathrm{div} = \SI{3.3e21}{s^{-1}}$, or $Q_{\mathrm{pump}} = \SI{12.4}{Pa.m^3.s^{-1}}$. If we compare these values to a standard 50:50 D-T burn scheme, the total flow rate through the exhaust system is $\dot{N}_\mathrm{tot}^\mathrm{div} = \SI{1.4e22}{s^{-1}}$, or $Q_{\mathrm{pump}} = \SI{51.1}{Pa.m^3.s^{-1}}$.

Therefore, the throughput is reduced by $\sim 5\times$ with asymmetric D-T transport and pumping, which is also reflected in the lower $I_\mathrm{startup}$. From the perspective of pumping requirements, the reduced tritium content enhances radiological safety, while the lower throughput extends the pump's operating time before regeneration is necessary (both \Cref{eq:pump_saturation} and \Cref{eq:pump_flammability} shows a $Q_{\mathrm{pump}}^{-1}$ dependence for $\Delta t$). Similarly, turbomolecular pumps may be capable of providing the required throughput when operating with asymmetric D-T transport and pumping, provided they can be made tritium compatible.

It should be noted that the required throughput must account for additional factors related to plasma control and operations (e.g., gas injection in the divertor region). Therefore, the throughput reported here does not guarantee that these technologies can be applied under the conditions described in \Cref{tab:tab1} (machine-specific considerations must be made), but it does illustrate how asymmetric D-T transport and pumping can ease some of the demands on the torus pumping system.

\subsection{Direct Internal Recycling}
The efficacy of DIR depends on its ability to directly recycle D and T with minimal processing time, delivering them to the gas distribution system before fueling them back in the fusion system. Similar D-T ratios at injection and exhaust allows for a short processing time, with the option of adjusting the mixture in the gas distribution system \cite{day2022pre}. However, if the exhaust D-T ratio differs significantly from that at fueling, an isotope rebalancing system may be needed, increasing the processing time of the DIR, limiting its effectiveness. For fuel pellet injection
\begin{equation}
    F_\mathrm{T}^\mathrm{in} = \frac{\dot{N}_\mathrm{T}^\mathrm{in}}{\dot{N}_\mathrm{Q}^\mathrm{in}} = \frac{n_\mathrm{T}^\mathrm{in}}{n_\mathrm{Q}^\mathrm{in}} = \frac{1}{1 + f_\mathrm{DT}^\mathrm{in}},
\end{equation}
because the speed of the two species (D and T) is the same, and equals the pellet speed. This means that the D-T fueling ratio is prescribed once $F_\mathrm{T}^\mathrm{in}$ is fixed. $f_\mathrm{DT,div}$ depends instead on both $F_\mathrm{T}^\mathrm{div}$ and $\Sigma_\mathrm{DT}$:
\begin{equation}
    F_{\mathrm{T}}^{\mathrm{div}} = \frac{1}{1 + f_{\mathrm{DT,div}} \Sigma_{\mathrm{DT}}}.
    \label{eq:FTdiv_fDT_SigmaDT}
\end{equation}
When $\Sigma_\mathrm{DT} \simeq 1$, DIR operations are mostly unaffected if $F_\mathrm{T}^\mathrm{in} \simeq F_\mathrm{T}^\mathrm{div}$ (i.e., $H_\mathrm{T} \simeq 1$). However, when $\Sigma_\mathrm{DT}$ differs significantly from 1, the optimal value of $H_\mathrm{T}$ for DIR operations changes. For the values in \Cref{tab:tab1} ($\Sigma_\mathrm{DT} = 1/5$, $f_\mathrm{T}^\mathrm{sep} = 0.43$, and $H_\mathrm{T} = 0.13/0.17 = 0.76$). However, because DIR requires $F_\mathrm{T}^\mathrm{in} = F_\mathrm{T}^\mathrm{sep}$, using $F_\mathrm{T}^\mathrm{in} \simeq F_\mathrm{T}^\mathrm{div}$, DIR operates with $H_\mathrm{T} \simeq 1.0$. Therefore, it is not certain whether the tritium-lean plasmas described here are compatible with DIR.

In terms of operations, the D-T ratio may or may not affect the performance of candidate technologies \cite{day2013direct, yamaguchi2022evaluation}. For metal foil pumps, plasma-driven permeation through Nb foils has been shown to exhibit no isotopic effect between H$_2$ and D$_2$ \cite{day2022pre}. Conversely, proton-conducting pumps do show isotope effects, indicating that their efficiency may vary depending on the D-T fuel mix \cite{iwahara1999hydrogen}.

\subsection{Tritium processing plant}

Operating at very high TBE changes some of the paradigms in fuel cycle dynamics. The critical role of the tritium processing plant (referred to as the inner fuel cycle in \cite{abdou2020physics, meschini2023modeling}) in achieving tritium self-sufficiency is evident since at low TBE, the tritium fueling flow rate significantly exceeds the tritium burned, with most of it being exhausted. For instance, at TBE = 2\%, $\dot{N}_\mathrm{T}^\mathrm{in} = 50 \dot{N}_\mathrm{T,burn}$ and $\dot{N}_\mathrm{T}^\mathrm{div} = 49 \dot{N}_\mathrm{T,burn}$. This results in high $I_\mathrm{startup}$, and, consequently, in high $\mathrm{TBR_\mathrm{r}}$. Strategies that shortens the residence (processing) time of tritium in the tritium plant prove to be very effective because they impact $\sim 98\%$ of the total tritium re-circulating in the plant. 

Operating at high TBE with asymmetric D-T transport and pumping (assume TBE = 60\% as in \Cref{tab:tab1}) has two major effects: (a) the amount of tritium injected is comparable to the tritium burnt ($\dot{N}_\mathrm{T}^\mathrm{in} = 1.2 \dot{N}_\mathrm{T}^\mathrm{burn}$), greatly reducing $I_\mathrm{startup}$, and (b) the amount of tritium exhaust is even lower than the tritium burnt ($\dot{N}_\mathrm{T}^\mathrm{div} = 0.6 \dot{N}_\mathrm{T,burn}$).
Considering that the tritium extracted from the outer fuel cycle (i.e., the breeding blanket, tritium extraction system, and, potentially, heat exchangers) is, roughly speaking, $\mathrm{TBR} \cdot \dot{N}_\mathrm{T,burn}$, the impact of the design parameters of outer fuel cycle components (residence times, tritium extraction efficiency, tritium losses, etc.) becomes comparable to those of the inner fuel cycle, which is not the case for low TBE operations (see for instance the parametric analysis in \cite{abdou2020physics, meschini2023modeling}). In general, reducing the tritium flow rate through the reprocessing plant enables a broader set of technologies, including those with longer residence times, should they provide benefits in terms of other performance metrics (e.g., purity level of the streams).

In this section, we have reviewed the current technological capabilities of fueling systems, exhaust pumping, direct internal recycling, and tritium processing. Developing pumps with asymmetric D-T pumping is a high priority, and we present such a scheme in \Cref{sec:future_tech_develop}. Direct internal recycling may not be needed with asymmetric D-T transport and pumping. Generally, operating at higher TBE allows a wider range of tritium processing technologies to be considered.

\section{ARC-class Study} \label{sec:ARC}

\begin{figure}
    \centering
    \includegraphics[width=\linewidth]{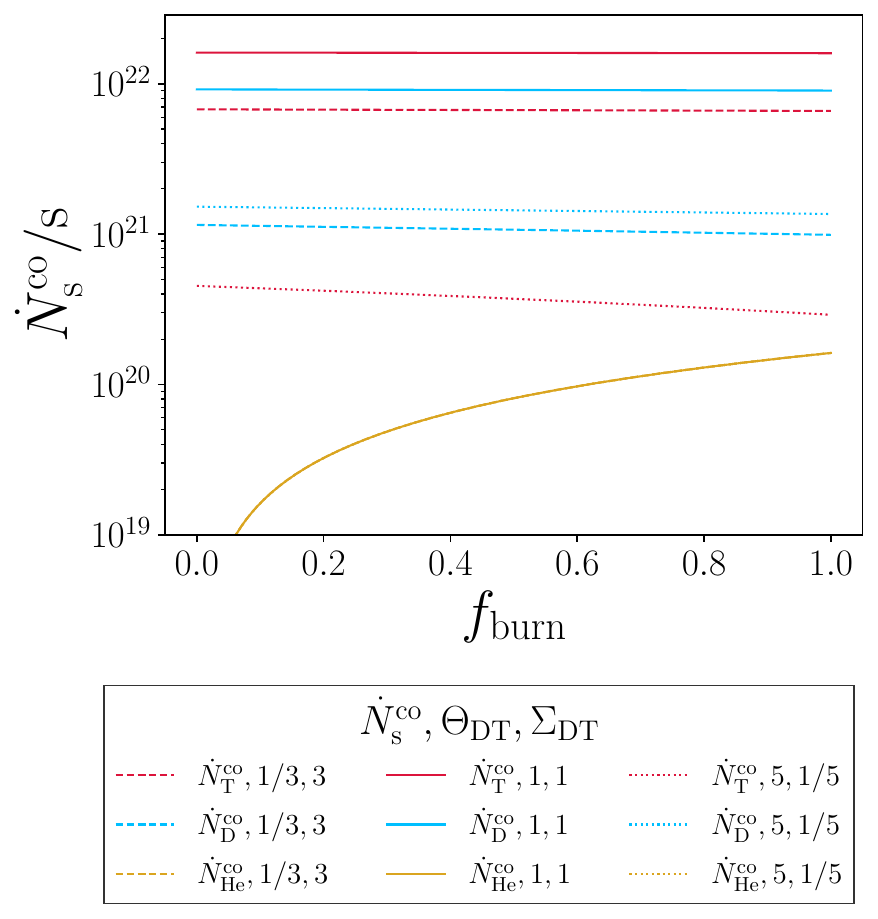}
    \caption{Radial dependence of deuterium, tritium, and helium flows for three ARC-class cases at fixed fusion power described in \Cref{sec:ultrahighPf_regimes}.}
    \label{fig:profiles_constant_power}
\end{figure}
In this section, we show the effect of differential deuterium-tritium transport and pumping on an ARC-class power plant. We consider two main regimes: high fusion power and high TBE. We then perform dedicated scans in $\Sigma_{\mathrm{DT}}$, $\Theta_{\mathrm{DT}}$, and $\Sigma_{\mathrm{DT}}$ and $\Theta_{\mathrm{DT}}$ simultaneously in order to clarify the separate effects of $\Sigma_{\mathrm{DT}}$ and $\Theta_{\mathrm{DT}}$. 

\begin{figure}[!tb]
    \centering
    \begin{subfigure}[t]{0.9\textwidth}
    \centering
    \includegraphics[width=1.0\textwidth]{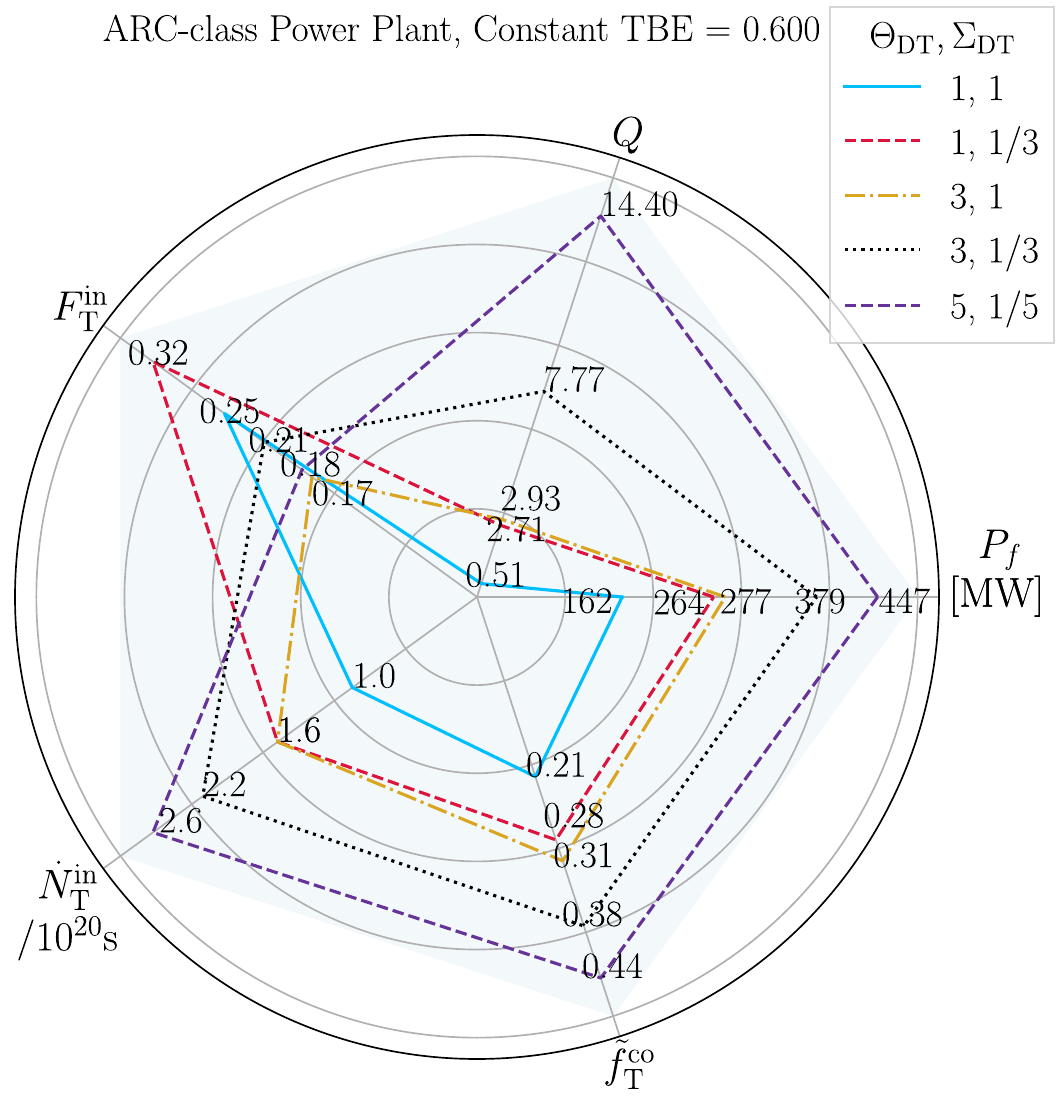}
    \caption{TBE=0.60.}
    \end{subfigure}
     ~
    \begin{subfigure}[t]{0.9\textwidth}
    \centering
    \includegraphics[width=1.0\textwidth]{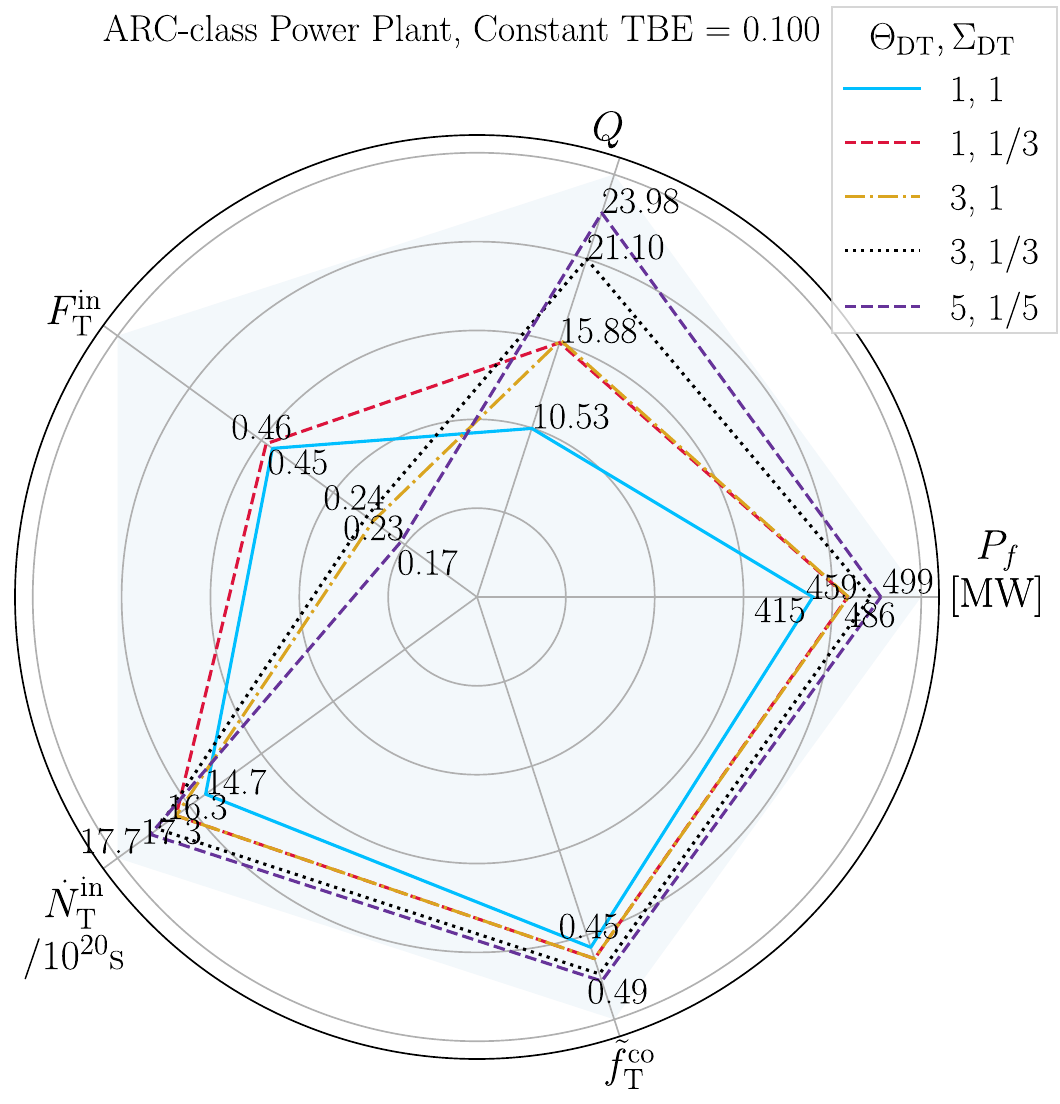}
    \caption{TBE=0.10.}
    \end{subfigure}
    \caption{Tritium self-sufficiency and fusion power parameters of an ARC-class device for five cases with very high tritium burn efficiency (TBE) = 0.60 (a) and moderate TBE = 0.10 (b): The D-T transport ratio is $\Theta_{\mathrm{DT}}$ (\Cref{eq:ThetaDTratio}), the D-T pumping speed ratio is $\Sigma_{\mathrm{DT}}$ (\Cref{eq:SigmaDT}), the fusion power is $P_f$, the tritium flow rate injection fraction is $F_{\mathrm{T}}^{\mathrm{in}}$ (\Cref{eq:fueling_ratio}), the plasma gain is $Q_{\mathrm{plas}}$, the tritium fueling rate is $\dot{N}_{\mathrm{T} }^{\mathrm{in} }$, and the typical tritium fraction is $\tilde{f}_{\mathrm{T}}^{\mathrm{co}}$ (\Cref{eq:fTcotilde}). All axes except for $Q$ have a value of zero at the origin and are linearly scaled.}
    \label{fig:arcclass_spider2}
\end{figure}

Our nominal case for the ARC-class fusion system has TBE = 0.026, $P_f = 481$ MW, $Q = 20$, and a startup tritium inventory of 0.40 kg \cite{Meschini2023}. Note that this is a higher TBE than the nominal case described in \cite{Parisi_2024} because in this work we use a more optimistic value of $\Sigma_{\mathrm{HeT}} = 1.0$ whereas in \cite{Parisi_2024} $\Sigma_{\mathrm{HeT}} = 0.63$. For all of the cases in this section, either the TBE or fusion power is maximized by finding the optimal value of $F_{\mathrm{T}}^\mathrm{in}$.

\subsection{High Fusion Power} \label{sec:ultrahighPf_regimes}

In this subsection, we study a high power ARC-class fusion system with $P_f = 481$ MW. By high power, we mean that the power degradation is relatively small ($p_{\Delta} = 0.95$), which corresponds to lower TBE values. The tritium input fraction $F_{\mathrm{T}}^{\mathrm{in}}$ is chosen to maximize the TBE at fixed power. \Cref{fig:arcclass_spider} shows the corresponding spider diagram. There is a 30-fold decrease in the required tritium fueling from the nominal to the leanest tritium case ($\Theta_{\mathrm{DT}} = 5$, $\Sigma_{\mathrm{DT}}=1/5$), and the TBE increases from 0.01 to 0.36. Previous results have shown that TBE increases with lower $\tilde{f}_{\mathrm{T}}^{\mathrm{co}}$ \cite{Boozer2021,Parisi_2024}. This work shows that by preferentially transporting deuterium and pumping tritium, one obtains higher $\tilde{f}_{\mathrm{T}}^{\mathrm{co}}$ while still running a tritium-lean power plant. Shown in \Cref{tab:tab1}, at very large $\Theta_{\mathrm{DT}}$ and small $\Sigma_{\mathrm{DT}}$, the amount of injected and divertor-pumped tritium and deuterium \textit{both} decrease. This is indicates that tritium-lean power plants may also be more deuterium-efficient.

To further illustrate the significant differences between these cases with constant fusion power, in \Cref{fig:profiles_constant_power} we plot the radial dependence of the deuterium, tritium, and helium particle particles flows. There is an almost two order of magnitude reduction in the tritium flows in the tritium-lean case compared with the nominal and tritium-abundant cases. Because the fusion power is constant, there is no difference in the helium flows.

Further details are listed in \Cref{tab:tab1}. $I_{\mathrm{startup,min}}$ is the minimum tritium startup inventory according to \cite{Meschini2023}.

In summary, at high fusion power (481 MW, $p_{\Delta} = 0.95$) we find that while the nominal ARC-class plant only achieves TBE = 0.026, with significant decreases in particle transport and divertor pumping, much higher TBE is attainable; we showed a tritium-lean case achieving TBE = 0.29.

\subsection{High TBE} \label{sec:ultrahighTBE_regimes}

In this subsection, we study the effect of $\Theta_{\mathrm{DT}}$ and $\Sigma_{\mathrm{DT}}$ at high TBE values. At such high TBE values, the importance of $\Theta_{\mathrm{DT}}$ and $\Sigma_{\mathrm{DT}}$ increases significantly. 

\Cref{fig:arcclass_spider2} shows two spider diagrams for an ARC-class case with very high TBE = 0.60 (\Cref{fig:arcclass_spider2}(a)) and moderate TBE = 0.10 (\Cref{fig:arcclass_spider2}(b)).

For very high TBE = 0.60, the fusion power and gain are very sensitive to $\Theta_{\mathrm{DT}}$ and $\Sigma_{\mathrm{DT}}$: as $\Theta_{\mathrm{DT}}$ increases and $\Sigma_{\mathrm{DT}}$ decreases, the plasma gain and fusion power increase. The blue line shows the case with $\Theta_{\mathrm{DT}} = \Sigma_{\mathrm{DT}} = 1$ and TBE = 0.60. This is not a viable plasma for a power plant: the gain is only $Q = 0.58$ and the power is only 162 MW. Although not shown in the table, with $\Theta_{\mathrm{DT}} = 1/3$ and $\Sigma_{\mathrm{DT}} = 3$, the plasma produces only 38 MW of fusion power. Therefore, it is important that $\Theta_{\mathrm{DT}} \gtrsim 1$ and $\Sigma_{\mathrm{DT}} \lesssim 1$, otherwise the TBE and/or fusion power will be very poor.

Increasing $\Theta_{\mathrm{DT}} \to 5$ and decreasing $\Sigma_{\mathrm{DT}} \to 1/5$, the power increases to 447 MW. In this case, the deuterium divertor density is 51 times higher than the tritium divertor density, $f_{\mathrm{DT,div}} = 51$ and the tritium and deuterium injection rates are relatively modest at $\dot{N}_\mathrm{T}^{\mathrm{in}} = 2.7 \times 10^{20} /\mathrm{s}$ and $\dot{N}_\mathrm{D}^{\mathrm{in}} = 12.5 \times 10^{20} /\mathrm{s}$. We can estimate the hydrogen divertor density from \Cref{eq:neutralpump},
\begin{equation}
n_{\mathrm{Q}}^\mathrm{div} = \frac{\dot{N}_{\mathrm{D} }^{\mathrm{div} }} { S_{\mathrm{D} }} + \frac{\dot{N}_{\mathrm{T} }^{\mathrm{div} }} { S_{\mathrm{T} }}.
\end{equation}
We use the value of $S_\mathrm{tot} = \SI{50}{m^3.s^{-1}}$ from \Cref{sec:vacuum_pumping} to motivate using a tritium pumping speed $S_{\mathrm{T} } = \SI{50}{m^3.s^{-1}}$. Given that $\Sigma_{\mathrm{DT} } = 1/5$, the deuterium pumping speed is $S_\mathrm{tot} = \SI{10}{m^3.s^{-1}}$. Using the values of $\dot{N}_{\mathrm{D} }^{\mathrm{div} } = 10.9 \times 10^{20}$/s and $\dot{N}_{\mathrm{T} }^{\mathrm{div} } = 1.1 \times 10^{20}$/s from \Cref{tab:tab1}, we find $n_{\mathrm{Q}}^\mathrm{div} = 1.1 \times 10^{20}/\mathrm{m}^3$. This separatrix density is comparable to the MANTA-class device design with $n_{\mathrm{e,sep}} = 0.9 \times 10^{20}/\mathrm{m}^3$ \cite{Rutherford_2024} -- the main difference being that the for the MANTA-class device $f_{\mathrm{DT,div}} \simeq 1$ whereas in our example $f_{\mathrm{DT,div}} = 51$.

However, at moderate TBE values, TBE = 0.10, $\Theta_{\mathrm{DT}}$ and $\Sigma_{\mathrm{DT}}$ have a much weaker effect on the fusion power. The difference in fusion power for the most pessimistic and optimistic cases we consider, $\Theta_{\mathrm{DT}} = 1/3$ and $\Sigma_{\mathrm{DT}} = 3$ and $\Theta_{\mathrm{DT}} = 5$ and $\Sigma_{\mathrm{DT}} = 1/5$, is much smaller: the former has $P_f = 351$ and the latter $P_f = 499$ MW. 

In summary, we have shown that power plants with extremely high TBE -- here, TBE = 0.60, may be attained with very large $\Theta_{\mathrm{DT}}$ and very small $\Sigma_{\mathrm{DT}}$. Compared with the nominal transport and pumping parameters $\Theta_{\mathrm{DT}} = 1$ and $\Sigma_{\mathrm{DT}} = 1$, a power plant with $\Theta_{\mathrm{DT}} = 5$ and $\Sigma_{\mathrm{DT}} = 1/5$ is predicted to achieve a fusion power that is nearly three times higher (447 MW versus 162 MW).

\begin{figure}[!tb]
    \centering
    \begin{subfigure}[t]{0.77\textwidth}
    \centering
    \includegraphics[width=1.0\textwidth]{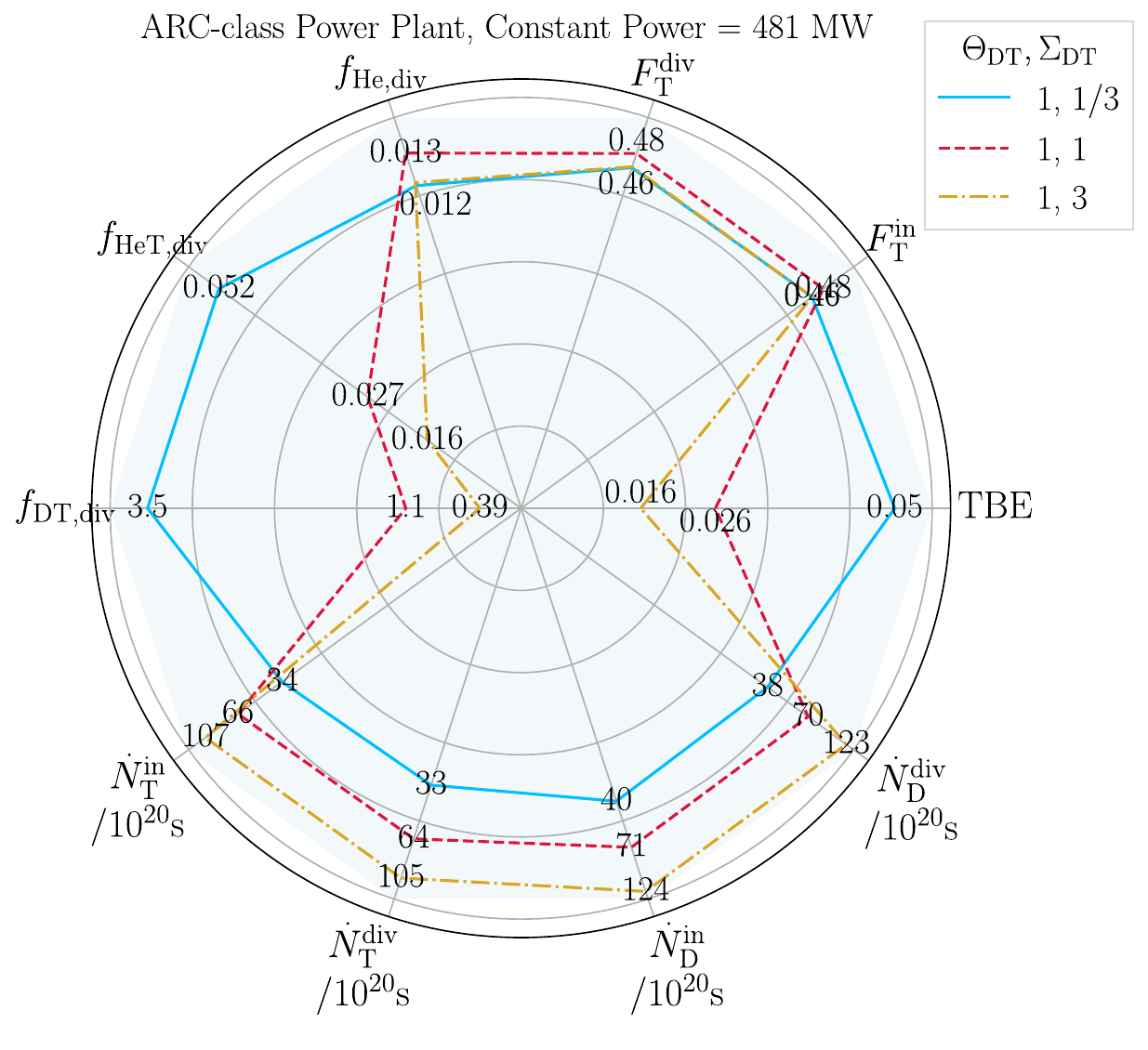}
    \caption{Scan in $\Sigma_{\mathrm{DT} }$.}
    \end{subfigure}
     ~
    \begin{subfigure}[t]{0.77\textwidth}
    \centering
    \includegraphics[width=1.0\textwidth]{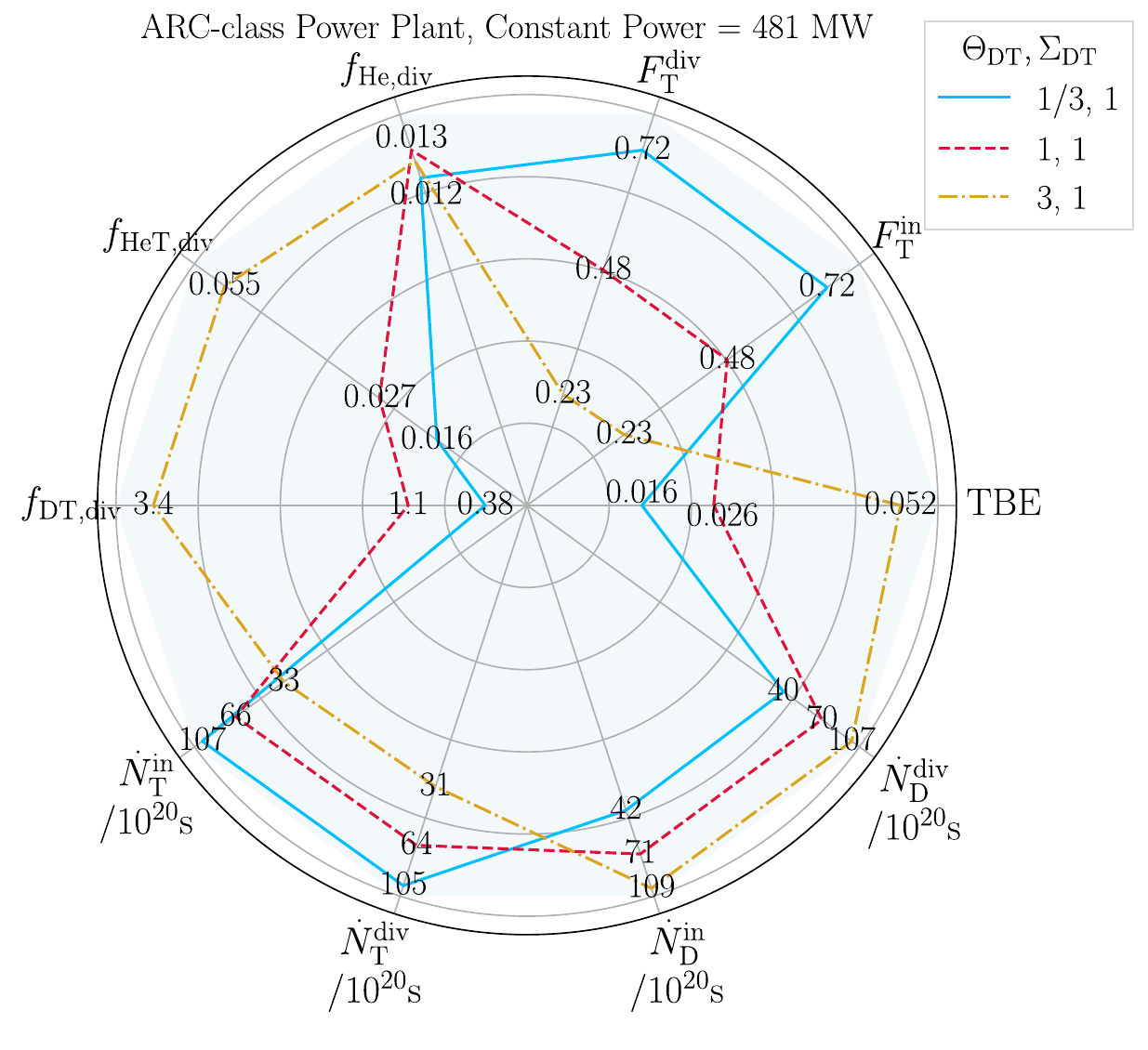}
    \caption{Scan in $\Theta_{\mathrm{DT} }$.}
    \end{subfigure}
     ~
    \begin{subfigure}[t]{0.77\textwidth}
    \centering
    \includegraphics[width=1.0\textwidth]{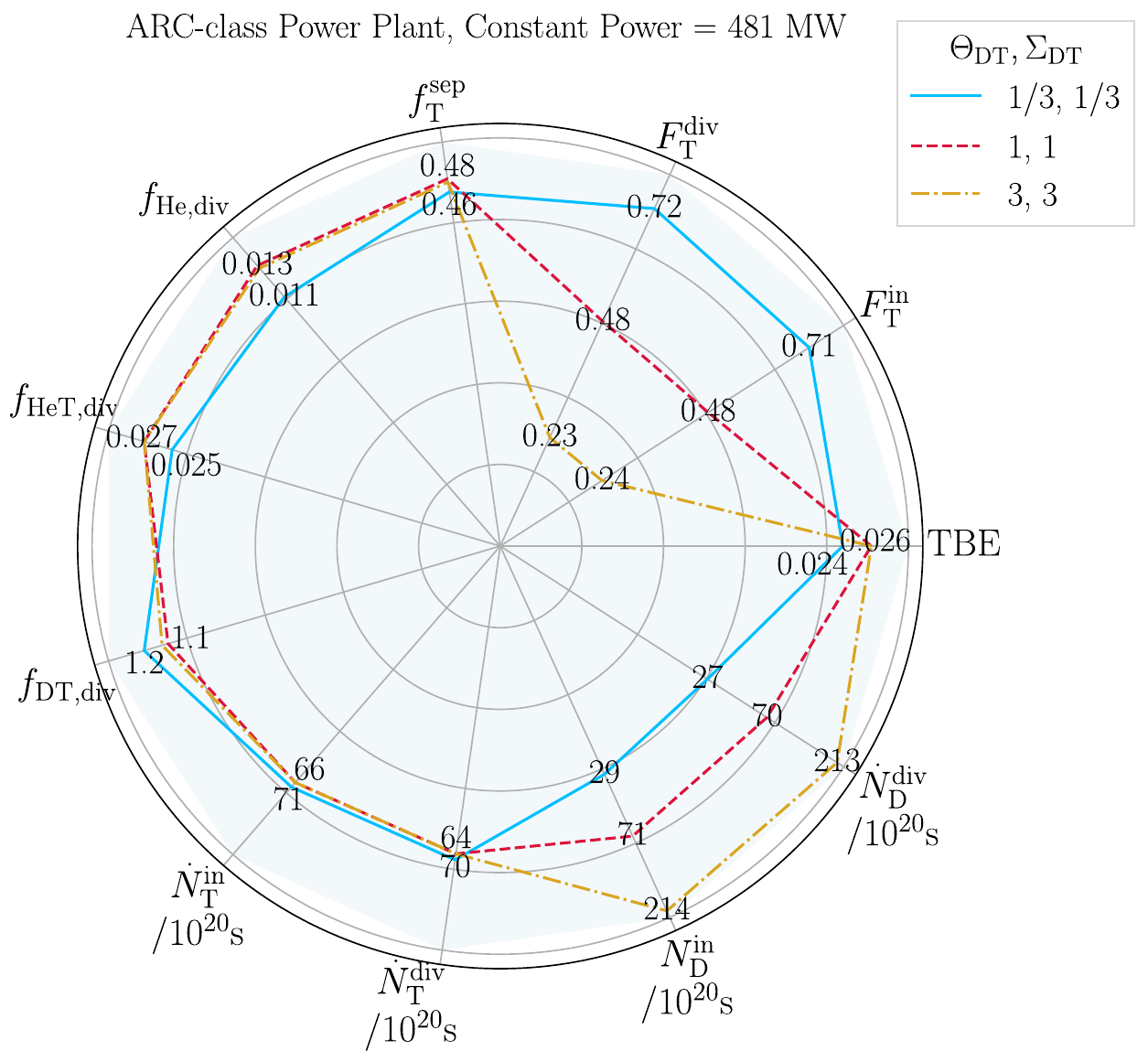}
    \caption{Scan in $\Theta_{\mathrm{DT} }$ and $\Sigma_{\mathrm{DT} }$.}
    \end{subfigure}
    \caption{Tritium self-sufficiency and fusion power parameters of an ARC-class device for a scan in (a) $\Sigma_{\mathrm{DT} }$, (b) $\Theta_{\mathrm{DT} }$, (c) both $\Sigma_{\mathrm{DT} }$ and $\Theta_{\mathrm{DT} }$. In (a), $D_\mathrm{D}$ and $D_\mathrm{T}$ simultaneously decrease as $\Sigma_{\mathrm{DT} }$ decreases. In (b), $D_\mathrm{D}$ increases and $D_\mathrm{T}$ decreases as $\Theta_{\mathrm{DT} }$ increases. In (c), $D_\mathrm{D}$ increases and $D_\mathrm{T}$ is constant as both $\Sigma_{\mathrm{DT} }$ and $\Theta_{\mathrm{DT} }$ increase.}
    \label{fig:SigmaDT_ThetaDT}
\end{figure}
\subsection{$\Sigma_{\mathrm{DT} }$ scan.}

In this subsection, we show the effects of changing $\Sigma_{\mathrm{DT} }$ at fixed $\Theta_{\mathrm{DT} }$ for an ARC-class plant. Shown in \Cref{fig:SigmaDT_ThetaDT} (a), decreasing $\Sigma_{\mathrm{DT} }$ from 3 to 1/3 increases the TBE by a factor of 3. Curiously, by noting the change in $\dot{N}_{\mathrm{T} }^{\mathrm{in} }$ and $\dot{N}_{\mathrm{D} }^{\mathrm{in} }$, there is roughly a factor of 3 decrease in both $D_{\mathrm{T} }$ and $D_{\mathrm{T} }$ as $\Sigma_{\mathrm{DT} }$ decreases from 3 to 1/3 (assuming that $n_{\mathrm{T}}^{\mathrm{co}}$ and $n_{\mathrm{D}}^{\mathrm{co}}$ are fixed). That is, at fixed power, obtaining the maximum TBE by varying $\Sigma_{\mathrm{DT} }$, $D_{\mathrm{D} }$, and $D_{\mathrm{T} }$ approximately satisfy
\begin{equation}
\Sigma_{\mathrm{DT} } \sim D_{\mathrm{T} } D_{\mathrm{D} }, \;\;\; D_{\mathrm{T} } \sim D_{\mathrm{D} }.
\end{equation}
The TBE also satisfies
\begin{equation}
\mathrm{TBE} \sim 1/D_{\mathrm{T} },
\end{equation}
as expected from \Cref{eq:TBEnew_expanded2}. Therefore, by decreasing both deuterium and tritium particle transport at fixed fusion power, the required D-T fueling rates decrease and therefore so the TBE increases. In the $\Sigma_\mathrm{DT} = 1/3$ regime, the deuterium divertor density is much higher than tritium, $f_\mathrm{DT,div} = 3.5$, but the volumetric tritium removal rate is much faster. In this case, both $D_{\mathrm{T} }$ and $D_{\mathrm{D} }$ had to decrease.

In the next subsection, we show that we obtain similar benefits to TBE by degrading the deuterium transport but improving the tritium transport. %

\subsection{$\Theta_{\mathrm{DT} }$ scan.}

In this subsection, we show the effects of improved tritium particle confinement but degraded deuterium particle confinement. We do this by changing $\Theta_{\mathrm{DT} }$ at fixed $\Sigma_{\mathrm{DT} }$ for an ARC-class plant. Shown in \Cref{fig:SigmaDT_ThetaDT} (b), increasing $\Theta_{\mathrm{DT} }$ from 1/3 to 3 increases the TBE by a factor of 3. As in the previous subsection, by inspecting $\dot{N}_{\mathrm{T} }^{\mathrm{in} }$ and $\dot{N}_{\mathrm{D} }^{\mathrm{in} }$ we deduce that $D_{\mathrm{D} }$ increased and $D_{\mathrm{T} }$ decreased as $\Theta_{\mathrm{DT} }$ increased,
\begin{equation}
\Theta_{\mathrm{DT} } \sim D_{\mathrm{D} } / D_{\mathrm{T} },
\end{equation}
The TBE also satisfies
\begin{equation}
\mathrm{TBE} \sim 1/D_{\mathrm{T} },
\label{eq:TBEscal2}
\end{equation}
as expected from \Cref{eq:TBEnew_expanded2}. While we see similar gains in TBE by increasing $\Theta_{\mathrm{DT} }$ and we did by decreasing $\Sigma_\mathrm{DT}$ (previous subsection), there is an important difference. Decreasing $\Sigma_\mathrm{DT}$ decreased the required fueling rate for both deuterium and tritium. Increasing $\Theta_{\mathrm{DT} }$ decreased the required fueling rate for tritium but increased it for deuterium. Hence, differential D-T particle transport can be beneficial to TBE, even if the pumping scheme does not differentially pump D-T, but it requires increasing the deuterium fueling rate while commensurately decreasing the tritium fueling rate.

\subsection{$\Theta_{\mathrm{DT} }$ and $\Sigma_{\mathrm{DT} }$ scan.} \label{sec:ThetaDTSigmaDTscan}

Finally, we show that TBE does not improve if deuterium particle confinement is degraded but tritium particle confinement is fixed. We vary $\Theta_{\mathrm{DT} }$ and $\Sigma_{\mathrm{DT} }$ simultaneously such that $\Theta_{\mathrm{DT} } \sim \Sigma_{\mathrm{DT} }$. Shown in \Cref{fig:SigmaDT_ThetaDT} (c), the TBE is roughly constant, while $D_{\mathrm{D} }$ varies by a factor of 9. Thus, the deuterium particle diffusivity scales as
\begin{equation}
D_{\mathrm{D} } \sim \Theta_{\mathrm{DT} } \Sigma_{\mathrm{DT} }.
\end{equation}
Comparing with \Cref{eq:TBEscal2}, the TBE satisfies
\begin{equation}
\mathrm{TBE} \sim  \Theta_{\mathrm{DT} } / \Sigma_{\mathrm{DT} }.
\label{eq:TBEscal3}
\end{equation}
Therefore, at fixed power we have increased deuterium particle diffusivity by a factor of 9 but the TBE is still approximately constant because the tritium particle diffusivity is approximately constant.

\subsection{TBE and Particle Confinement} \label{sec:TBE_confinement}

\textit{In our model, TBE is independent of deuterium particle transport ($D_{\mathrm{D}}$) as long as the divertor pumping scheme can handle the resulting flow rates of deuterium and tritium. Thus, increasing $D_{\mathrm{D}}$ while also proportionally increasing pumping deuterium speed does not increase TBE -- increased $D_{\mathrm{D}}$ must result in a corresponding increase in} $f_{\mathrm{DT,div}}$ \textit{in order for TBE to increase.}

This is a useful result because $D_\mathrm{D}$ and $D_\mathrm{T}$ are coupled by ambipolarity $\sum_s Z_s \Gamma_s = 0$. For a three-species plasma with deuterium, tritium, and electrons,
\begin{equation}
\Gamma_{\mathrm{T} } = \Gamma_{\mathrm{e} } - \Gamma_{\mathrm{D} }.
\label{eq:GammaT_ambi_eq}
\end{equation}
Therefore, if one can decrease $\Gamma_{\mathrm{e} } - \Gamma_{\mathrm{D} }$, for example by increasing $\Gamma_{\mathrm{D} }$ at fixed $\Gamma_{\mathrm{e} }$, one decreases $\Gamma_{\mathrm{T} }$. Depending on the density profiles, this may correspond to lower $D_{\mathrm{T} }$ values, and hence higher TBE. More possibilities for manipulating tritium particle transport arise in the presence of impurities, where ambipolarity satisfies
\begin{equation}
\Gamma_{\mathrm{T} } = \Gamma_{\mathrm{e} } - \Gamma_{\mathrm{D} } - \sum_j Z_j \Gamma_j,
\label{eq:GammaT_ambi_eq_impurities}
\end{equation}
where the last term is a sum over impurity species.

\textit{Therefore, if the fueling and pumping systems can handle particle flow rates corresponding to changes in deuterium and tritium transport, it may be desirable to increased deuterium particle transport in order to decrease tritium particle transport and therefore increase tritium burn efficiency.}

Heuristically, this argument can be demonstrated as follows. Writing \Cref{eq:GammaT_ambi_eq} as
\begin{equation}
D_{\mathrm{T} } = \frac{\Gamma_{\mathrm{e} } - \Gamma_{\mathrm{D} }}{-\nabla n_{\mathrm{T} }^{\mathrm{co} }}, 
\end{equation}
and using $\mathrm{TBE} \sim 1 / D_{\mathrm{T} }$, we find
\begin{equation}
\mathrm{TBE} \sim \frac{-\nabla n_{\mathrm{T} }^{\mathrm{co} }}{\Gamma_{\mathrm{e} } - \Gamma_{\mathrm{D} }}.
\end{equation}
Writing the electron and deuterium particle transport diffusively,
\begin{equation}
\mathrm{TBE} \sim \frac{1}{ \frac{\nabla n_{\mathrm{e} }^{\mathrm{co} }}{\nabla n_{\mathrm{T} }^{\mathrm{co} }} \left( D_\mathrm{e} + D_\mathrm{D} \right)  - D_\mathrm{D}}.  
\end{equation}
Therefore, we arrive at a heuristic prescription for higher TBE: increase $D_\mathrm{D}$ and decrease $D_\mathrm{T}$. Actively prevent deuterium from being pumped too quickly from the divertor so that the deuterium density fraction rises, increasing the TBE. As pointed out in \cite{Boozer2024stellarators}, deuterium is orders of magnitude cheaper than tritium and engineering requirements for handling deuterium are much simpler than for tritium.

There are many caveats to this statement. In this work, we have assumed that diffusive turbulent transport dominates particle transport. We have assumed that the ratio of $D_{\mathrm{D} }/ D_{\mathrm{T} }$ does not have a radial dependence. We have not paid close attention to particle transport in the divertor \cite{Masline2024}. Higher fidelity models are required to validate the intuition gained from the simple models presented in this work and techniques for intentionally increasing $D_{\mathrm{D} }/ D_{\mathrm{T} }$ need to be developed.

\begin{table*}[bt]
\caption{Key fusion power and tritium self-sufficiency parameters for different operating scenarios in an ARC-class device. In addition to the listed parameters, all cases have $\eta_{\mathrm{He}} = 1.00$, $\Sigma_{\mathrm{HeT}} = 1.00$, $\tau_{\mathrm{IFC}} = 4$h, $\tau_{\mathrm{OFC}} = 24$h, and TBR = 1.08. Note that $I_{\mathrm{startup,min}}$ values below a certain value are likely not well-described by the model in this work.}
\begin{ruledtabular}
  \begin{tabular}{|c|c|c|c|c|c|c|c|c|c|c|c|c |c|c|c| }
   $\Theta_{\mathrm{DT}}, \Sigma_{\mathrm{DT}}$ & \shortstack{$P_f$ \\ (MW)} & $Q$ & TBE & \shortstack{$I_\mathrm{startup,min}$ \\ (kg)} & $\tilde{f}_{\mathrm{T}}^{\mathrm{co}}$  & $f_{\mathrm{T}}^{\mathrm{sep}}$ & $f_{\mathrm{He},\mathrm{div}}$ &  \shortstack{$\dot{N}_{\mathrm{T}}^{\mathrm{burn}}$ \\ ($10^{20}$/s)} & $F_{\mathrm{T}}^{\mathrm{div}}$ & $f_{\mathrm{DT,div}}$ & \shortstack{$\dot{N}_{\mathrm{T}}^{\mathrm{div}}$ \\ ($10^{20}$/s)} & \shortstack{$\dot{N}_{\mathrm{D}}^{\mathrm{div}}$ \\ ($10^{20}$/s)} & $F_{\mathrm{T}}^{\mathrm{in}}$ & \shortstack{$\dot{N}_{\mathrm{T}}^{\mathrm{in}}$ \\ ($10^{20}$/s)} & \shortstack{$\dot{N}_{\mathrm{D},\mathrm{in}}$ \\ ($10^{20}$/s)}  \\
    \hline
    1/3,3 & 481 & 20 & 0.01 & $0.80$ & 0.48 & 0.48 & 0.01 & 1.7 & 0.74 & 0.12 & 120 & 43 & 0.73& 122& 45 \\
    1,1  & 481 & 20 & 0.03 & $0.40$ & 0.48 & 0.48 & 0.01& 1.7 & 0.48 & 1.1 & 64 & 70 & 0.48 & 66 & 71 \\
    5,1/5  & 481 & 20 & 0.29 & $<0.01$ & 0.47 & 0.43 & 0.01 & 1.7 & 0.13 & 33 & 4.1 & 27.1 & 0.17 & 5.8 & 29 \\ \hline
    1/3,3 & 38 & - & 0.60 & $<0.01$ & 0.12 & 0.09 & 0.69 & 0.13 & 0.22 & 1.2 & 0.09 & 0.31 & 0.33 & 0.22 & 0.45 \\
    1,1  & 162 & $0.51$ & 0.60 & $<0.01$ & 0.21 & 0.15 & 0.22 & 0.58 & 0.15 & 5.9 & 0.38 & 2.3 & 0.25 & 0.96 & 2.84 \\
    5,1/5  & 447 & $14$ & 0.60 & $<0.01$ & 0.44 & 0.33 & 0.03 & 1.6 & 0.09 & 51 & 1.1 & 10.9 & 0.18 & 2.7 & 12.5 \\ \hline
    1/3,3  & 351 & $5.3$ & 0.10 & $0.05$ & 0.48 & 0.51 & 0.10 & 1.3 & 0.76 & 0.11 & 11.2 & 3.6 & 0.72 & 12.5 & 4.87 \\
    1,1  & 415 & $11$ & 0.10 & $0.06$ & 0.45 & 0.45 & 0.05 & 1.5 & 0.45 & 1.3 & 13.3 & 16.5 & 0.45 & 14.7 & 18.0 \\
    5,1/5  & 499 & $24$ & 0.10 & $0.07$ & 0.49 & 0.48 & 0.00 & 1.8 & 0.16 & 27 & 16 & 86.3 & 0.17 & 18 & 88 \\ \hline
\end{tabular}
\end{ruledtabular}
\label{tab:tab1}
\end{table*}

\section{Development of Pumping Technology} \label{sec:future_tech_develop}

In this section, we describe potential avenues for developing pumping technologies that could facilitate the D-T asymmetric scenarios in this work.

\subsection{Rapid Hydrogen Re-injection}

\begin{figure}[bt]
    \centering
    \includegraphics[width=0.9\textwidth]{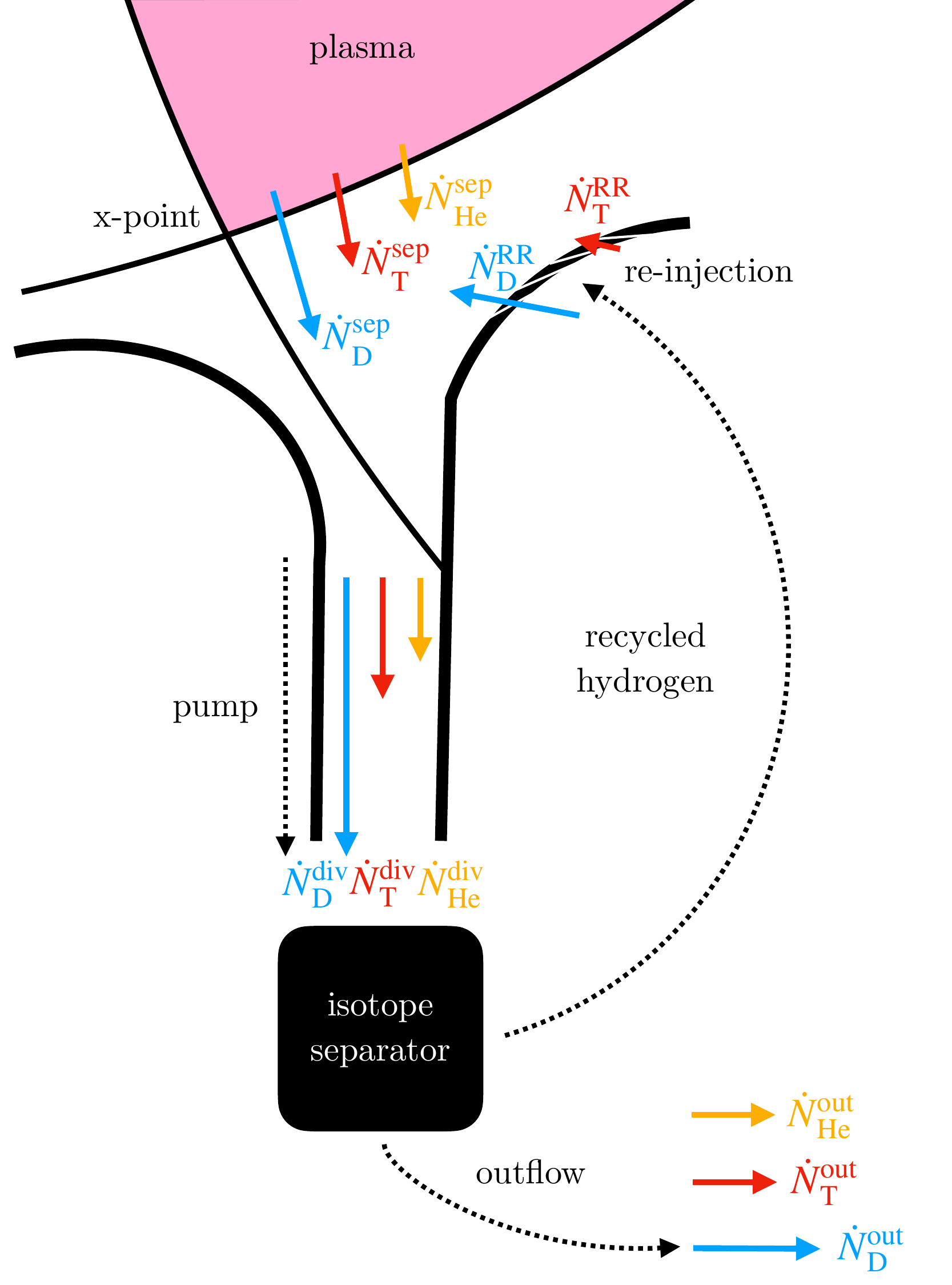}
    \caption{Schematic of a rapid divertor re-injection scheme. An isotope separator quickly outputs a deuterium and tritium flow with a tritium fraction $F_{\mathrm{T} }^{\mathrm{RR} } = \dot{N}_{\mathrm{T} }^{\mathrm{RR} }/\dot{N}_{\mathrm{Q} }^{\mathrm{RR} }$, re-injecting particles back into the divertor region.}
    \label{fig:rapidrecycling_schematic}
\end{figure}

\begin{figure}[!tb]
    \centering
    \begin{subfigure}[t]{0.9\textwidth}
    \centering
    \includegraphics[width=0.9\textwidth]{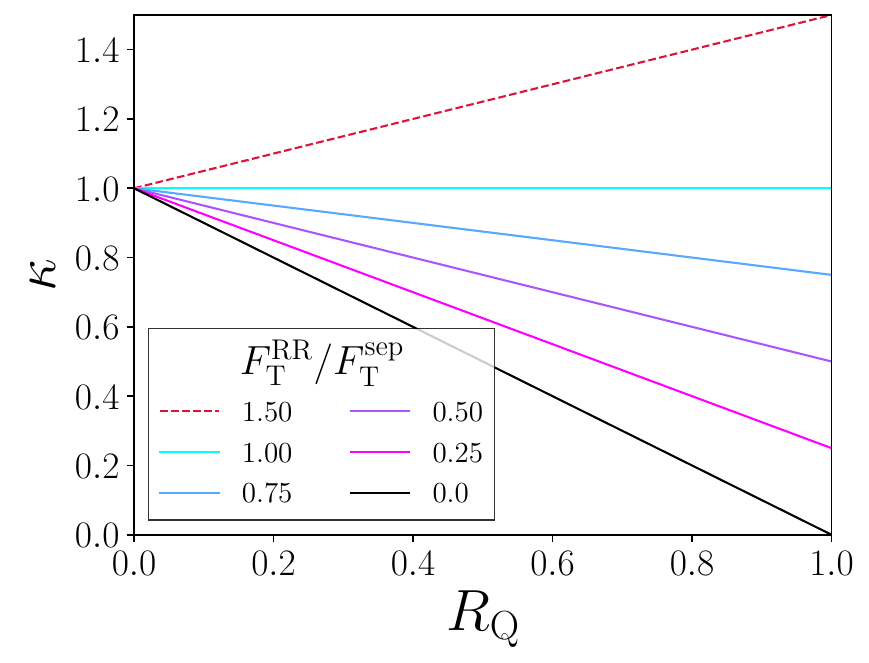}
    \caption{$\kappa$.}
    \end{subfigure}
     ~
    \begin{subfigure}[t]{0.9\textwidth}
    \centering
    \includegraphics[width=1.0\textwidth]{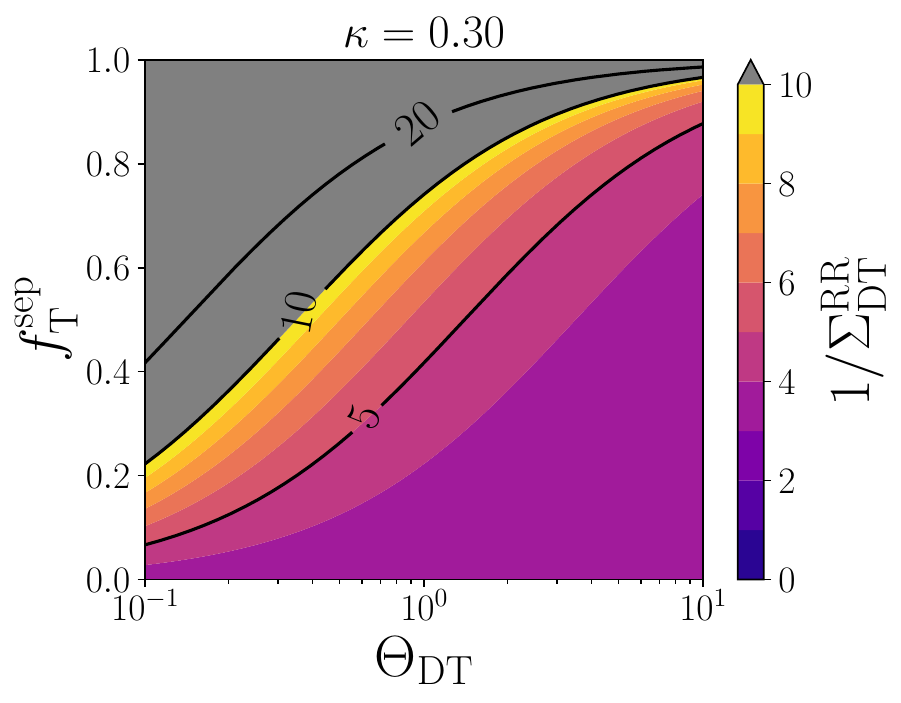}
    \caption{$1/\Sigma_{\mathrm{DT} }^{\mathrm{RR} }$ for $\kappa = 0.30$.}
    \end{subfigure}
     ~
    \begin{subfigure}[t]{0.9\textwidth}
    \centering
    \includegraphics[width=1.0\textwidth]{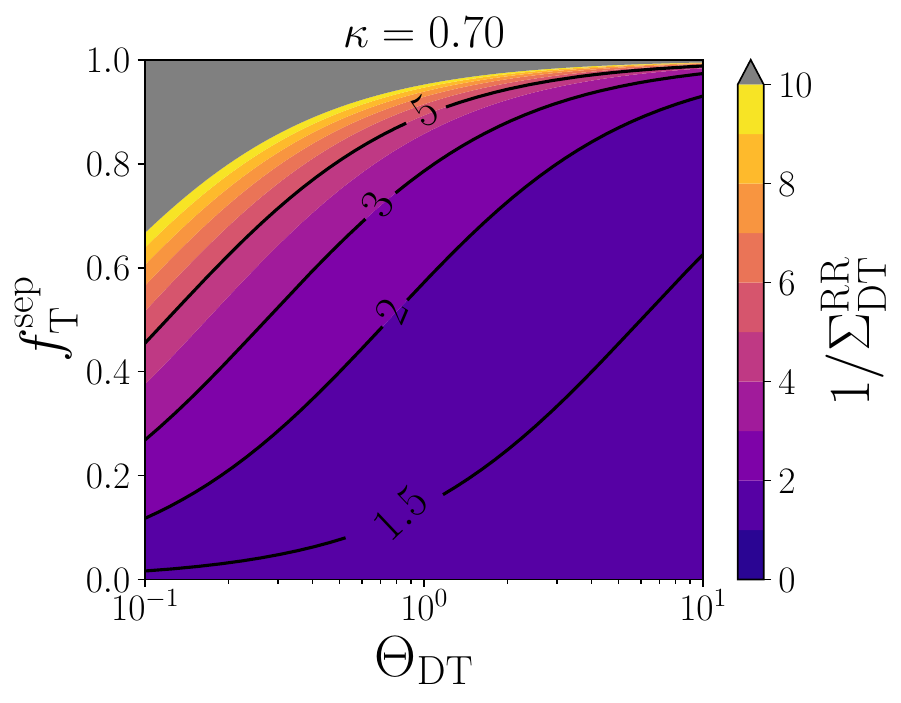}
    \caption{$1/\Sigma_{\mathrm{DT} }^{\mathrm{RR} }$ for $\kappa = 0.70$.}
    \end{subfigure}
    \caption{Key quantities in a rapid re-injection (RR) system: (a) $\kappa$ (\Cref{eq:kappa}) versus $R_\mathrm{Q} $ for several $F_{\mathrm{T} }^{\mathrm{RR} }/F_{\mathrm{T} }^{\mathrm{sep} }$ values, and $1/\Sigma_{\mathrm{DT} }^{\mathrm{RR} }$ (\Cref{eq:SigmaRRDT}) versus $f_{\mathrm{T} }^{\mathrm{sep} }$ and $\Theta_{\mathrm{DT} }$ for $\kappa = 0.30$ in (b) and $\kappa = 0.70$ in (c). $\Lambda = \Sigma_{\mathrm{HeT}} =\eta_{\mathrm{He}} = 1.0$.}
    \label{fig:rapid_reinjection}
\end{figure}

Here, we discuss a rapid hydrogen re-injection (RR) technique for achieving an effect analogous to $\Sigma_{\mathrm{DT} } < 1$ without changing the relative D-T pumping speed $\Sigma_{\mathrm{DT} }$. As in previous sections, the goal is to increase the relative density of deuterium to tritium at the divertor, $f_{\mathrm{DT,div}}$. As with improvements in $\Sigma_{\mathrm{DT} }$ discussed earlier, it is important to clarify the causal chain: decreasing $\Sigma_{\mathrm{DT} }$ does not automatically increase TBE, rather, decreases in particle transport can be translated to higher TBE values by deploying technologies that obtain lower $\Sigma_{\mathrm{DT} }$. We describe this system as `rapid' because the hydrogen is re-injected back into the scrape-off layer much faster than hydrogen leaving through the divertor, being processed in the fuel cycle, and then re-fuelling the plasma.

RR is designed to increase the TBE by increasing the relative helium-tritium divertor density $f_{\mathrm{HeT,div}}$. Suppose that once hydrogen is removed from the divertor, the deuterium and tritium can be rapidly separated, and if desired, re-injected back into the divertor at a desired tritium flow rate fraction $F_{\mathrm{T} }^{\mathrm{RR} }$. A technology capable of achieving this -- the partial ionization plasma centrifuge -- is described in \Cref{sec:PIPC}.

A schematic of the RR setup is shown in \Cref{fig:rapidrecycling_schematic}. By re-injecting hydrogen back into the divertor, this modifies the boundary condition $F_{\mathrm{T} }^{\mathrm{div}}=F_{\mathrm{T} }^{\mathrm{sep}}$ (\Cref{eq:BC_sepdiv}). To determine the new boundary condition, consider the tritium and total hydrogen flows around the separatrix and the divertor
\begin{equation}
\dot{N}_{\mathrm{T} }^{\mathrm{div} } = \dot{N}_{\mathrm{T} }^{\mathrm{sep} } + \dot{N}_{\mathrm{T} }^{\mathrm{RR} },
\label{eq:NTdiv_new}
\end{equation}
\begin{equation}
\dot{N}_{\mathrm{Q} }^{\mathrm{div} } = \dot{N}_{\mathrm{Q} }^{\mathrm{sep} } + \dot{N}_{\mathrm{Q} }^{\mathrm{RR} },
\label{eq:NQdiv_new}
\end{equation}
where $\dot{N}_{\mathrm{T} }^{\mathrm{RR} }$ and $\dot{N}_{\mathrm{Q} }^{\mathrm{RR} }$ are the rapidly recycled tritium and total hydrogen flow rates resulting from being re-injected into the divertor. Here, $\dot{N}_{\mathrm{T} }^{\mathrm{div} }$ refers to the tritium flow rate through the pump into the isotope separator. The tritium flow rate of rapidly recycled hydrogen is
\begin{equation}
F_{\mathrm{T} }^{\mathrm{RR} } \equiv \dot{N}_{\mathrm{T} }^{\mathrm{RR} }/\dot{N}_{\mathrm{Q} }^{\mathrm{RR} },
\label{eq:FTRR}
\end{equation}
and the fraction of hydrogen entering the divertor that is rapidly recycled is
\begin{equation}
R_{\mathrm{Q} } \equiv \dot{N}_{\mathrm{Q} }^{\mathrm{RR} }/\dot{N}_{\mathrm{Q} }^{\mathrm{div} }.
\label{eq:RQ}
\end{equation}
Combining \Cref{eq:NTdiv_new,eq:NQdiv_new,eq:FTRR,eq:RQ}, the new boundary condition is
\begin{equation}
F_{\mathrm{T} }^{\mathrm{div} } = F_{\mathrm{T} }^{\mathrm{sep} } \kappa,
\label{eq:new_FTdiv_FTsep_BC}
\end{equation}
where
\begin{equation}
\kappa \equiv 1 - R_{\mathrm{Q} } \left( 1 - \frac{F_{\mathrm{T} }^{\mathrm{RR} }}{F_{\mathrm{T} }^{\mathrm{sep} }}  \right).
\label{eq:kappa}
\end{equation}
$\kappa$ is plotted versus $R_{\mathrm{Q} }$ for several $F_{\mathrm{T} }^{\mathrm{RR} }/F_{\mathrm{T} }^{\mathrm{sep} }$ values in \Cref{fig:rapid_reinjection}(a). For $F_{\mathrm{T} }^{\mathrm{RR} }/F_{\mathrm{T} }^{\mathrm{sep} } = 1$, it is intuitive that $\kappa$ remains equal to 1 for all $R_{\mathrm{Q} }$ values because the same tritium flow rate fraction is injected back into the divertor. Using \Cref{eq:FTco_ftco} for $F_{\mathrm{T} }^{\mathrm{co} }$ evaluated at the separatrix,
\begin{equation}
F_{\mathrm{T} }^{\mathrm{sep} } = \frac{ f_{\mathrm{T}  }^{\mathrm{sep} }}{ \Theta_{\mathrm{DT}} \left(\Lambda - f_{\mathrm{T}  }^{\mathrm{sep}} \right) + f_{\mathrm{T}  }^{\mathrm{sep}} },
\end{equation}
and
\begin{equation}
F_{\mathrm{T} }^{\mathrm{div} } = \frac{1}{1 + f_{\mathrm{DT,div} }\Sigma_{\mathrm{DT} }},
\end{equation}
we find that the modified tritium-to-deuterium divertor density ratio is
\begin{equation}
\begin{aligned}
& f_{\mathrm{DT,div} } = \\
& \frac{1}{\Sigma_{\mathrm{DT} }} \frac{\Theta_{\mathrm{DT}} \left(\Lambda - f_{\mathrm{T}  }^{\mathrm{sep}} \right) + f_{\mathrm{T}  }^{\mathrm{sep} } \left( 1 - \kappa \right) }{f_{\mathrm{T}  }^{\mathrm{sep}} \kappa}.
\end{aligned}
\label{eq:fTdiv_RR_simple_new}
\end{equation}
Finally, the hydrogen flow rate out of the system (see \Cref{fig:rapidrecycling_schematic}) is equal to the hydrogen flow across the separatrix
\begin{equation}
\dot{N}_{\mathrm{Q} }^{\mathrm{out} } = \dot{N}_{\mathrm{Q} }^{\mathrm{div} } - \dot{N}_{\mathrm{Q} }^{\mathrm{RR} } = \dot{N}_{\mathrm{Q} }^{\mathrm{sep} }.
\end{equation}
However, because in general $F_{\mathrm{T}  }^{\mathrm{RR} } \neq F_{\mathrm{T}  }^{\mathrm{sep} }$, the tritium output flow $\dot{N}_{\mathrm{T} }^{\mathrm{out} }$ will be different to $\dot{N}_{\mathrm{T} }^{\mathrm{sep} }$ and $\dot{N}_{\mathrm{T} }^{\mathrm{div} }$. In order to recast the effect of rapid hydrogen recycling as an effective D-T pumping ratio, we write \Cref{eq:fTdiv_RR_simple_new} as
\begin{equation}
\begin{aligned}
& f_{\mathrm{DT,div} } = \\
& \frac{1}{\Sigma_{\mathrm{DT} }^{\mathrm{RR} }} \frac{1}{\Sigma_{\mathrm{DT} }} \frac{\Theta_{\mathrm{DT}} \left(\Lambda - f_{\mathrm{T}  }^{\mathrm{sep}} \right) }{f_{\mathrm{T}  }^{\mathrm{sep}}},
\end{aligned}
\label{eq:fTdiv_RR_simple}
\end{equation}
where
\begin{equation}
\begin{aligned}
& \Sigma_{\mathrm{DT} }^{\mathrm{RR} } = \kappa \left[ 1 + \frac{f_{\mathrm{T}  }^{\mathrm{sep} } \left( 1 - \kappa \right)}{\Theta_{\mathrm{DT}} \left( \Lambda - f_{\mathrm{T}  }^{\mathrm{sep}} \right) }  \right]^{-1}  \\
& = \frac{1 + R_{\mathrm{Q}} \left[ F_{\mathrm{T} }^{\mathrm{RR} } \Theta_{\mathrm{DT}} \left( \frac{\Lambda}{f_{\mathrm{T}  }^{\mathrm{sep}}}  - 1 \right)  + F_{\mathrm{T} }^{\mathrm{RR} } - 1 \right]  }{1 - F_{\mathrm{T} }^{\mathrm{RR} } R_{\mathrm{Q}} - \frac{f_{\mathrm{T}  }^{\mathrm{sep}} R_{\mathrm{Q} } (F_{\mathrm{T} }^{\mathrm{RR} }-1) }{  \Theta_{\mathrm{DT}} \left(\Lambda - f_{\mathrm{T}  }^{\mathrm{sep}} \right) } }.
\end{aligned}
\label{eq:SigmaRRDT}
\end{equation}
In \Cref{fig:rapid_reinjection}(b) and (c), we plot $1/\Sigma_{\mathrm{DT} }^{\mathrm{RR} }$ versus $f_{\mathrm{T}  }^{\mathrm{sep}}$ and $\Theta_{\mathrm{DT}}$ for two $\kappa$ values. Increasing $f_{\mathrm{T}  }^{\mathrm{sep}}$ and $\Theta_{\mathrm{DT}}$ at fixed $\Lambda = 1.0$ always increases $\Sigma_{\mathrm{DT} }^{\mathrm{RR} }$, which means there is a tradeoff between increasing $\Theta_{\mathrm{DT}}$ and decreasing $\Sigma_{\mathrm{DT} }^{\mathrm{RR} }$. However, for the parameters considered in \Cref{fig:rapid_reinjection}(b) and (c), the maximum $\Sigma_{\mathrm{DT} }^{\mathrm{RR} }$ is $\kappa$, which if less than one, can still be very beneficial for TBE and fusion power. Analogous to the earlier discussion on $\Sigma_{\mathrm{DT} }$ and transport, decreasing $\Sigma_{\mathrm{DT} }^{\mathrm{RR} }$ also corresponds to a decrease in deuterium and tritium particle transport.

\begin{figure}[bt]
    \centering
    \includegraphics[width=0.85\textwidth]{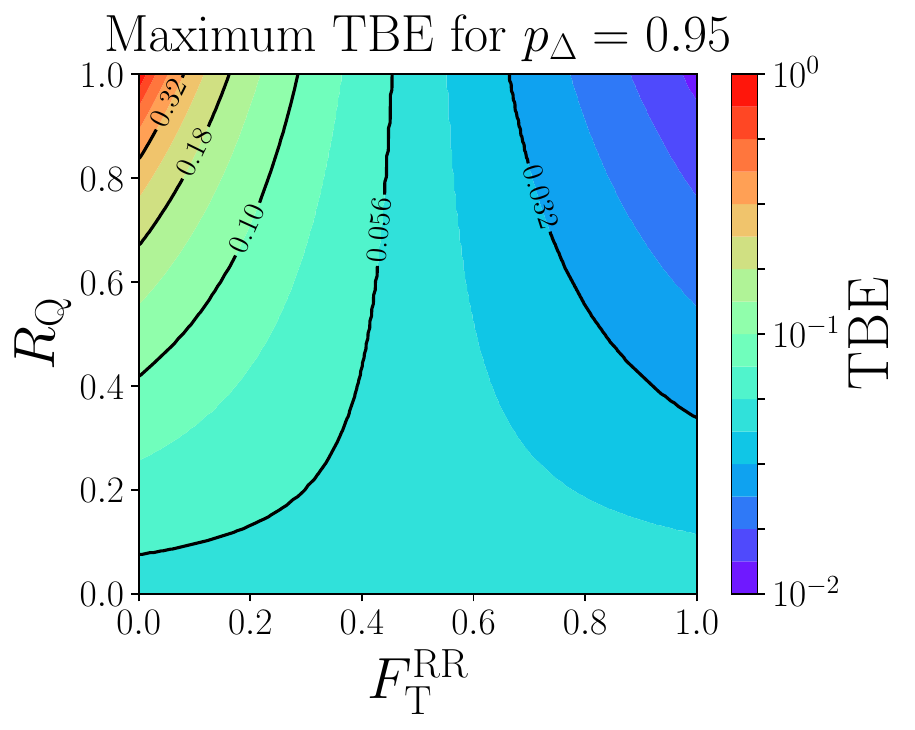}
    \caption{Maximum TBE versus $F_{\mathrm{T} }^{\mathrm{RR} }$ and $R_{\mathrm{Q}}$ for $p_{\Delta} = 0.95$, $\Lambda = \Sigma_{\mathrm{HeT}} =\eta_{\mathrm{He}} = 1.0$. See TBE expression in \Cref{eq:TBE_new}.}
    \label{fig:maxTBE_RQ_FTRR}
\end{figure}

\begin{figure*}[bt]
    \centering
    \includegraphics[width=0.73\textwidth]{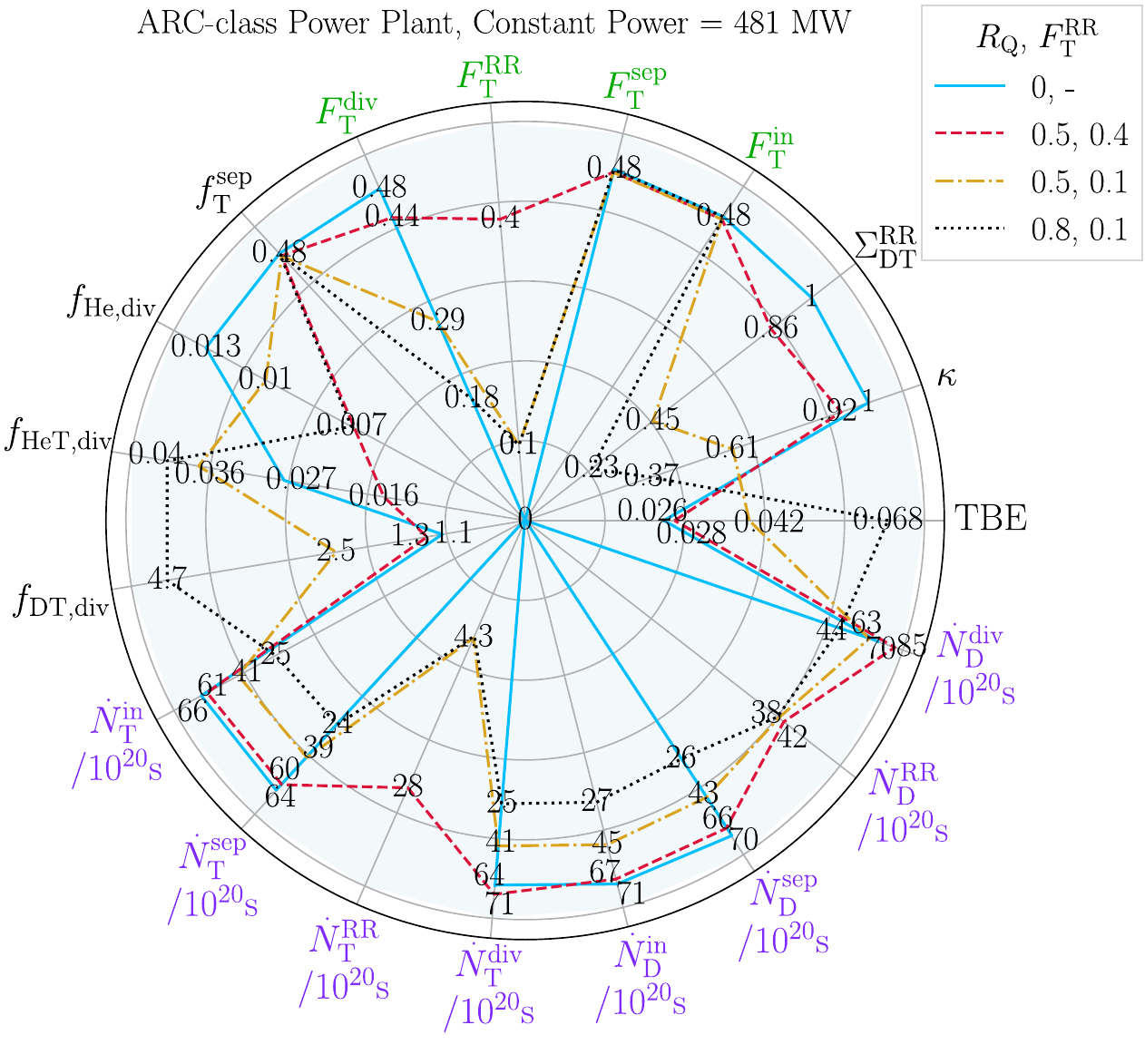}
    \caption{Tritium self-sufficiency and fusion power parameters of an ARC-class device for four cases with different rapid re-injection parameters and fixed fusion power $P_f = 481$ MW. Axes with titles with font colors different to black are scaled the same as other axis titles of the same color. Some axes are log-scaled, some are linearly scaled. $\Theta_{\mathrm{DT}} = \Sigma_{\mathrm{DT}} = \Sigma_{\mathrm{HeT}} = \Lambda = \eta_{\mathrm{He} } = 1.0$.}
    \label{fig:arcclass_spider_rapid_recycling}
\end{figure*}

We now study the effects of $R_{\mathrm{Q} } \neq 0$ on ARC-class cases. We assume that $\Sigma_{\mathrm{DT} } = 1$, and that all the changes in the TBE and and fusion power instead arise from $\Sigma_{\mathrm{DT} }^{\mathrm{RR} } \neq 1$. The results are shown in \Cref{fig:arcclass_spider_rapid_recycling} for a case with fixed fusion power.

The expression for tritium burn efficiency in \Cref{eq:TBEnew} is also modified to
\begin{equation}
\mathrm{TBE} = \frac{f_{\mathrm{HeT,div}} \Sigma_{\mathrm{HeT}} }{ \left( 1-  \frac{F_{\mathrm{T}}^{\mathrm{RR}} }{ F_{\mathrm{T}} ^{\mathrm{div}} } R_Q \right) + f_{\mathrm{HeT,div}} \Sigma_{\mathrm{HeT}} }.
\label{eq:TBE_new}
\end{equation}
In \Cref{fig:maxTBE_RQ_FTRR} we plot the maximum TBE (again, optimized over $F_{\mathrm{T}}^{\mathrm{in}}$) versus $F_{\mathrm{T}}^{\mathrm{RR}}$ and $R_Q$. In this case, for $F_{\mathrm{T}}^{\mathrm{RR}} < F_{\mathrm{T}}^{\mathrm{div}} \simeq 1/2$, increasing $R_Q$ always increases TBE. However, for $F_{\mathrm{T}}^{\mathrm{RR}} > F_{\mathrm{T}}^{\mathrm{div}} \simeq 1/2$ increasing $R_Q$ always decreases TBE. Very high $R_Q$ values and very small $F_{\mathrm{T}}^{\mathrm{RR}}$ values can support a very high TBE of TBE$\gtrsim 0.3$.
Requiring that the denominator of \Cref{eq:TBE_new} is positive gives the physical constraint that the flow rate of rapidly recycled tritium cannot exceed the tritium injection rate
\begin{equation}
    \dot{N}_{\mathrm{T}}^{\mathrm{RR}} < \dot{N}_{\mathrm{T}}^{\mathrm{in}},
\end{equation}
and requiring $\mathrm{TBE} < 1$ gives $\dot{N}_{\mathrm{T}}^{\mathrm{RR}} < \dot{N}_{\mathrm{T}}^{\mathrm{div}}$.

\begin{table*}[bt]
\caption{Key fusion power and tritium self-sufficiency parameters for different operating scenarios in an ARC-class device with rapid re-injection. In all four cases, $\Theta_{\mathrm{DT}} = \Lambda = \Sigma_{\mathrm{HeT}} = \Sigma_{\mathrm{DT}} = 1.0$. More quantities are shown in \Cref{fig:arcclass_spider_rapid_recycling}.}
\begin{ruledtabular}
  \begin{tabular}{|c|c|c|c|c|c|c|c|c|c|c|c|c|c|c|c| }
   $ R_{Q}, F_{\mathrm{T}}^{\mathrm{RR}}$ & \shortstack{$P_f$ \\ (MW)} & Q & TBE & \shortstack{$I_\mathrm{startup,min}$ \\ (kg)} & $\tilde{f}_{\mathrm{T}}^{\mathrm{co}}$  & $f_{\mathrm{T}}^{\mathrm{sep}}$ & $f_{\mathrm{He},\mathrm{div}}$ &  \shortstack{$\dot{N}_{\mathrm{T}}^{\mathrm{burn}}$ \\ ($10^{20}$/s)} & $F_{\mathrm{T}}^{\mathrm{div}}$ & $f_{\mathrm{DT,div}}$ & \shortstack{$\dot{N}_{\mathrm{T}}^{\mathrm{div}}$ \\ ($10^{20}$/s)} & \shortstack{$\dot{N}_{\mathrm{D}}^{\mathrm{div}}$ \\ ($10^{20}$/s)} & $F_{\mathrm{T}}^{\mathrm{in}}$ & \shortstack{$\dot{N}_{\mathrm{T}}^{\mathrm{in}}$ \\ ($10^{20}$/s)} & \shortstack{$\dot{N}_{\mathrm{D},\mathrm{in}}$ \\ ($10^{20}$/s))}  \\
    \hline
    0,     -  & 481 & 20 & 0.026 & $0.40$ & 0.48 & 0.48 & 0.01& 1.7 & 0.48 & 1.1 & 64 & 70 & 0.48 & 66 & 71 \\
    0.5, 0.4 & 481 & 20 & 0.028 & $0.37$ & 0.48 & 0.48 & 0.01 & 1.7 & 0.48 & 1.3 & 71 & 85 & 0.48& 61& 67 \\
    0.5, 0.1  & 481 & 20 & 0.042 & $0.23$ & 0.48 & 0.48 & 0.01& 1.7 & 0.48 & 2.5 & 41& 63 & 0.48& 41& 45 \\
    0.8, 0.1 & 481 & 20 & 0.068 & $0.12$ & 0.48 & 0.48 & 0.01& 1.7 & 0.48 & 4.7 & 25& 44 & 0.48& 25& 27 \\ \hline
\end{tabular}
\end{ruledtabular}
\label{tab:tabRR}
\end{table*}

Shown in \Cref{fig:arcclass_spider_rapid_recycling}, we consider four cases with fixed fusion power. The nominal case has zero rapid re-injection and TBE = 0.01. Case A has $R_{\mathrm{Q} } = 0.5$ and $F_{\mathrm{T}}^{\mathrm{RR}} = 0.4$ only marginally increases the TBE from 0.026 to 0.028. For case B, $R_{\mathrm{Q} } = 0.5$ and $F_{\mathrm{T}}^{\mathrm{RR}} = 0.1$, the TBE increases further to $\mathrm{TBE}=0.042$ through a combination of $\Sigma_{\mathrm{DT} }^{\mathrm{RR} } < 1$ and an increase in $f_{\mathrm{HeT},\mathrm{div}}$. For case C, $R_{\mathrm{Q} } = 0.8$ and $F_{\mathrm{T}}^{\mathrm{RR}} = 0.1$, the TBE increases to $\mathrm{TBE}=0.068$, an almost three-fold improvement over the nominal case.

There are important caveats and future questions to address. As discussed in \cite{Whyte2023,Parisi_2024}, increasing the helium to hydrogen density assumes that the helium particle confinement time increases. A transport model coupling the plasma core, pedestal, scrape-off-layer, and divertor is needed to more definitively assess the effects of a rapid re-injection scheme.

In summary, we have modeled the effects of a rapid divertor re-injection scheme, shown in \Cref{fig:rapidrecycling_schematic}. By re-injecting deuterium and tritium from the divertor pumped outflow, we demonstrated a factor of three increase in the tritium burn efficiency could be achieved while maintaining constant fusion power. Theoretically, even higher TBE values are possible with extreme $R_Q$ and $F_{\mathrm{T}}^{\mathrm{RR}}$ values. Such improvements are mainly obtained by increasing the helium and deuterium density relative to the tritium density at the divertor, while also reducing the tritium fueling fraction.

\subsection{Mass-Selective Pumping with a Partial Ionization Plasma Centrifuge} \label{sec:PIPC}

\begin{figure*}[bt]
    \centering
    \hspace*{-0.2\textwidth} 

    \includegraphics[trim = 0 75 0 75,clip,width=\textwidth]{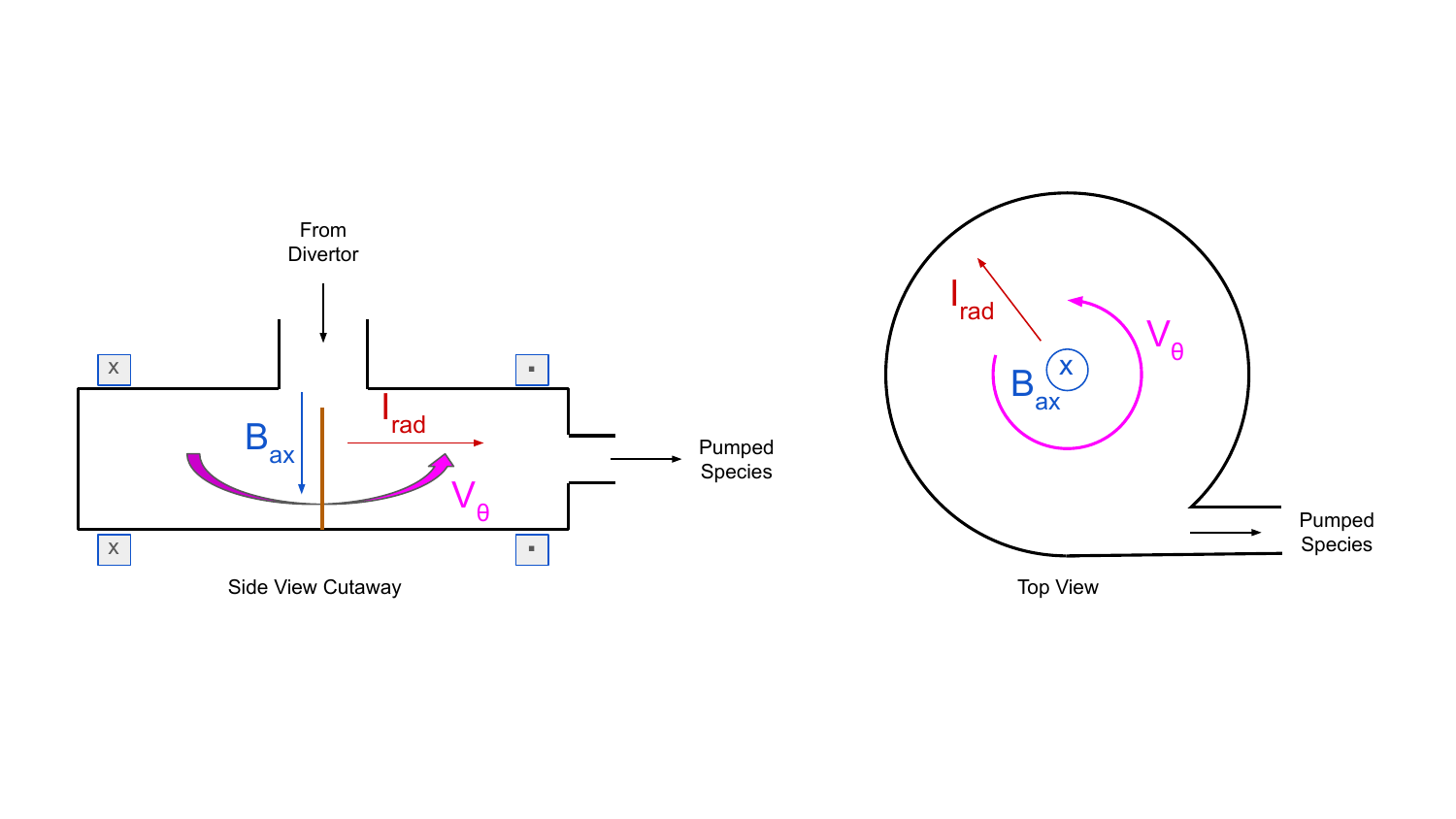}
    \caption{Schematic view of a plasma centrifuge operating based on azimuthal rotation achieved through a radial current and axial magnetic field.}
    \label{fig:plasmaCentrifuge}
\end{figure*}

\begin{figure}[bt]
    \centering
    \includegraphics[width=0.9\textwidth]{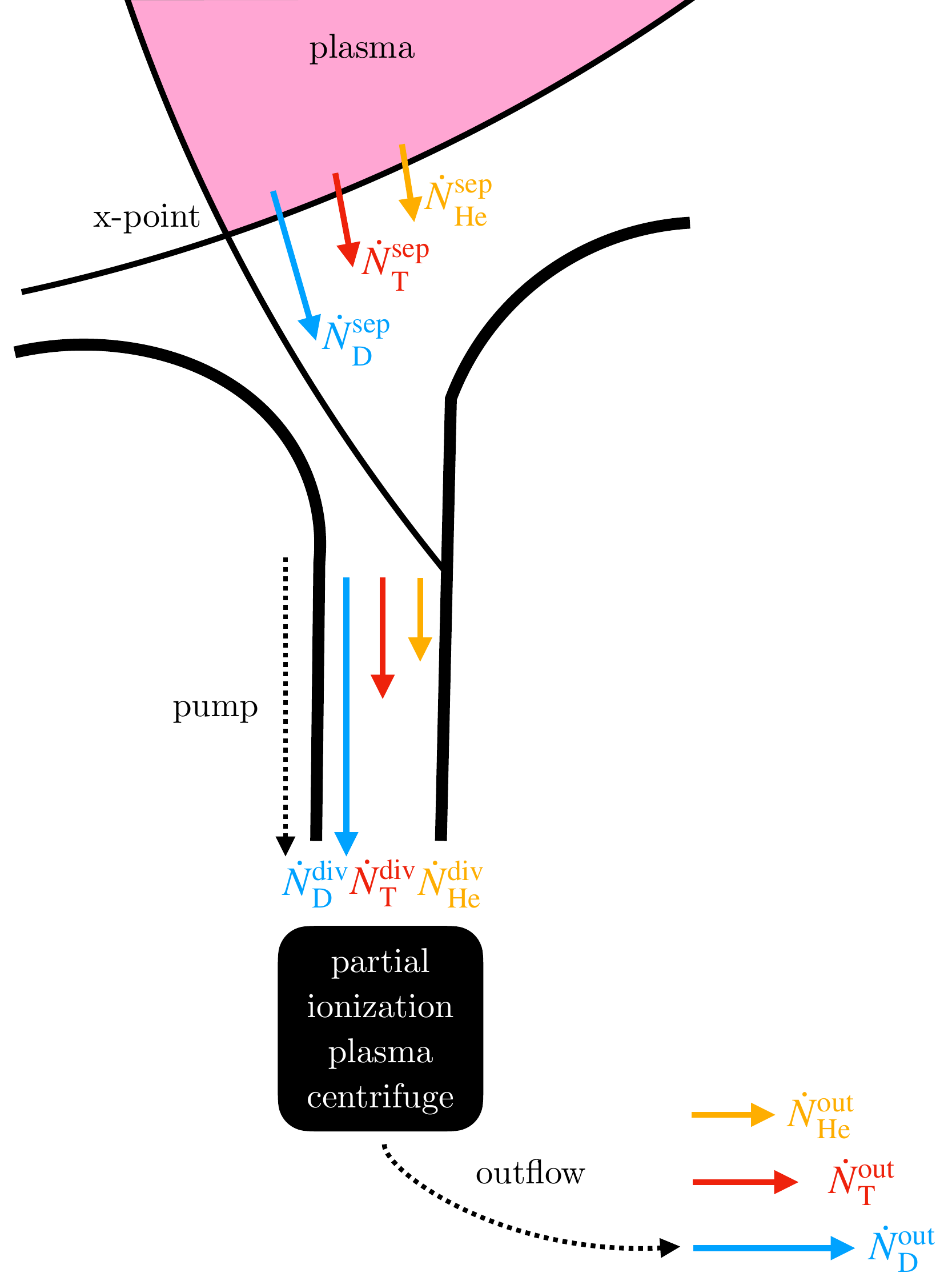}
    \caption{Schematic of a partial ionization plasma centrifuge (PIPC) scheme. The PIPC differentially outputs deuterium, tritium, and helium-4 according to \Cref{eq:separationfactorHeT,eq:separationfactor}.}
    \label{fig:PIPC_schematic}
\end{figure}

In this section, we describe a scheme for efficiently separating species by their mass. The separation of hydrogen isotopes is a challenging task due to small differences in the chemical properties between  isotopes. Additional constraints are imposed by the requirement to have minimal tritium inventory in the system, which requires short tritium processing times.

The challenge of differential pumping is further increased by the presence of helium, which ideally would be pumped at least at the same rate as  tritium, and if possible should be pumped at a higher rate. However, the fact that fully-ionized helium-4 and deuterium nuclei have the same charge-to-mass ratio limits the range of possible separation approaches, especially for concepts operating in a fully ionized plasma.  Furthermore, the fact that helium and the deuterium molecule D2 have approximately the same mass means that mass-based separation is infeasible for a fully neutral gas. Therefore, mass-based separation of deuterium and helium is only possible with atomic, not molecular, deuterium.

The partial ionization plasma centrifuge (PIPC) could enable mass-based separation of hydrogenic species from helium-4 as well as the higher-Z noble gases. It could also enable differential in-situ pumping in the divertor. By primarily operating on atomized neutral hydrogen, a PIPC operates on deuterium and helium atoms, separated by a factor of two in mass and therefore straightforwardly separable. 

It can be shown that in an azimuthally symmetric rotating gas, the radial pressure profile $p(r)$ in the gas can be described by \cite{grossman_plasma_1991} 
\begin{equation}
    p(a)/p(r) = \exp\left(\frac{m}{k_BT}\int_r^a \frac{v_{\theta}(r')^2}{r'} dr'\right),
\end{equation}
where p is the partial pressure of the species of mass $m$ in atomic mass units, $a$ is the coordinate of the outermost radius of the device, and $v_\theta(r)$ is the azimuthal velocity of the gas. 

We further simplify the problem by assuming that for a rigid body rotation profile, $\omega = v_{\theta}(r)/r$ = constant. The maximum compression between $r = 0$ and the outer radius at $ r = a$ is then
\begin{equation}
    p(a)/p(0) = \exp\left(\frac{mv_{\theta}(a)^2}{2k_BT}\right).
\end{equation}
Evidently, high compression ratios can be achieved at high rotational velocities and low temperatures, with an especially strong dependence on the rotational velocity. 

In a mechanical ultracentrifuge, rotational speeds are ultimately limited by material strength; in contrast, plasma centrifuges operate by driving a radial current through a partially ionized gas inside a background magnetic field, using the ion population as a kind of rotor to spin up the neutrals through collisions.  As a result, there are no moving parts and no structural limits to gas velocity.  By operating in a partially ionized state, this approach achieves lower temperatures and the relevant physics of the process are simpler than fully ionized plasma centrifuges.

While there are no basic structural limits on operating parameters, there is a velocity limit imposed by the Alfven ionization limit (or alternatively, the `Critical Ionization Velocity') \cite{Alfven1960}, given by the velocity at which the neutral gas kinetic energy equals the ionization energy of that species.  

Furthermore, as noted by prior authors \cite{wijnakker_limitations_1980} another limit on achievable separation factors in this process is imposed by plasma heating through various mechanisms, including viscous shear, resistive heating, and electron collisions with ions and neutrals.

\begin{figure*}[bt!]
    \begin{subfigure}[t]{0.45\textwidth}
    \centering
    \includegraphics[width=1.0\textwidth]{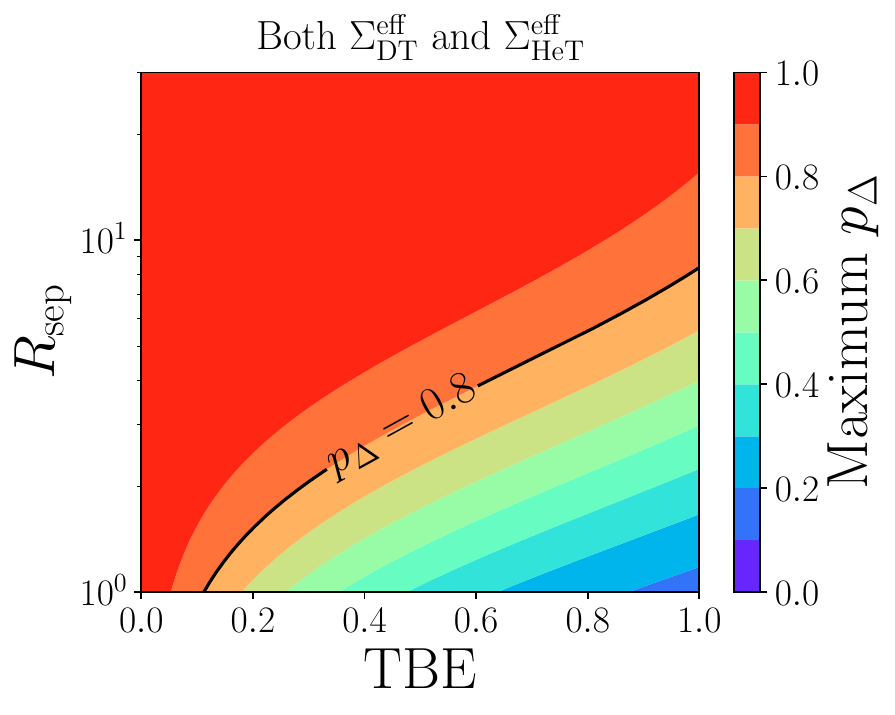}
    \caption{}
    \end{subfigure}
     ~
    \begin{subfigure}[t]{0.45\textwidth}
    \centering
    \includegraphics[width=1.0\textwidth]{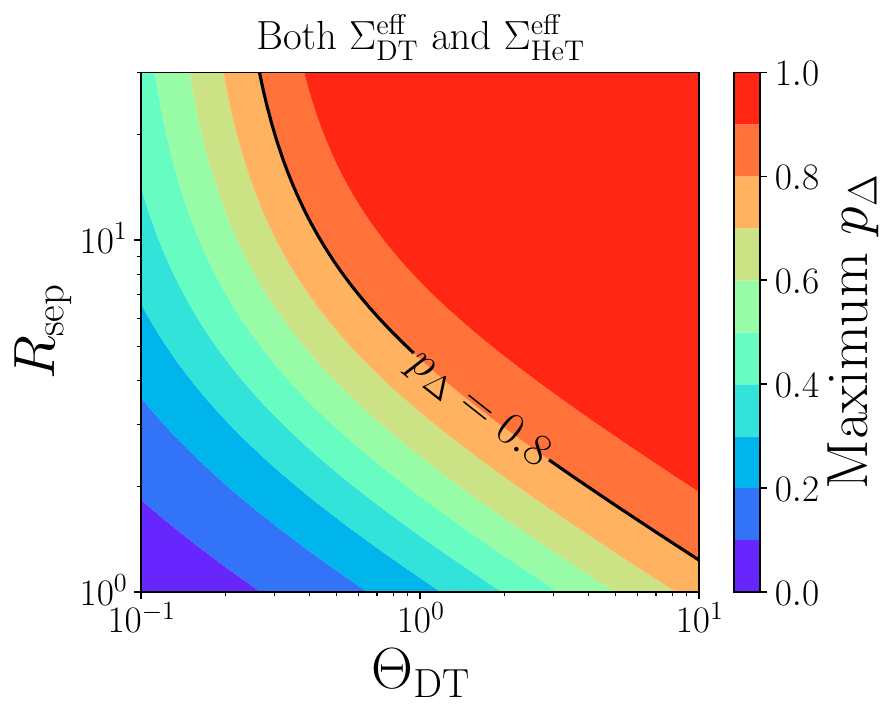}
    \caption{}
    \end{subfigure}
    \begin{subfigure}[t]{0.45\textwidth}
    \centering
    \includegraphics[width=1.0\textwidth]{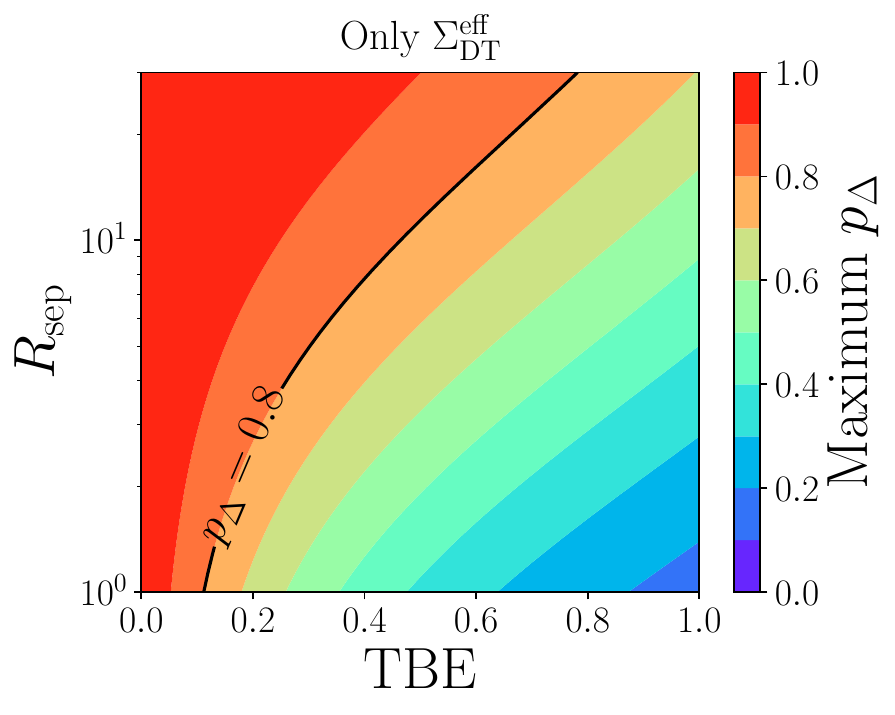}
    \caption{}
    \end{subfigure}
     ~
    \begin{subfigure}[t]{0.45\textwidth}
    \centering
    \includegraphics[width=1.0\textwidth]{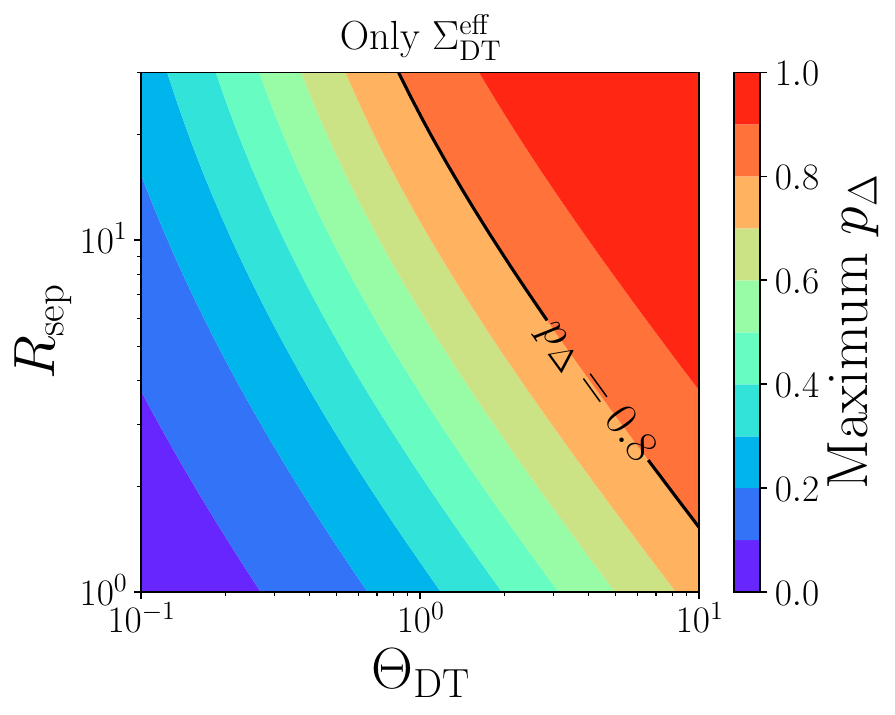}
    \caption{}
    \end{subfigure}
    \begin{subfigure}[t]{0.45\textwidth}
    \centering
    \includegraphics[width=1.0\textwidth]{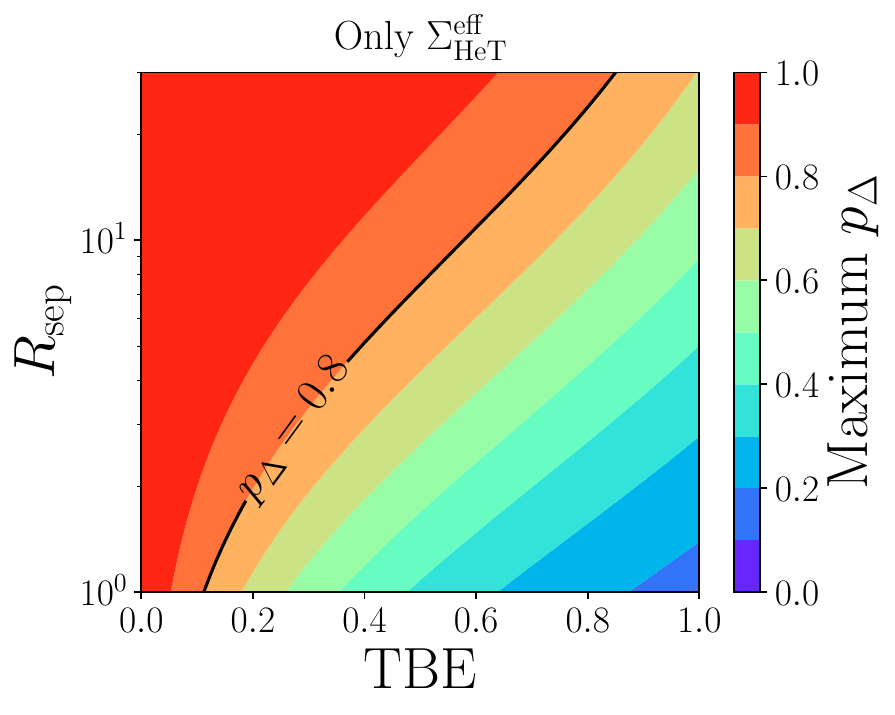}
    \caption{}
    \end{subfigure}
     ~
    \begin{subfigure}[t]{0.45\textwidth}
    \centering
    \includegraphics[width=1.0\textwidth]{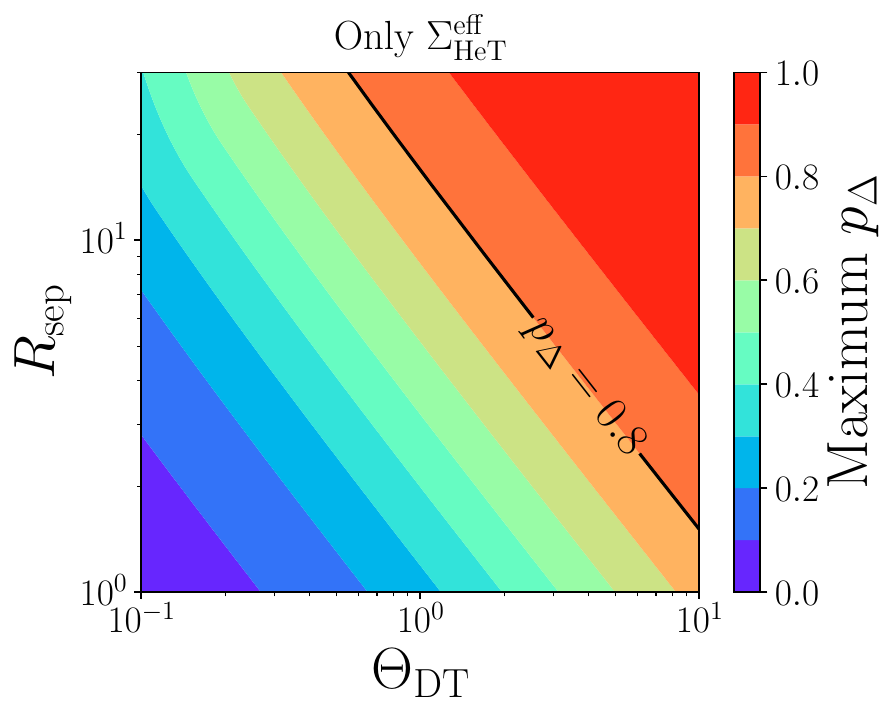}
    \caption{}
    \end{subfigure}
    \caption{$p_{\Delta}$ versus $R_\mathrm{sep}$ and TBE (a,c,e) and $\Theta_{\mathrm{DT} }$ (b,d,f) for a PIPC, where all pumping speed ratios are scaled appropriately (\Cref{eq:effective_pumping}). We assume nominal divertor pumping speed ratios of $\Sigma_{\mathrm{HeT} } = \Sigma_{\mathrm{DT} } = 1$, $\Lambda = \eta_{\mathrm{He} } = 1$, in (a,c,e) $\Theta_{\mathrm{DT} } = 1$ and in (b,d,f) TBE = 0.70. For (a,b), the PIPC differentially pumps all three ion species, whereas in (c,d) and (e,f), it only differentially pumps deuterium and tritium, and helium and tritium, respectively.}
    \label{fig:PIPC_consistent}
\end{figure*}

\begin{figure*}[bt]
    \centering
    \includegraphics[width=0.73\textwidth]{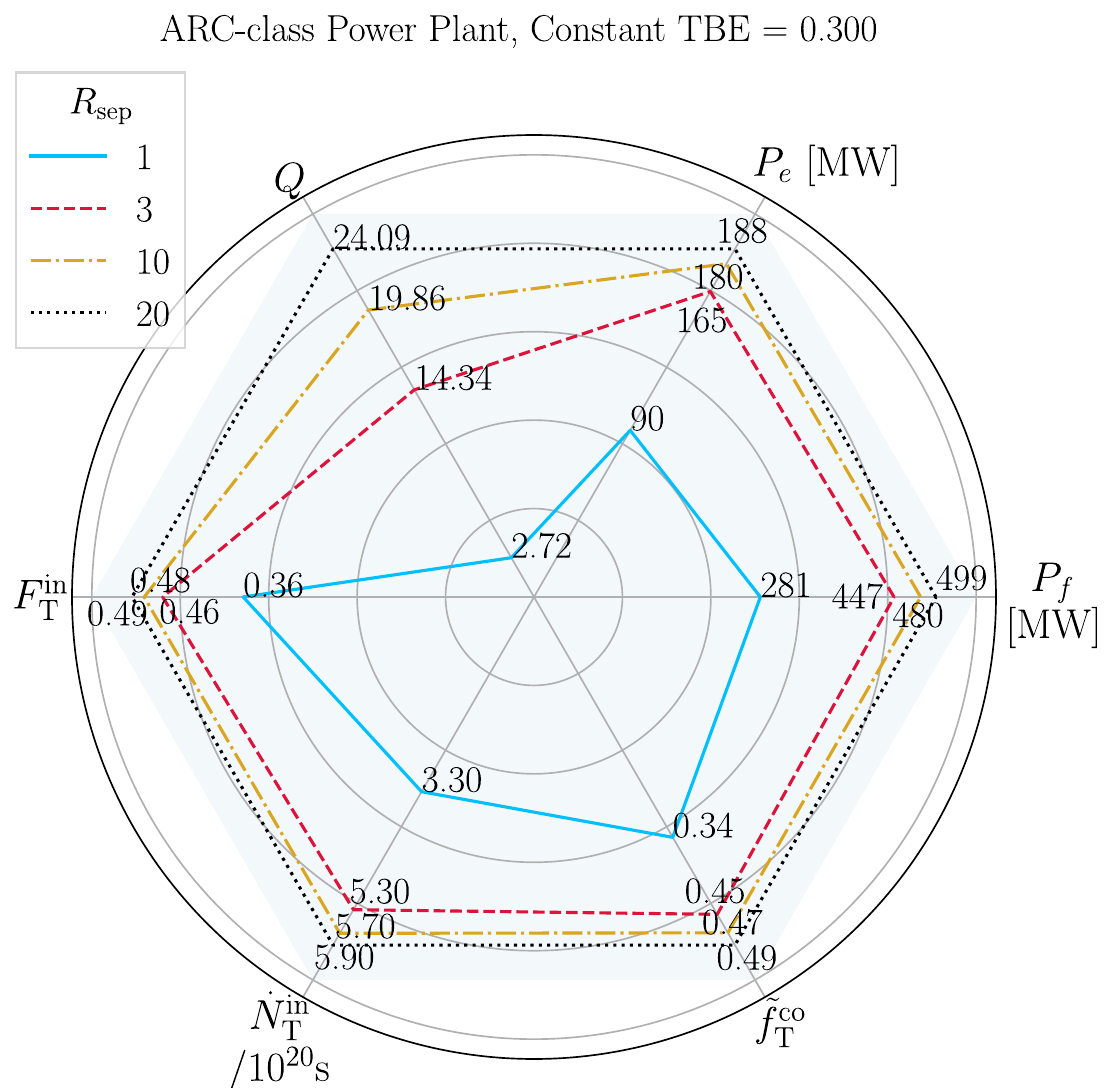}
    \caption{Tritium self-sufficiency and fusion power parameters of an ARC-class device for four cases with different $R_{\mathrm{sep}}$ values fixed TBE =0.30. All axes are linearly scaled. $\Theta_{\mathrm{DT}} = \Lambda = \eta_{\mathrm{He} } = 1.0$. The nominal pumping ratios are $\Sigma_{\mathrm{DT}} = \Sigma_{\mathrm{HeT}} =1.0$.}
    \label{fig:PIPC_spider}
\end{figure*}

Still, large differences in compression can be achieved between the relevant isotopes while assuming realistic plasma temperatures, resulting in the capability for differential pumping.  As noted in \Cref{eq:pumpthroughput}, the flow rate of a gas is proportional to both upstream partial pressure of a species and the pumping speed for that species, meaning that species-dependent increases in partial pressure function identically to increases in species-dependent pumping speed.  
The ratio of tritium to deuterium pumping rates can be written, using \Cref{eq:pumpthroughput}, as 
\begin{equation}
\frac{\dot{N}_{\mathrm{ T}}^{\mathrm{out}}}{\dot{N}_{\mathrm{ D}}^{\mathrm{out}}} = \frac{1}{\Sigma_{\mathrm{DT}} f_{\mathrm{DT, div}}} R_{\mathrm{sep,TD} },
\label{eq:NdotTNdotD_wPIPC}
\end{equation}
where $R_{\mathrm{sep,TD} }$ is the separation factor between tritium and deuterium \cite{grossman_plasma_1991}
\begin{equation}
R_{\mathrm{sep,TD} } = \exp\left(\frac{(m_\mathrm{T} -m_\mathrm{D} )v_{\theta}(a)^2}{2k_BT}\right).
\label{eq:separationfactor}
\end{equation}
There is also a separation factor between helium-4 and tritium,
\begin{equation}
R_{\mathrm{sep,HeT} } = \exp\left(\frac{(m_{\mathrm{He} }-m_\mathrm{T})v_{\theta}(a)^2}{2k_BT}\right),
\label{eq:separationfactorHeT}
\end{equation}
where in \Cref{eq:separationfactor,eq:separationfactorHeT} we assume that deuterium, tritium, and helium-4 all have the same temperature. Therefore, we will write the separation factor 
\begin{equation}
R_{\mathrm{sep} } = R_{\mathrm{sep,TD} } = R_{\mathrm{sep,HeT} }.
\label{eq:Rsep}
\end{equation}

Prior work in the field of plasma chemistry has already demonstrated large separation factors in light isotopes, providing some validation that this could be a promising approach to explore. In reference \cite{korobtsev_chemical_1986}, the investigators measured separation factors up to $R_{\mathrm{sep} }\sim25$  between \ce{H2} and \ce{D2} molecules; because of relative recombination rates this corresponds to a separation factor of ~5 between \ce{H} and \ce{D} atoms.  In the present case where we are concerned with differential pumping of \ce{D} and \ce{T}, the same mass difference is present and so the same separation factor could be expected.  To compare with a specific case identified here, the high-TBE case presented in the third row of Table \ref{tab:tab1} requires a separation factor of $\sim6.1$ to satisfy the pumping requirement $\Sigma_{\mathrm{DT}}^{\mathrm{PIPC}} \Sigma_{\mathrm{DT}} = 1/5$, demonstrating that the required performance for this application has nearly been achieved in similar experimental conditions.

The PIPC is similar to the rapid hydrogen re-injection system in the last subsection but with zero flow rate returning to the divertor region and with the option of also differentially pumping helium-4. The TBE is
\begin{equation}
\mathrm{TBE} = \frac{\dot{N}_\mathrm{T}^\mathrm{burn}}{\dot{N}_\mathrm{T}^\mathrm{in}} = \left( 1 + \frac{\dot{N}_\mathrm{T}^\mathrm{out}}{\dot{N}_\mathrm{He}^\mathrm{out}}  \right)^{-1},
\end{equation}
where we used
\begin{equation}
\dot{N}_\mathrm{T}^\mathrm{in} = \dot{N}_\mathrm{T}^\mathrm{burn} + \dot{N}_\mathrm{T}^\mathrm{out}
\end{equation}
and 
\begin{equation}
\dot{N}_\mathrm{He}^\mathrm{out} = \dot{N}_\mathrm{T}^\mathrm{burn}.
\end{equation}
Using 
\begin{equation}
\frac{\dot{N}_{\mathrm{ He}}^{\mathrm{out}}}{\dot{N}_{\mathrm{ T}}^{\mathrm{out}}} = \Sigma_{\mathrm{HeT}} f_{\mathrm{HeT, div}} R_\mathrm{sep},
\label{eq:NdotHeNdotTout}
\end{equation}
the TBE is
\begin{equation}
\mathrm{TBE} = \frac{ \Sigma_\mathrm{HeT} f_\mathrm{HeT,div} R_\mathrm{sep} }{1 + \Sigma_\mathrm{HeT} f_\mathrm{HeT,div} R_\mathrm{sep}}.
\label{eq:TBE_PIPC}
\end{equation}
Since $R_\mathrm{sep} > 1 $, the PIPC can be seen as enhancing $\Sigma_\mathrm{HeT}$ and $\Sigma_\mathrm{DT}$. Thus we define effective pumping speeds
\begin{equation}
\Sigma_\mathrm{HeT}^\mathrm{eff} \equiv \Sigma_\mathrm{HeT} R_\mathrm{sep}, \;\;\; \Sigma_\mathrm{DT}^\mathrm{eff} \equiv \frac{\Sigma_\mathrm{DT}}{R_\mathrm{sep}}.
\label{eq:effective_pumping}
\end{equation}
Therefore, all equations in \Cref{sec:diff_part_transp} through to the end of \Cref{sec:ARC} are modified with a PIPC by substituting
\begin{equation}
\Sigma_\mathrm{HeT} \to \Sigma_\mathrm{HeT}^\mathrm{eff}, \;\;\; \Sigma_\mathrm{DT} \to \Sigma_\mathrm{DT}^\mathrm{eff}.
\end{equation}
In \Cref{fig:PIPC_consistent}, we plot the fusion power degradation $p_{\Delta}$ versus $R_\mathrm{sep}$ and $\mathrm{TBE}$ (a) and $\Theta_{\mathrm{DT} }$ (b) for a system using a PIPC. To calculate $p_{\Delta}$ we modify $p_{\Delta}$ in \Cref{eq:pDeltaformG} to include the effective pumping speeds $\Sigma_\mathrm{HeT}^\mathrm{eff}$ and $\Sigma_\mathrm{DT}^\mathrm{eff}$ from \Cref{eq:effective_pumping},
\begin{equation}
p_{\Delta} = \frac{ 4 \tilde{f}_{\mathrm{T}}^{\mathrm{co}} (1-\tilde{f}_{\mathrm{T}}^{\mathrm{co}})}{ \left[1 -  \frac{2}{ \eta_{\mathrm{He}} \Sigma_{\mathrm{HeT}}^{\mathrm{eff}} \left( 1 + f_{\mathrm{DT},\mathrm{div}} \right) \left( 1 - \frac{1}{\mathrm{TBE}} \right) }  \right]^{2}}.
\label{eq:pDeltaformG2}
\end{equation}
\Cref{fig:PIPC_consistent}(a) and (b) show that even increasing $R_{\mathrm{sep} }$ from $R_{\mathrm{sep} }=1$ to $R_{\mathrm{sep} }=5$ would allow significant gains in the fusion power and TBE. For completeness, we also show the effect of a PIPC only differentially pumping two sets of ion species while allowing the third ion species to pass unhindered: deuterium and tritium in \Cref{fig:PIPC_consistent}(c) and (d) and helium and tritium in \Cref{fig:PIPC_consistent}(e) and (f). There is a significant beneficial combined effect of allowing the PIPC to differentially pump all three species rather than just a set of two species.

The residence time of tritium in the PIPC can be written as
\begin{equation}
\tau_\mathrm{T} = \frac{N_\mathrm{T}^\mathrm{PIPC} }{\dot{N}_\mathrm{T}^\mathrm{out}} = \frac{n_\mathrm{T}^\mathrm{PIPC} V^\mathrm{PIPC} }{\dot{N}_\mathrm{T}^\mathrm{out}},
\end{equation}
where $N_\mathrm{T}^\mathrm{PIPC}$ is the total number of tritium particles in the PIPC, $n_\mathrm{T}^\mathrm{PIPC}$ is the mean tritium density in the PIPC, and $V^\mathrm{PIPC}$ is the PIPC chamber volume. The ratio of the helium and tritium residence times in the PIPC is
\begin{equation}
\frac{\tau_\mathrm{He} }{\tau_\mathrm{T} } =  \frac{n_\mathrm{He}^\mathrm{PIPC}  }{n_\mathrm{T}^\mathrm{PIPC} } \left( \frac{1}{\mathrm{TBE} } -1 \right).
\end{equation}
Rearranging for TBE gives
\begin{equation}
\mathrm{TBE} = \frac{1}{ 1 + \frac{\tau_\mathrm{He} }{\tau_\mathrm{T} } \frac{n_\mathrm{T}^\mathrm{PIPC}  }{n_\mathrm{He}^\mathrm{PIPC} } }.
\label{eq:TBE_PIPC2}
\end{equation}
Equating \Cref{eq:TBE_PIPC2} to \Cref{eq:TBE_PIPC} gives an expression for $R_{\mathrm{sep, HeT} }$,
\begin{equation}
R_{\mathrm{sep, HeT} } = \frac{1}{\Sigma_\mathrm{HeT} } \frac{f_\mathrm{HeT,PIPC}}{f_\mathrm{HeT,div}}  \frac{\tau_\mathrm{T} }{\tau_\mathrm{He}},
\end{equation}
where
\begin{equation}
f_\mathrm{HeT,PIPC} = \frac{n_\mathrm{He}^\mathrm{PIPC}}{n_\mathrm{T}^\mathrm{PIPC}}.
\end{equation}
The effective pumping speed is
\begin{equation}
\Sigma_\mathrm{HeT}^\mathrm{eff} = \frac{f_\mathrm{HeT,PIPC}}{f_\mathrm{HeT,div}} \frac{\tau_\mathrm{T} }{\tau_\mathrm{He}}.
\end{equation}
Therefore, an increase in $\Sigma_\mathrm{HeT}^\mathrm{eff}$ and therefore TBE is achieved by a relative increase in helium-to-tritium density in the PIPC to the divertor, and a relative increase in the tritium-to-helium residence time in the PIPC.

We study the effect of $R_{\mathrm{sep}}$ using a PIPC on an ARC-class power plant. We include an estimate for the new electric power output $P_{e}$ using the constant recirculating power approximation where the recirculating power $P_{\mathrm{recirc}}$ is held constant. The electric power is
\begin{equation}
P_{e} = P_{e,\mathrm{gross} } - P_{\mathrm{recirc}},
\end{equation}
where $P_{e,\mathrm{gross} } = \eta P_f$ is the gross electric power output and $\eta$ is the thermal-to-electric conversion efficiency. In \cite{Meschini2023}, the nominal ARC-class power plant with $P_{f} = 525$ MW had a net electric power of $P_{e} = 200$ MW. Assuming $P_{\mathrm{recirc}} = 36.3$ MW, this gives $\eta = 0.45$, which is the value that we use. In \Cref{fig:PIPC_spider}, we show the effect of four different $R_\mathrm{sep}$ values for a power plant with TBE = 0.30 -- a modest increase in $R_\mathrm{sep}$ from $R_\mathrm{sep} = 1.0$ to $R_\mathrm{sep} = 3.0$ nearly doubles the electric power output and increases plasma gain from $Q = 2.7$ to $Q = 14.3$. Further increases in $R_\mathrm{sep}$ allow further increases in $P_e$ and $Q$.

It should be noted that higher differential pumping of both tritium and helium relative to deuterium is achieved in this process, with the helium-to-deuterium pumping speed being enhanced by $R_\mathrm{sep}^2$ and the tritium-to-deuterium pumping speed being enhanced by $R_\mathrm{sep}$. The present work has identified the significance of the former, while earlier work \cite{Whyte2023} emphasized the critical importance of higher helium pumping rates in achieving high TBE at high fusion power density.

Further work could be done to improve these separation factors through improved engineering design to increase azimuthal velocities while limiting the plasma temperature in the device. A future paper will explore some of these methods and discuss early experimental work in progress to develop this device for applications in fusion energy \cite{rutkowski_adam_partial_nodate}.

In addition to the benefits for differential pumping, the value of a tritium compatible divertor pump with no moving parts should not be understated.  Cryosorption pumps have large disadvantages in fusion systems operating for long periods of time including large tritium inventories, inherently pulsed operation, and large recirculating power cost.  While diffusion pumps have been explored as primary pumps, mercury usage as the working fluid is prohibited by the Minamata convention \cite{noauthor_decision_2022} and while other authors \cite{baus_kyoto_2023} have proposed using liquid lithium as a working fluid, there have been reports of other alkali metals as the working fluid (K and Na) \cite{yagi_potassium_2024}, but not yet demonstrated with Li.  

\begin{table*}
\caption{Key quantities used in this work.}
\begin{ruledtabular}
\centering
  \begin{tabular}{ cccc  }
   Name & Quantity & Units & Equation  \\
    \hline
    Deuterium, tritium, hydrogen flow rate (plasma core) & $\dot{N}_\mathrm{D}^{\mathrm{co}}, \dot{N}_\mathrm{T}^{\mathrm{co}}, \dot{N}_\mathrm{Q}^{\mathrm{co}}$ &s$^{-1}$&  \cref{eq:NdotTco} \\
    Tritium particle flux density & $\Gamma_{\mathrm{T}}$ &m$^{-2}$s$^{-1}$&   \cref{eq:NdotTco} \\
    Deuterium, tritium particle diffusivity & $D_\mathrm{D}, D_\mathrm{T}$ &m$^{2}$s$^{-1}$&   \cref{eq:diffusivetransp} \\
    Deuterium, tritium, hydrogen, electron density (plasma core) & $n_{\mathrm{D}}^{\mathrm{co}}, n_{\mathrm{T}}^{\mathrm{co}}, n_{Q}^{\mathrm{co}}, n_{e}^{\mathrm{co}}$ &m$^{-3}$& \cref{eq:diffusivetransp} \\
    Core tritium density fraction & $f_{\mathrm{T} }^{\mathrm{co} }$ &&  \cref{eq:fTco_def} \\
    Deuterium-tritium particle diffusivity ratio & $\Theta_{\mathrm{DT} } \equiv D_\mathrm{D}/ D_\mathrm{T}$ &&   \cref{eq:ThetaDTratio} \\
    Tritium-hydrogen logarithmic density ratio & $\Lambda$ &&   \cref{eq:Lambda_main} \\
    Tritium injection flow rate fraction & $F_{\mathrm{T} }^{\mathrm{in} } \equiv \dot{N}_\mathrm{T}^{\mathrm{in}}/\dot{N}_\mathrm{Q}^{\mathrm{in}}$ &&  \cref{eq:fueling_ratio} \\
    Tritium divertor removal flow rate fraction & $F_{\mathrm{T} }^{\mathrm{div} } \equiv \dot{N}_\mathrm{T}^{\mathrm{div}}/\dot{N}_\mathrm{Q}^{\mathrm{div}}$ &&  \cref{eq:divertor_ratio} \\
    Tritium, hydrogen flow rate (divertor) & $\dot{N}_\mathrm{T}^{\mathrm{div}}, \dot{N}_\mathrm{Q}^{\mathrm{div}}$ &s$^{-1}$&  \cref{eq:fueling_ratio} \\
    Tritium flow rate enrichment & $H_{\mathrm{T}}$ && \cref{eq:HTfirst} \\
    Fusion power & $P_f$ &W& \\
    D-T fusion energy release & $E$ & J &  \cref{eq:pfform2} \\
    Fusion power density & $p_f$ & W m$^{-3}$&   \cref{eq:pfform2} \\
    D-T fusion reactivity & $\langle v \overline{ \sigma} \rangle$ & m$^{3}$s$^{-1}$&   \cref{eq:pfform2} \\
    Tritium burn rate & $\dot{N}_{\mathrm{T}}^{\mathrm{burn}}$ &s$^{-1}$&  \cref{eq:NTco_2} \\
    Tritium burn fraction & $f_{\mathrm{burn}}$ &&  \cref{eq:NTco_2} \\
    Helium ash removal rate & $\dot{N}_{\mathrm{He}}^{\mathrm{div}}$ &s$^{-1}$&  \cref{eq:TBEzeroth} \\
    Tritium burn efficiency & TBE $\equiv \dot{N}_\mathrm{T}^{\mathrm{burn}}/\dot{N}_\mathrm{T}^{\mathrm{in}}$ && \cref{eq:TBEzeroth} \\
    Divertor neutral pumping speed for species $x$ & $S_x$ &m$^{3}$s$^{-1}$& \cref{eq:neutralpump} \\
    Helium-to-fuel divertor density ratio & $f_{\mathrm{He,div} }$ && \cref{eq:ashtofuel} \\
    Helium-to-electron core density ratio & $f_{\mathrm{dil} }$ && \cref{eq:dil1} \\
    Power density multiplier & $p_{\Delta}$ && \cref{eq:pDeltaform0} \\
    Helium density enrichment & $\eta_{\mathrm{He} }$ && \cref{eq:etaHe} \\
    D-T pumping speed ratio & $ \Sigma_{\mathrm{DT}} \equiv S_{\mathrm{D} }/S_{\mathrm{T} }$ && \cref{eq:SigmaDT} \\
    D-T divertor density ratio & $f_{\mathrm{DT,div}}$ && \cref{eq:fDTdiv} \\
    He-T pumping speed ratio & $ \Sigma_{\mathrm{HeT}} \equiv S_{\mathrm{He} }/S_{\mathrm{T} }$ && \cref{eq:Sigma_HeT} \\
    He-T divertor density ratio & $f_{\mathrm{HeT,div}}$ && \cref{eq:fHeTdiv} \\
    Plasma gain & $Q$ && \cref{eq:corepower} \\
    Energy confinement time & $\tau_E$ &s& \cref{eq:corepower} \\
    Constant plasma gain multiplication factor & $C$ && \cref{eq:C} \\
    Separatrix to divertor tritium fraction flow rate ratio & $\kappa$ &  & \cref{eq:kappa} \\
    Tritium rapid re-injection flow rate fraction & $F_{\mathrm{T} }^{\mathrm{RR} } \equiv \dot{N}_\mathrm{T}^{\mathrm{RR}}/\dot{N}_\mathrm{Q}^{\mathrm{RR}}$ &  & \cref{eq:kappa} \\
    Hydrogen rapid re-injection to divertor flow rate fraction & $R_{\mathrm{Q} } \equiv \dot{N}_\mathrm{Q}^{\mathrm{RR}}/\dot{N}_\mathrm{Q}^{\mathrm{div}}$ &  & \cref{eq:kappa} \\
    Centrifuge separation factor for tritium and deuterium & $R_{\mathrm{sep,TD} }$ &  & \cref{eq:separationfactor} \\
    Centrifuge separation factor for helium-4 and tritium & $R_{\mathrm{sep,HeT} }$ &  & \cref{eq:separationfactorHeT} \\
    Centrifuge separation factor & $R_{\mathrm{sep} }$ &  & \cref{eq:Rsep} \\
  \end{tabular}
\end{ruledtabular}
\label{tab:tab0}
\end{table*}

\section{Discussion} \label{sec:DISCUSS}

We have introduced two new avenues for operating magnetic confinement fusion power plants with very high tritium burn efficiency, fusion power, and fusion gain: (1) controlling the physics parameter $\Theta_{\mathrm{DT}}$ (\Cref{eq:ThetaDTratio}) and (2) the engineering parameter $\Sigma_{\mathrm{DT}}$ (\Cref{eq:SigmaDT}). Increasing $\Theta_{\mathrm{DT}}$ causally improves the tritium burn efficiency without compromising fusion power. Decreasing $\Sigma_{\mathrm{DT}}$ is required when both deuterium and tritium particle transport decrease and a high tritium burn efficiency is desired. Without lower values of $\Sigma_{\mathrm{DT}}$, tritium cannot be exhausted sufficiently quickly for a high tritium burn efficiency to be achieved. This results in a relatively high tritium-to-helium divertor density, corresponding to lower TBE. The result for $\Sigma_{\mathrm{DT}}$ is complementary to \cite{Whyte2023}, which quantified the importance of differential helium-hydrogen pumping for TBE.

Combining high $\Theta_{\mathrm{DT}}$ and low $\Sigma_{\mathrm{DT}}$ values resulted in an eleven-fold improvement in TBE for an ARC-class power plant with constant fusion power (\Cref{fig:arcclass_spider}). While this tritium-lean design has a separatrix density comparable to other fusion systems such as MANTA \cite{Rutherford_2024}, the tritium-lean design's ratio of separatrix deuterium to tritium density is 51:1 -- an extremely high value -- whereas the ratio is roughly 1:1 in most fusion system designs. At very high fixed TBE = 0.600, the high $\Theta_{\mathrm{DT}}$ and low $\Sigma_{\mathrm{DT}}$ values resulted in a fusion power of 447 MW versus 162 MW for the nominal case. Introducing a mass-selective pump with a partial ionization plasma centrifuge could enable significant differences in deuterium and tritium pumping speeds, which is helpful for attaining very high TBE without significant power degradation. We showed how differentially pumping deuterium, tritium, and helium-4 with a rapid re-injection scheme and partial ionization plasma centrifuge can allow significant increases in fusion (and net electric) power for an ARC-class power plant.

We have also shown that low $\Theta_{\mathrm{DT}}$ and high $\Sigma_{\mathrm{DT}}$ values give very poor TBE and fusion power -- avoiding operating such regimes is important for commercially viable fusion power systems.

Combined with other proposals for increasing tritium burn efficiency \cite{Xie_2020,Abdou2021,Boozer2021,Whyte2023,Parisi_2024}, magnetic confinement power plants with very high TBE could be attainable. While we have focused on magnetic confinement fusion, the ideas presented here might lend themselves to other fusion concepts.

It is worth reiterating that while no conclusive evidence exists for very strong asymmetries in particle transport between the main ion species in magnetically confined plasmas, some experimental results hint that strong asymmetries could exist. \cite{Maggi2018} reported a mass-dependent particle diffusivity for hydrogen and deuterium in JET experiments, using a diffusive transport model. Furthermore, \cite{Horvath2021} identified even greater differences in hydrogen and deuterium diffusivities at JET, both at the plasma edge and extending toward the core, with hydrogen diffusivity $D_{\mathrm{H}}$ being up to ten times larger than $D_{\mathrm{D}}$. This finding is encouraging for D-T operations, as it suggests that achieving $\Theta_{\mathrm{DT}} \gg 1$ may be possible with current experimental techniques. Additionally, \cite{Tala2023} found that gyrokinetic transport models could not accurately predict the edge tritium density profiles in JET plasmas, potentially hinting at a mass-dependent particle diffusivity for deuterium and tritium. Analysis of relative D-T transport in the first D-T experiments on JET was inconclusive, though it did not exclude the possibility of a mass dependence in the diffusivities of deuterium and tritium. Although \cite{scott1995isotopic} did not explicitly analyze deuterium and tritium particle diffusivities, it did reveal a strong dependence of electron particle diffusivity $D_e$ on the D-T fuel mixture. Upcoming D-T burning plasma experiments, such as SPARC and ITER \cite{Creely2020,Aymar2002}, as well as ongoing magnetic confinement experiments with multiple main ion species, offer potential opportunities to explore and drive asymmetric D-T transport. The values of $\Theta_{\mathrm{DT}} \gg 1$ shown to be desirable in this paper would requires an extremely strong (and unrealistic with current techniques) dependence on the relative D-T masses, likely requiring new techniques to achieve. In \Cref{sec:TBE_confinement} we presented a heuristic argument that $D_{\mathrm{D}}$ and $D_{\mathrm{T}}$ are coupled, and therefore higher TBE might be achieved by intentionally degrading $D_{\mathrm{D}}$ and/or improving $D_{\mathrm{T}}$.

Finally, it is important to re-emphasize the causal direction between pumping and TBE -- we do not claim that differential divertor pumping of different ion species causes TBE to increase. Rather, differential divertor pumping is a necessary condition for higher TBE if core tritium plasma particle confinement improves. A recent validation of the helium transport model in \cite{Whyte2023}, used in this work, was demonstrated recently using the SOLPS-ITER code \cite{Masline2024}.

\section{Acknowledgements} \label{sec:acknowledgements}

We are grateful to M. Parsons and J. A. Schwartz for helpful discussions. The US Government retains a non-exclusive, paid-up, irrevocable, world-wide license to publish or reproduce the published form of this manuscript, or allow others to do so, for US Government purposes. This work was supported by the U.S. Department of Energy under contract numbers DE-AC02-09CH11466, DE-SC0022270, DE-SC0022272.

\section{Data Availability Statement}

The data that support the findings of this study will be made openly available upon publication.

\appendix

\section{Effect of $\Lambda$} \label{sec:approximation1}

\begin{figure}[!tb]
    \begin{subfigure}[t]{0.8\textwidth}
    \centering
    \includegraphics[width=1.0\textwidth]{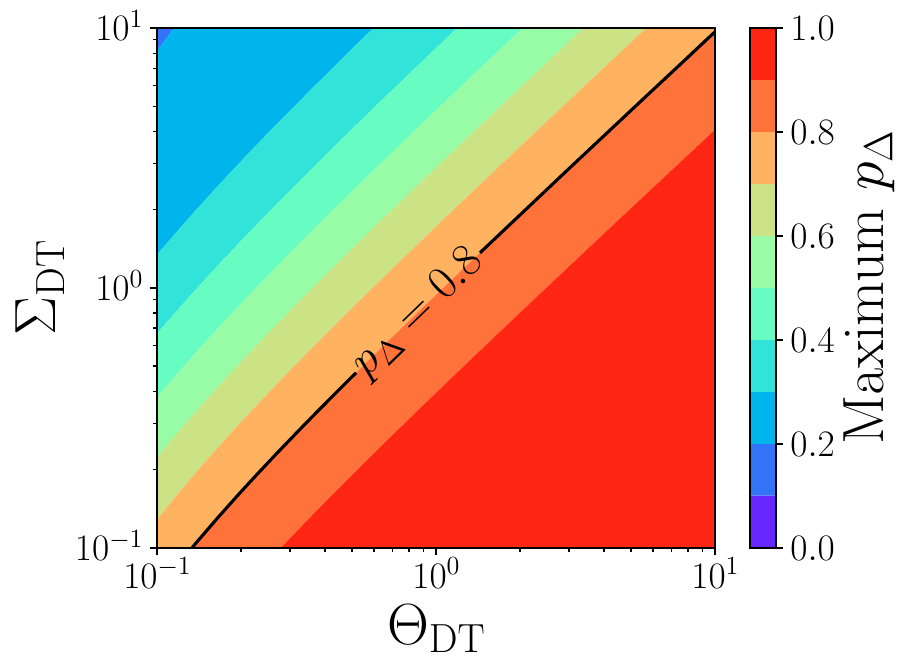}
    \caption{$\Lambda = 4.0$.}
    \end{subfigure}
     ~
    \begin{subfigure}[t]{0.8\textwidth}
    \centering
    \includegraphics[width=1.0\textwidth]{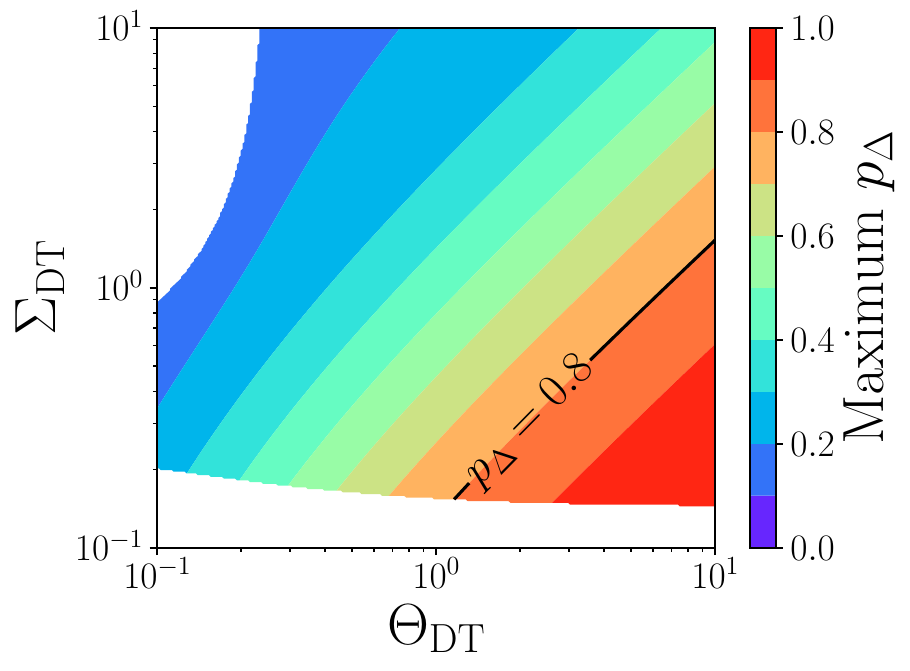}
    \caption{$\Lambda = 1.0$.}
    \end{subfigure}
     ~
    \begin{subfigure}[t]{0.8\textwidth}
    \centering
    \includegraphics[width=1.0\textwidth]{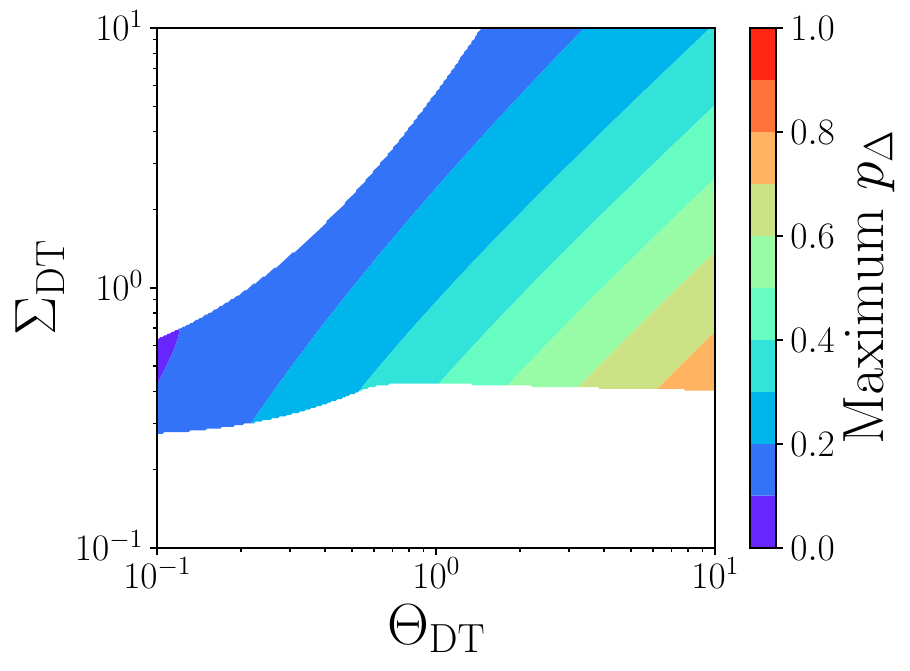}
    \caption{$\Lambda = 0.5$.}
    \end{subfigure}
    \caption{Effect of $\Lambda$ (\Cref{eq:Lambda_main}), $\Theta_{\mathrm{DT}}$, and $\Sigma_{\mathrm{DT}}$ on fusion power multiplier $p_{\Delta}$ at TBE = 0.40. Here, $p_{\Delta}$ is maximized over all $F_{\mathrm{T}}^{\mathrm{in}}$ values.}
    \label{fig:TBE_pDelta_Lambda}
\end{figure}

In this appendix, show the effect of different $\Lambda$ (\Cref{eq:Lambda_main}) values on the total fusion power. Recall that $\Lambda \equiv L_{n,T}/L_{n,Q}$. Therefore, we generally expect that plasmas with lower tritium fraction satisfy $\Lambda \gtrsim 1$ because the total hydrogen densities need to be larger than just for tritium. The maximum $p_{\Delta}$ value is found over $F_{\mathrm{T}}^{\mathrm{in}}$. Shown in \Cref{fig:TBE_pDelta_Lambda}, the maximum $p_{\Delta}$ value is plotted versus $\Theta_{\mathrm{DT}}$ and $\Sigma_{\mathrm{DT}}$. Higher values of $\Lambda$ always lead to higher $p_{\Delta}$ values.

\bibliographystyle{apsrev4-2} %
\bibliography{EverythingPlasmaBib,fuelcycle,DifferentialPumping_AR3} %

\end{document}